\documentclass[apj]{emulateapj}
\usepackage{bookmark}
\usepackage{footmisc}  
\usepackage[usenames,dvipsnames]{xcolor}
\usepackage[flushleft]{threeparttable}
\usepackage{graphicx}	% Including figure files
\usepackage{epsfig}
\graphicspath{ {} {} }
\usepackage{multirow}
\usepackage{amsmath}	% Advanced maths commands
\usepackage{amssymb}	% Extra maths symbols
\usepackage{times}
\usepackage{newtxtext,newtxmath}
\usepackage[T1]{fontenc}
\usepackage{ae,aecompl}
\usepackage{tabularx}% http://ctan.org/pkg/tabularx
\usepackage {enumerate} 
\usepackage[normalem]{ulem}
\usepackage{listings}
\usepackage{rotating,color,array,threeparttable,booktabs,multirow,upgreek}
\def\HII{{H \sc ii}}

\newcommand{\hii}{H{\sc ii}}
\newcommand{\uchii}{UC\,H{\sc ii}}
\newcommand{\hchii}{HC\,H{\sc ii}}

\newcommand{\kms}{\,km\,s$^{-1}$} % kilometres per second
\shorttitle{Massive outflows associated with ATLASGAL clumps}
\shortauthors{Aiyuan Yang et al.}
\begin{document}

\title{Massive Outflows Associated with ATLASGAL Clumps}

\author{
A. Y. Yang\altaffilmark{1,2,3},
M. A. Thompson\altaffilmark{3},
J. S. Urquhart\altaffilmark{4},
W. W. Tian\altaffilmark{1,2,5},
}
\altaffiltext{1}{Key Laboratory of Optical Astronomy, National Astronomical Observatories, Chinese Academy of Sciences, Beijing~100012, China; ayyang@bao.ac.cn}
\altaffiltext{2}{University of Chinese Academy of Science, 19A Yuquan Road, Beijing 100049, China}
\altaffiltext{3}{Centre for Astrophysics Research, School of Physics Astronomy $\&$ Mathematics, University of Hertfordshire, College Lane, Hatfield, AL10 9AB, UK; m.a.thompson@herts.ac.uk}
\altaffiltext{4}{Centre for Astrophysics and planetary Science, University of Kent, Canterbury, CT2 7NH, UK}
\altaffiltext{5}{Department of Physics $\&$ Astronomy, University of Calgary, Calgary, Alberta T2N 1N4, Canada}
\begin{abstract}

We have undertaken the largest survey for outflows within the Galactic plane using simultaneously observed $\rm{^{13}CO}$ and $\rm C^{18}O$ data. Out of a total of 919 ATLASGAL clumps, 325 have data suitable to identify outflows, and 225 ($69\pm3\%$) show high-velocity outflows. The clumps with detected outflows show significantly higher clump masses ($ M_{\rm clump}$), bolometric luminosities ($ L_{\rm bol}$), luminosity-to-mass ratios ($L_{\rm bol}/M_{\rm clump}$ ), and peak $\rm H_2$ column densities ($ N_{\rm H_2}$) compared to those without outflows. Outflow activity has been detected within the youngest quiescent clump (i.e., $70\rm \mu m$ weak) in this sample, and we find that the outflow detection rate increases with $ M_{\rm clump}$, $ L_{\rm bol}$, $L_{\rm bol}/M_{\rm clump}$ , and $ N_{\rm H_2}$, approaching 90\% in some cases  (\uchii\ regions = $93\%\pm3\%$; masers = $86\%\pm4\%$; \hchii\ regions = 100\%). This high detection rate suggests that outflows are ubiquitous phenomena of MSF (MSF). 
The mean outflow mass entrainment rate implies a mean accretion rate of $\rm \sim10^{-4}\,M_\odot\,yr^{-1}$, in full agreement with the accretion rate predicted by theoretical models of MSF. Outflow properties are tightly correlated with $ M_{\rm clump}$, $ L_{\rm bol}$ and $L_{\rm bol}/M_{\rm clump}$ , and show the strongest relation with the bolometric clump luminosity. This suggests that outflows might be driven by the most massive and luminous source within the clump. The correlations are similar for both low-mass and high-mass outflows over 7 orders of magnitude, indicating that they may share a similar outflow mechanism. 
Outflow energy is comparable to the turbulent energy within the clump; however, we find no evidence that outflows increase the level of clump turbulence as the clumps evolve. This implies that the origin of turbulence within clumps is fixed before the onset of star formation.

\end{abstract}


\section{Introduction}
%%%%%
Star formation is an intrinsically complex process involving the collapse and accretion of matter onto protostellar objects but also the loss of mass from the star-forming system in the form of bipolar outflows \citep{Lada1985ARAA23}. Outflows from newly formed stars inject momentum and energy into the surrounding molecular cloud at distances ranging from a few au to up to tens of pc away from the star \citep{Arce2007prplconf245A}.  Molecular outflows are thus one of the earliest observable signatures of both low- and high-mass star formation \citep{Shepherd1996ApJ472S,Kurtz2000prpl299K,Beuther2002AA383B,Molinari2002ApJ570M,Wu2004AA426W}. The first detection of outflows was in 1976 \citep{Kwan1976ApJ210L,Zuckerman1976ApJ209L}. Since then, carbon monoxide (CO) emission lines from single-dish and interferometer observations have been widely used to identify outflows \citep[e.g.,][]{Arce2007prplconf245A,deVilliers2014MNRAS444,Maud2015MNRAS645M}. Outflows can be identified as CO lines showing high-velocity wings, with two spatially separated lobes that are respectively blue and red velocity shifted \citep{Snell1980ApJ239L}. 
%%%

Molecular outflows are thus a useful tool to improve our understanding of the underlying formation process of stars of all masses \citep{Arce2007prplconf245A}, in particular for high-mass stars ($\rm >8\,M_\odot$). For low-mass stars, 
bipolar outflows driven by accretion disks are basic building blocks of the formation process verified in theoretical models \citep{Shu1987ARAA23S} and in observations \citep[e.g.,][]{Bachiller1996ARAA111B,Bontemps1996AA311,Richer2000867R,Arce2007prplconf245A,Hatchell2007AA187H}. 
 However, the formation process of massive stars is still very much under debate \citep{Tan2014prplconf149T}, with two major competing models: (i) core accretion via disk \citep{Yorke2002ApJ846Y,McKee2003ApJ850M} 
 and (ii) competitive accretion \citep{Bonnell2001MNRAS785B}.  The former can be subdivided into two main categories: (a) increased spherical accretion rates via turbulent cores to overcome the radiation pressure \citep{McKee2003ApJ850M} and (b) accretion via a disk that allows the beaming of photons to escape along the polar axis (the so-called flashlight effect) to alleviate the limit of radiation \citep{Yorke2002ApJ846Y}.
The easiest way to discriminate the two models  of ``accretion via disk" and ``competitive accretion'' might be the detection of the accretion disk around massive protostars; however, these can be difficult to detect, as the accretion disk is small, short-lived, and easily confused with the circumstellar envelope \citep{Kim2006ApJ643}. If massive stars do form via an accretion disk, as low-mass stars do, they should generate massive and powerful outflows similar to those seen toward low mass stars \citep{Zhang2001ApJ552L,deVilliers2014MNRAS444}. Thus, observing outflows toward massive young stellar objects (YSOs) can be directly used to help shed light on the debate \citep{Kim2006ApJ643}. 
While detailed high angular resolution interferometry is ultimately required to study outflows at sufficient resolution to distinguish between theoretical models, large outflow surveys using heterodyne focal plane arrays \citep[e.g.,][]{deVilliers2014MNRAS444} provide statistically significant samples and are useful finder charts for later interferometric studies.
 
Outflow feedback can also improve our understanding of the origin of turbulence in clouds, but it remains a challenge to quantify the cumulative impact of  the outflow-driven turbulence on molecular clouds \citep{Frank2014prplconf451F}. Observations and simulation have both suggested that outflow-driven turbulence can and cannot have a significant effect on natal core \citep[e.g.,][]{Cunningham2009ApJ816C,Arce2010ApJ1170A,Krumholz2012ApJ71K,Mottram2012MNRAS420M}. Some simulation results indicated that outflow feedback has a smaller impact on high-mass star forming regions \citep[e.g.,][]{Krumholz2012ApJ71K}, but others have suggested that outflows can act to maintain the turbulence in a cloud \citep[e.g.,][]{Cunningham2009ApJ816C}. There exists evidence that outflows have enough power to drive turbulence in the local environment \citep{Arce2010ApJ1170A,Mottram2012MNRAS420M} but not to contribute significantly to the turbulence of the clouds \citep{Arce2010ApJ1170A,Maud2015MNRAS645M,Plunkett2015ApJ22P}. \citet{Frank2014prplconf451F} have reviewed that impact driven by outflows on length scales of disks, envelopes, and clouds. A statistical sample of outflow-harboring cores at different evolutionary stages is needed to understand the effect of outflows on their  parent clumps \citep{Arce2007prplconf245A}. 

Outflow activities have been detected at different evolutionary stages of young stellar objects (YSOs): low-mass YSOs from Class 0 \citep[e.g.,][] {Bontemps1996AA311,Bally2016ARAA491B} to FU Orionis \citep[e.g.,][] {Evans1994ApJ793E,Konigl2011MNRAS757K} and high-mass YSOs from pre-ultracompact \,\hii\ regions \citep[e.g.,][]{Kim2006ApJ643,deVilliers2014MNRAS444} to ultracompact (UC)\, \hii\ regions \citep[e.g.,][]{Qin2008AA361Q,Maud2015MNRAS645M}. With four evolutionary phases of low-mass YSOs from Class 0 to III \citep{Lada1984ApJ610L,Andre1993ApJ122A}, the most powerful CO outflows are detected around the youngest (Class 0) objects \citep{Bachiller1992AARv257B}, and the outflow energy is found to decrease with YSO evolutionary stages \citep{Bontemps1996AA311, Curtis2010MNRAS1516C, Bally2016ARAA491B}. According to an early evolutionary sequences of MSF (MSF): from hot cores to hypercompact regions\, (\hchii) and\, \uchii\  regions \citep[e.g.,][]{Churchwell2002ARAA, Zinnecker2007ARAA481Z}, outflows are thought to be developed from the\, ``hot core" phase \citep{Kurtz2000prpl299K}, just before the\, \uchii\ phase \citep{Shepherd1996ApJ472S,Wu1999732W,Zhang2001ApJ552L, Beuther2002AA383B,Molinari2002ApJ570M}. These early phases of MSF are frequently associated with water and methanol masers \citep[e.g.,][]{Urquhart2011MNRAS1689U,Urquhart2015MNRAS3461U,Caswell2013IAUS79C}, which supports a close association between these masers and outflow activity \citep[e.g.,][] {Codella2004AA615C, deVilliers2014MNRAS444}. Referring to our sample, \cite{koenig2017} and \cite{Urquhart2017arXiv392U} identified an early evolutionary sequence for MSF clumps based on their infrared-to-radio spectral energy distribution (SED), including the youngest quiescent phase (i.e., a starless or prestellar phase with weak $\rm 70\mu m$ emission), protostellar (i.e., clumps with mid-infrared $\rm 24\mu m$ weak but far-infrared bright), YSO-forming clumps (YSO clump; i.e., mid-infrared $\rm 24\mu m$ bright clumps), and MSF clumps (MSF clumps; i.e., mid-infrared $\rm 24\mu m$ bright clumps with an MSF tracer). The earliest quiescent stage has been found to be associated with molecular outflows \citep{Traficante2017MNRAS3882T}. Discussing the outflow properties of a large sample of clumps at different evolutionary stages allows us to study outflow activity as a function of massive YSO (MYSO) evolutionary state.

%%%

%%table%%%%
\begin{table}
\caption {\rm Typical values for low-mass and high-mass outflows}
\begin{tabular}{lll}
\hline
\hline
Parameters  &  Low-mass Outflows{ $^{a}$}  & High-mass Outflows{  $^{b}$} \\
\hline
$ M_{\rm out}$ &  $0.1\sim\,1\,M_{\odot}$ & $\rm 10\sim10^{3}\,M_{\odot}$ \\
\hline
$\dot{ M}_{\rm out}$ &  $ 10^{-7}\sim\,10^{-6}\,M_{\odot}/yr$ & $ 10^{-5}\sim\,10^{-3}\,M_{\odot}/yr$ \\
\hline
$ F_{\rm out}$ &  $ 10^{-6}\sim\,10^{-5}\,\scriptsize M_\odot\,km/s/yr$ & $ 10^{-4}\sim10^{-2}\,\scriptsize M_\odot\,km/s/yr$ \\
\hline
$ L_{\rm out}$ &  $ 0.1\sim\,1\,L_{\odot}$         & $ 0.1\sim100\,L_{\odot}$ \\
\hline
$\ell_{\rm out}$ &  $\rm 0.1\sim\,1\,pc$ & $\rm 0.5\sim2.5\,pc$ \\
\hline
$t_{\rm d}$ &  $\rm (0.1\sim10)\times10^{5}\,yr$         & $\rm (0.1\sim10)\times10^{5}\,yr$ \\
\hline
\hline
\end{tabular}
\begin{tablenotes}
\item
{  Notes: a, E.g., \cite{Bontemps1996AA311,Wu2004AA426W,Arce2007prplconf245A,Hatchell2007AA187H}. 
b, E.g., \cite{Richer2000867R,Beuther2002AA383B,Wu2004AA426W,Zhang2005ApJ625,Kim2006ApJ643,Arce2007prplconf245A,deVilliers2014MNRAS444,deVilliers2015MNRAS119D,Maud2015MNRAS645M}.}
%These typical values are summarized from references in the text.
\end{tablenotes}
\label{tab_outflows_param_summary}
\end{table}    
%%%%%%%end table%%%%

 Bipolar outflows have been extensively studied in low-mass \citep[e.g.,][]{Bachiller1996ARAA111B,Bontemps1996AA311,Hatchell2007AA187H,Bjerkeli2013AA8B} and high-mass sources {  \citep[e.g.,][]{Beuther2002AA383B,Zhang2005ApJ625,deVilliers2014MNRAS444,Maud2015MNRAS645M}}. The typical values of outflow mass ($ M_{\rm out}$), outflow entrainment rate ($\dot{ M}_{\rm out}$), momentum rate ($ F_{\rm out}$), mechanical luminosity ($ L_{\rm out}$), dynamic timescale ($t_{\rm d}$), and average outflow size ($\ell_{\rm out}$) for low- and high-mass objects are summarized in Table\,\ref{tab_outflows_param_summary}. Outflows from massive protostars with typical values \citep[e.g.,][]{Richer2000867R,Beuther2002AA383B,Wu2004AA426W,Zhang2005ApJ625,Kim2006ApJ643,Arce2007prplconf245A,deVilliers2014MNRAS444,deVilliers2015MNRAS119D,Maud2015MNRAS645M} are approximately more than two orders of magnitude greater than typical outflows from low-mass YSOs \citep[e.g.,][]{Bontemps1996AA311,Wu2004AA426W,Arce2007prplconf245A}, with a similar dynamic timescale. The similar correlations between outflow properties and clump mass, and bolometric luminosity over several orders of magnitude suggest that a common driving mechanism may be responsible for all masses and luminosities  \citep{Bontemps1996AA311,Beuther2002AA383B,Wu2004AA426W,Zhang2005ApJ625,Lopez_Sepulcre2009AA499L,deVilliers2014MNRAS444,Matsushita2017MNRAS1026M}. 
%%%

High-velocity outflow structures are common in both low-mass and high-mass YSOs. The occurrence frequency of molecular outflows in low-mass {  YSOs} ranges between 70\% and 90\% \citep{Bontemps1996AA311,Bjerkeli2013AA8B}. 
For massive {  protostars}, \citet{Zhang2001ApJ552L,Zhang2005ApJ625} detected high-velocity gas in 57\% of 69 luminous IRAS sources, and \citet{Codella2004AA615C} showed a similar detection rate of 50\% (39/80) for masers. Higher detection rates of $70\%\sim90\%$ are found in MSF regions \citep[e.g.,][]{Shepherd1996ApJ457S,Beuther2002AA383B,Kim2006ApJ643,Maud2015MNRAS645M}.
Recent studies show  detection rates of 100\% for 11 very luminous YSOs \citep{Lopez_Sepulcre2009AA499L} and 44 methanol masers \citep{deVilliers2014MNRAS444}. This suggests that outflows are ubiquitous phenomena of high-mass and low-mass star formation. However, all of these studies have focused on selected samples, and therefore these high detection rates may not be representative of the general population of embedded massive protostellar sources.
%%%

The physical parameters of the outflows and their relations have also been investigated for massive protostars \citep{Cabrit1992AA274C,Shepherd1996ApJ457S,Beuther2002AA383B,Wu2004AA426W,deVilliers2014MNRAS444}. These studies have proposed a view that massive protostars can drive powerful outflows and further suggested that outflows can provide a link between low- and high-mass star formation scenarios. However, these correlations between outflow parameters are obtained from targeted observations for small samples of luminous or maser sources or massive star-forming regions  \citep{Beuther2002AA383B,Kim2006ApJ643,Lopez_Sepulcre2009AA499L}. 
\citet{Wu2004AA426W} undertook statistical analysis toward a large sample of 139 high-mass objects with outflow detection based on the compilation of data from the literature. However,  \citet{Cabrit1990ApJ530C} proposed that the estimation of outflow parameters could vary over 2$-$3 orders of magnitude depending on the procedures used. Recently, \citet{vanderMarel2013AA556A} proposed a scatter by up to a factor of 5 for the outflow force of low-luminosity embedded sources from different studies. Analyzing compiled data from the literature would thus have a large dispersion due to the differing procedures used by various authors. 
Therefore, a self-consistent statistical analysis toward a large homogeneous sample of molecular outflows is needed to further understand outflow characteristics. 
%%%%

In this paper, we undertake the largest and most unbiased survey of outflows yet carried out by combining the ATLASGAL and CHIMPS surveys. Our search covers all 919 ATLASGAL clumps in the CHIMPS survey region, i.e.~approximately 18 $\rm deg^{2}$ and comprising 325 clumps with known distances and suitable CHIMPS data.  We estimate the physical properties of outflows toward a large sample of massive clumps and discuss the correlations between these parameters, which are crucial in revealing the intrinsic properties and driving mechanism of outflows. Our study benefits from a homogeneous and self-consistent analysis which acts to minimize systematic errors and  allows us to investigate the relationship between outflows and their associated clumps in a much more unbiased manner than previous studies. This paper is organized as follows. Section\,2 describes the ATLASGAL and CHIMPS surveys and displays the sample selection process. Data analysis of the CO spectra, outflow detection, and mapping are described in Section\,3. In Section\,4, we examine the detection statistics of the detected outflows and calculate their physical properties. Differences between clumps that are associated and not associated with outflows are discussed in Section\,5 along with the physical properties of the clumps and their correlation with turbulence and outflow evolution of the clumps. We give a summary and our conclusions in Section 6. 

\section{The surveys and our sample of clumps}
\label{sect:surveys}

\subsection{CHIMPS}

The $^{13}$CO/C$^{18}$O ($J=3\rightarrow 2$) Heterodyne Inner Milky Way plane Survey (CHIMPS) covers a region of $28\degr \lesssim \ell \lesssim 46\degr$ and $|b| \leq 0 \fdg 5$ in the inner Galactic plane \citep{Rigby2016MNRAS456R}  has been carried out using the James Clerk Maxwell Telescope (JCMT). 
The observations have an angular resolution of 15\arcsec\ and velocity resolution of 0.5\kms\,, with a median rms of $\rm \sim 0.6\,K\,channel^{-1}$. This sensitivity corresponds to column densities of $N_{\mathrm{H}_{2}} \sim 3 \times 10^{20}$ 
and $N_{\mathrm{H}_{2}} \sim 4 \times 10^{21}\,$cm$^{-2}$ for $^{13}$CO and C$^{18}$O, respectively. The critical density of  $^{13}$CO and C$^{18}$O is $\gtrsim 10^{4}$\,cm$^{-3}$ at temperatures of $\leq 20$\,K, and so CHIMPS is a good tracer of the higher-density gas associated with star formation. The $^{13}$CO data from CHIMPS \footnote{http://dx.doi.org/10.11570/16.0001 \label{first_footnote}} can also be a useful tool to trace high-velocity structures, because it is less contaminated by other high-velocity motions within star-forming complexes and is less affected by emission from diffuse clouds along the line of sight. The simultaneously observed C$^{18}$O is optically thin compared to $^{13}$CO in the same clump; thus, its peak emission is most likely to associate with the most dense center of the star-forming clump and can therefore be a good tracer of emission emanating from the dense core at the center of the clump.  The CHIMPS data may therefore serve as an excellent resource for detecting molecular outflows toward clumps with MSF. 
%%%

\subsection{ATLASGAL} 

The APEX Telescope Large Area Survey of the Galaxy (ATLASGAL) is an unbiased 870\,$\rm \mu m$ submillimeter (submm) survey that covers the inner Galactic plane ($|\ell| \leq 60$ with $|b| \leq 1.5\degr$). 
ATLASGAL has a resolution of 19\arcsec\ and a typical noise level of 50$-$70\,mJy\,beam$^{-1}$ \citep{Schuller2009AA504}. This submm survey provides the largest unbiased database of dense clumps that can be used as a starting point for detailed studies of large numbers of massive prestellar and protostellar clumps in the Galactic plane. A comprehensive database of $\sim 10,163$ MSF clumps has been compiled \citep[ATLASGAL compact source catalog (CSC);][]{Contreras2013AA45C,Urquhart2014AA568A} that allows us to undertake a blind search for CO outflow activity toward star-forming clumps. Furthermore, the physical properties (e.g., distance, clump mass, column density, bolometric luminosity) of these MSF clumps have been measured by \citet{Urquhart2017arXiv392U}, which allows us to conduct a statistical analysis of correlations between outflow parameters and clump properties for a large and representative sample of MSF clumps.
%%%

\subsection{The clump sample}

The complete region covered by the two surveys is the sky region of CHIMPS 
spanning $28\degr \lesssim \ell \lesssim 46\degr$ and $|b| \leq 0 \fdg 5$. 
There are 919 ATLASGAL clumps in this region \citep{Contreras2013AA45C,Urquhart2014AA568A}. We extract the  $^{13}$CO and C$^{18}$O spectra toward all 919 clumps using data from CHIMPS\footref{first_footnote}. Our outflow search method requires detections in both $^{13}$CO and C$^{18}$O, and we found a final sample of 325 clumps that fulfilled this criterion. The physical properties of 10  clumps are given in Table\,\ref{clump_properties}, with total 325 clumps at Appendix Table\,\ref{total_clump_properties}.

\begin{figure}
%\centering
\begin{tabular}{c}
\includegraphics[width = 0.5\textwidth]{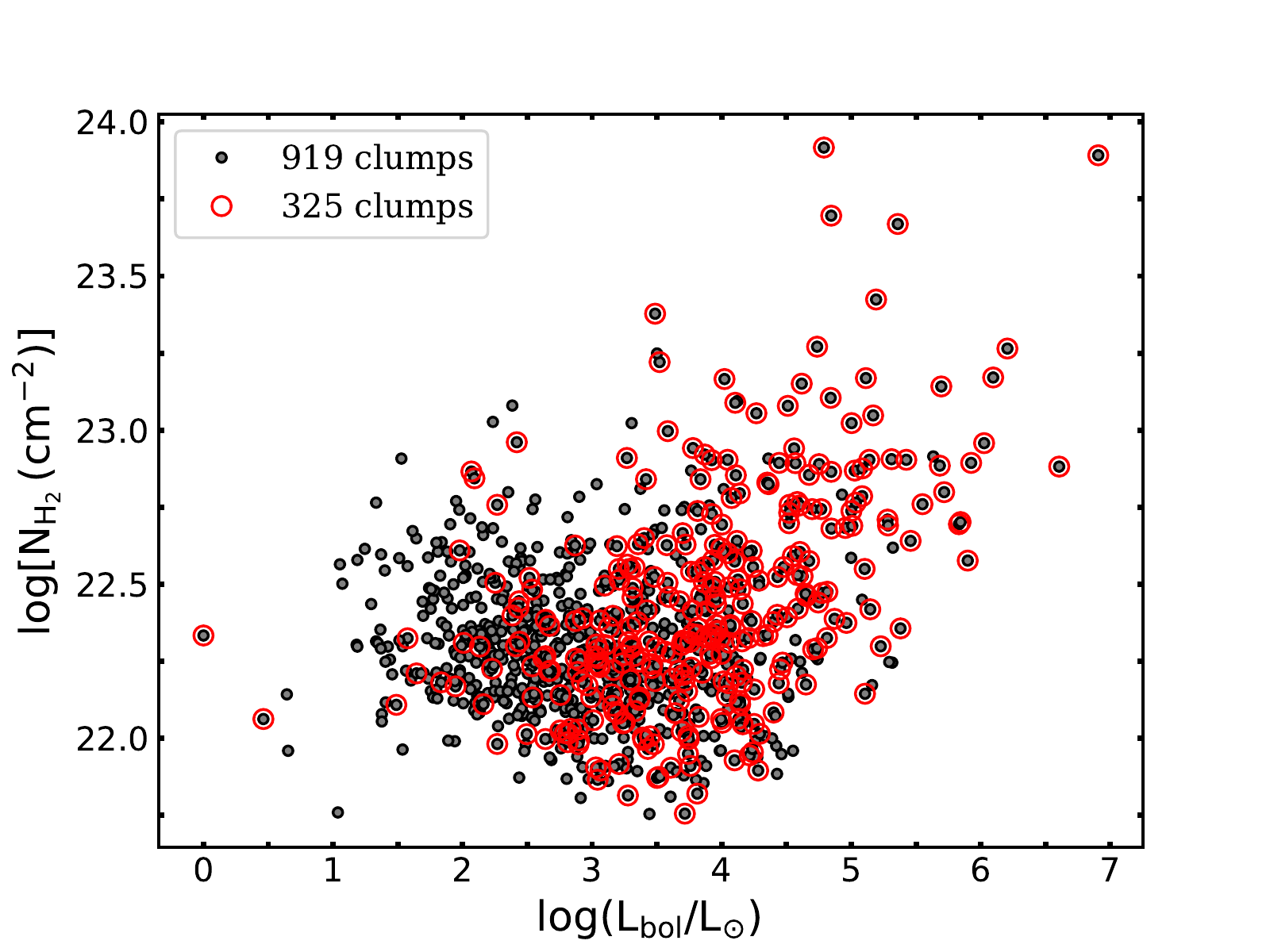}\\
\includegraphics[width = 0.5\textwidth] {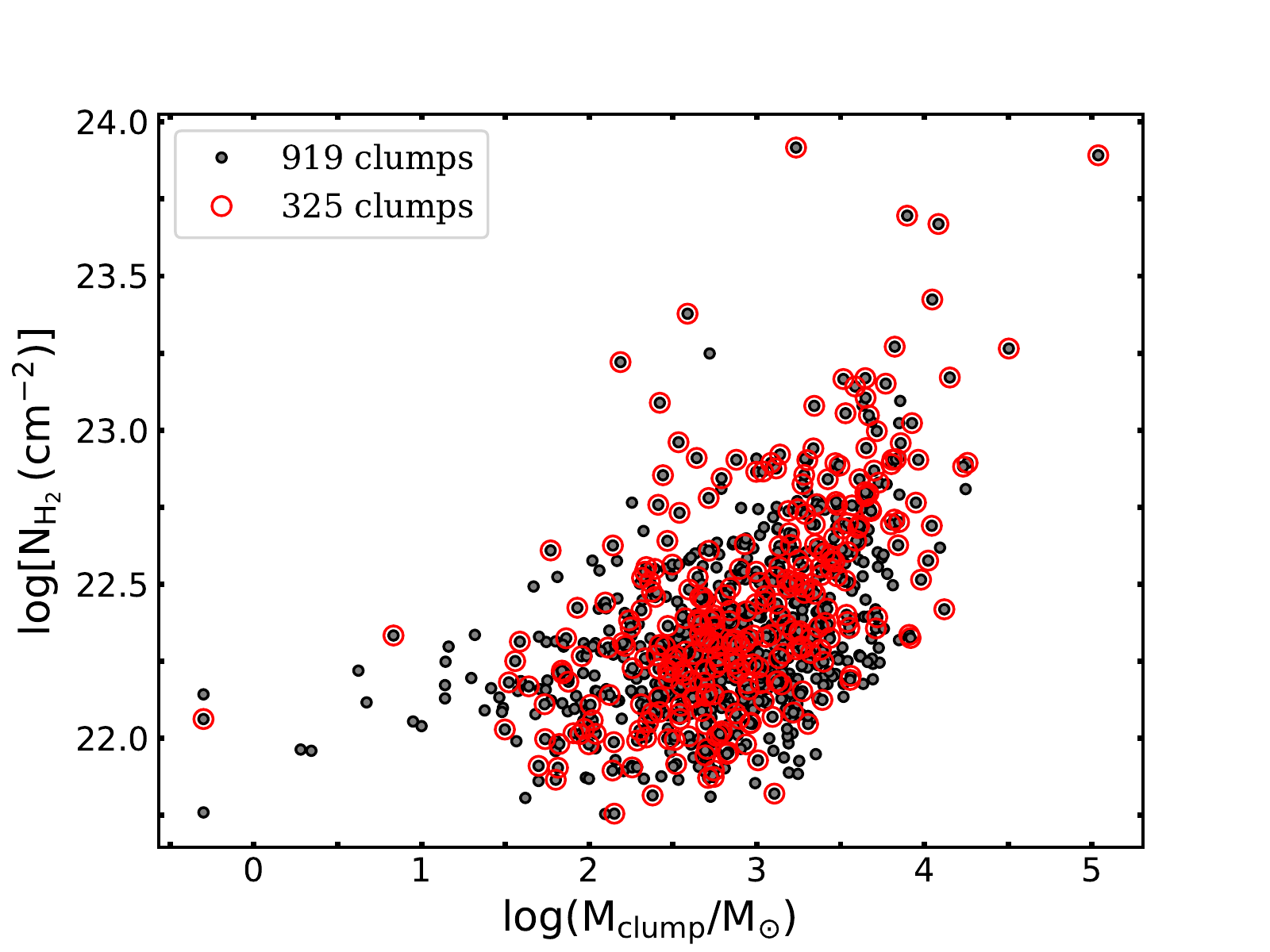}\\
\includegraphics[width = 0.5\textwidth]{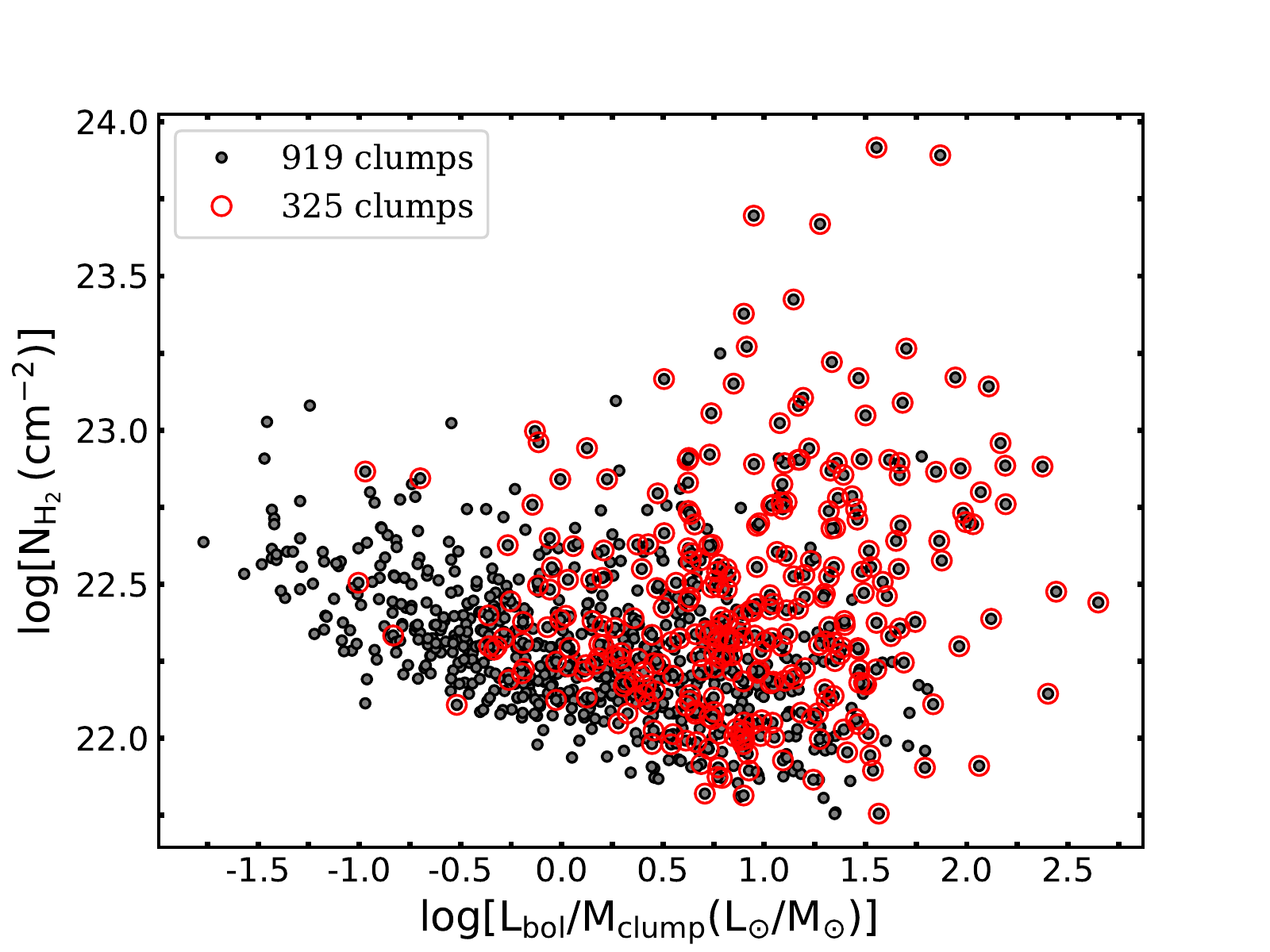}\\  
\end{tabular}
 \caption{ Distributions of $ N_{\rm H_2}$, $ L_{\rm bol}$, $ M_{\rm clump}$, and $L_{\rm bol}/M_{\rm clump}$  in logarithmic scale for the selected 325 ATLASGAL clumps compared to the 919 total clumps. The range of physical parameters of the selected 325 clumps are well covered compared to the whole 919 clumps.}
\label{subsam_distribution}
\end{figure}

In order to show that this sample of clumps is representative of the whole, we plot their physical properties in Figure\,\ref{subsam_distribution}. The plotted quantities are the peak $\rm H_2$ column density ($ N_{\rm H_2}$) against, respectively, clump mass ($ M_{\rm clump}$), bolometric luminosity ($ L_{\rm bol}$), and luminosity-to-mass ratio ($L_{\rm bol}/M_{\rm clump}$ ). These physical properties were measured by \citet{Urquhart2017arXiv392U}.  The average values of 325 clumps of $\rm log(N_{H_2}/cm^{2})=22.45\pm0.36$ with a spread of 21.76$-$23.92, $\rm log(M_{clump}/M_{\odot})=2.93\pm0.64$ with a spread of -0.30$-$5.04, $\rm log(L_{bol}/L_{\odot})=3.8\pm1.0$ with a spread of 1.64$-$6.21, 
and $\rm log[L_{bol}/M_{clump}(L_{\odot}/M_{\odot})]=0.89\pm0.62$ with a spread of -1.0 to 2.65 are shown.

Comparing the means of the two samples, we find that the average values of $ M_{\rm clump}$, $ L_{\rm bol}$, $L_{\rm bol}/M_{\rm clump}$  and $ N_{\rm H_2}$ for the 325 clumps detected in $^{13}$CO and C$^{18}$O  are moderately larger than those of all 919 clumps (see Table\,\ref{tab_summary_param}). Kolmogorov$-$Smirnov (K-S) tests for these two samples suggest that they are from different parent distributions for peak column density (statistic = 0.13, and $p$-value $\ll0.001$), bolometric luminosity (statistic = 0.29, $p$-value $\ll0.001$), and luminosity-to-mass ratio (statistic = 0.33, $p$-value $\ll0.001$). Distributions of the clump mass of the two samples show a much smaller difference and only an 11\% probability  that the two are drawn from the same distribution; i.e.~we cannot exclude the null hypothesis with significance.

Thus, the sample of clumps that forms the basis for our outflow search (i.e. detected in $^{13}$CO and C$^{18}$O) has moderately higher $ N_{\rm H_2}$, $ L_{\rm bol}$, and $L_{\rm bol}/M_{\rm clump}$ , but similar $ M_{\rm clump}$ compared to the total sample, which suggests that the selected clumps are associated with more evolved protostars \citep{Urquhart2017arXiv392U}. Inspecting Figure\,\ref{subsam_distribution} shows that our outflow search sample of clumps covers almost the full observed range of properties in the parent sample, as our sample has a comparable minimum and maximum value of physical parameters with the parent sample (see Table \ref{clump_properties}). We are thus relatively confident that the inferences we draw are valid across the full sample of clumps.

%Nevertheless we should draw to the attention of the reader that our sample are bear in mind the potential for bias towards clumps hosting more evolved star formation, given the higher mean values of $ N_{\rm H_2}$, $ L_{\rm bol}$, $L_{\rm bol}/M_{\rm clump}$  and the K-S test results.

%%%%figure%%%
%%%%%%figure%%
 
%%%%%%%%%%%%%%%%%%%%% 
\section{Data Analysis}

%%%%%%table%%%%%

\setlength{\tabcolsep}{2pt}
\begin{table}
\centering
 \scriptsize
\caption[]{{ \it \rm Clump Properties of All 325 ATLASGAL Clumps to Search for Outflows: Clump's Galactic Name and Coordinates, 
 Integrated Flux Density at 870$\mu m$ ($ F_{\rm int}$), Heliocentric Distance (Dist.), Peak $\rm H_2$ Column Density ($ N_{\rm H_2}$), 
 Bolometric Luminosity ($ L_{\rm bol}$), and Clump Mass ($ M_{\rm clump}$). These value are from \citet{Urquhart2017arXiv392U}. 
 Only a small part of the whole table is presented here, with full version at Appendix Table\,\ref{total_clump_properties}.  }}
\begin{tabular}{lllllllll}
\hline
\hline
ATLASGAL          & $\ell$  & $b$ & $ F_{\rm int}$ & Dist. &  $ logN_{\rm H_2}$   & $ logL_{\rm bol}$& $ logM_{\rm clump}$\\
CSC Gname            & ($\rm \degr$)  & ($\rm \degr$) & ($\rm Jy$) & (kpc) & ($\rm cm^{-2}$) & ($ L_\odot$)&($ M_{\odot}$) \\

\hline
  G027.784$+$00.057   &     27.784    &      0.057      &      9.11     &      5.9          &      22.578      &      3.9         &     3.2          \\
  G027.796$-$00.277   &     27.796    &      $-$0.277   &      4.48     &      2.9          &      22.36       &      3.1         &     2.2          \\
  G027.883$+$00.204   &     27.883    &      0.204      &      9.16     &      8.3          &      22.19       &      3.3         &     3.6          \\
  G027.903$-$00.012   &     27.903    &      $-$0.012   &      8.36     &      6.1          &      22.437      &      4.2         &     3.1          \\
  G027.919$-$00.031   &     27.919    &      $-$0.031   &      2.11     &      3.0          &      21.866      &      3.0         &     1.8          \\
  G027.923$+$00.196   &     27.923    &      0.196      &      7.02     &      8.3          &      22.125      &      3.4         &     3.4          \\
  G027.936$+$00.206   &     27.936    &      0.206      &      7.48     &      2.7          &      22.416      &      3.4         &     2.3          \\
  G027.978$+$00.077   &     27.978    &      0.077      &      9.49     &      4.5          &      22.381      &      4.2         &     2.8          \\
  G028.013$+$00.342   &     28.013    &      0.342      &      1.76     &      8.3          &      21.872      &      3.5         &     2.7          \\
  G028.033$-$00.064   &     28.033    &      $-$0.064   &      1.98     &      6.1          &      22.133      &      3.0         &     2.6          \\
\hline
\hline
\end{tabular}
\label{clump_properties}
\end{table}
\setlength{\tabcolsep}{3pt}
%%%%%%%%%%%%%%%%%%%%%end table%%%%%% 

%%%%%%%%%%%%%%%%%%%table%%%
\begin{table*}
%\scriptsize
\centering
\caption[]{\it \rm $\rm{^{13}CO}$ Outflow Calculations of All Blue and Red Wings for 225 ATLASGAL Clumps: 
 observed peak $\rm{^{13}CO}$ and $\rm C^{18}O$ velocities, the antenna temperatures 
 are corrected for main-beam efficiency (0.72), the velocity range $\rm \Delta{V_{b/r}}$ for blue and red wings of $\rm{^{13}CO}$ spectra, the maximum projected velocity for blue and red shifted $\rm V_{max_{b/r}}$ relative to the peak $\rm C^{18}O$ velocity. Only a small part of the table is presented here, with full version at Appendix Table\,\ref{total_outflow_wings}.}
\begin{tabular}{lllllllll}
\hline
\hline
ATLASGAL            &$\rm{^{13}CO}\,v_p$ & $\rm{^{13}CO}\,T_{mb}$ & $\rm C^{18}O\,v_p$ & $\rm C^{18}O\,T_{mb}$ & $\rm \Delta{V_b}$ &  $\rm \Delta{V_r}$ &  $\rm V_{max_{b}}$ & $\rm V_{max_{r}}$ \\
CSC Gname          &   ($\rm km\,s^{-1}$)    & (K)                                      &   ($\rm km\,s^{-1}$)    & (K) &   ($\rm km\,s^{-1}$) &   ($\rm km\,s^{-1}$)  &   ($\rm km\,s^{-1}$)    &   ($\rm km\,s^{-1}$) \\ 
\hline
    G027.784$+$00.057   &    101.2      &    5.9     &    100.8      &    1.8     &    [96.3,100.8]    &    [103.8,104.8]   &    4.5             &    4.0             \\
  G027.903$-$00.012   &    97.9       &    6.3     &    97.5       &    4.9     &    [95.3,96.8]     &    [98.8,100.3]    &    2.2             &    2.8             \\
  G027.919$-$00.031   &    47.6       &    6.1     &    47.7       &    3.7     &    [46.3,46.8]     &    [48.3,49.8]     &    1.4             &    2.1             \\
  G027.936$+$00.206   &    42.3       &    6.2     &    42.0       &    2.3     &    [37.3,40.3]     &    [43.8,46.8]     &    4.7             &    4.8             \\
  G027.978$+$00.077   &    74.7       &    4.2     &    75.3       &    2.9     &    [71.8,73.3]     &    [76.8,79.3]     &    3.5             &    4.0             \\
  G028.148$-$00.004   &    98.6       &    4.0     &    98.5       &    3.1     &    [96.3,97.8]     &    [99.8,100.8]    &    2.2             &    2.3             \\
  G028.151$+$00.171   &    89.7       &    4.8     &    89.6       &    2.1     &    [86.8,88.8]     &    [90.8,92.3]     &    2.8             &    2.7             \\
  G028.199$-$00.049   &    96.3       &    6.8     &    95.6       &    3.6     &    [89.3,95.8]     &    [98.3,107.3]   &    6.3             &    11.7            \\%
  G028.231$+$00.041   &    107.0      &    3.3     &    107.0      &    1.2     &    [104.8,105.8]   &    [107.3,110.3]   &    2.2             &    3.3             \\
  G028.234$+$00.062   &    107.1      &    4.9     &    107.0      &    1.8     &    [104.8,105.8]   &    [107.8,108.8]   &    2.2             &    1.8             \\

  \hline
\hline
\end{tabular}
\label{outflow_wings}
\end{table*}    
%%%%%%%%%%%%%%%%%%end table2%%%%
%%%%%%%%%%%%%%%%%%%%%%%%

\subsection{$^{13}$CO Spectrum Extraction and Outflow Wing Identification} 

There are several studies that have identified high-velocity outflows in $^{12}$CO toward MSF regions  
\citep{Shepherd1996ApJ457S,Beuther2002AA383B,Wu2004AA426W,Zhang2005ApJ625}. In addition, $^{13}$CO has been shown to be a useful tool to detect molecular outflows because it can trace high-velocity gas in croweded high-mass star-forming regions where $^{12}$CO can be seriously affected by confusion \citep{Codella2004AA615C,Arce2010ApJ1170A}. The simultaneous observation of $\rm{C^{18}O}$ emission,  which is more optically thin, and can be a good tracer of the dense cores of targets \citep{Codella2004AA615C,deVilliers2014MNRAS444}. 
In this work, we extract $^{13}$CO and C$^{18}$O spectra from CHIMPS data cubes {  of an area of clump size at peak emission of each ATLASGAL clump} to identify outflow activity. 
%%%

The detailed strategy of identifying high-velocity outflow wings used in this study is essentially the same as that described by \citet{deVilliers2014MNRAS444}, which has been developed 
from the work of \citet{vanderWalt2007AA464} and \citet{Codella2004AA615C}. 
Here we give a brief description of the method employed to identify outflow wings; for more details, please see \citet{deVilliers2014MNRAS444}. 

\begin{figure*}
 \centering
     \begin{tabular}{cc}
   \includegraphics[width = 0.50\textwidth]{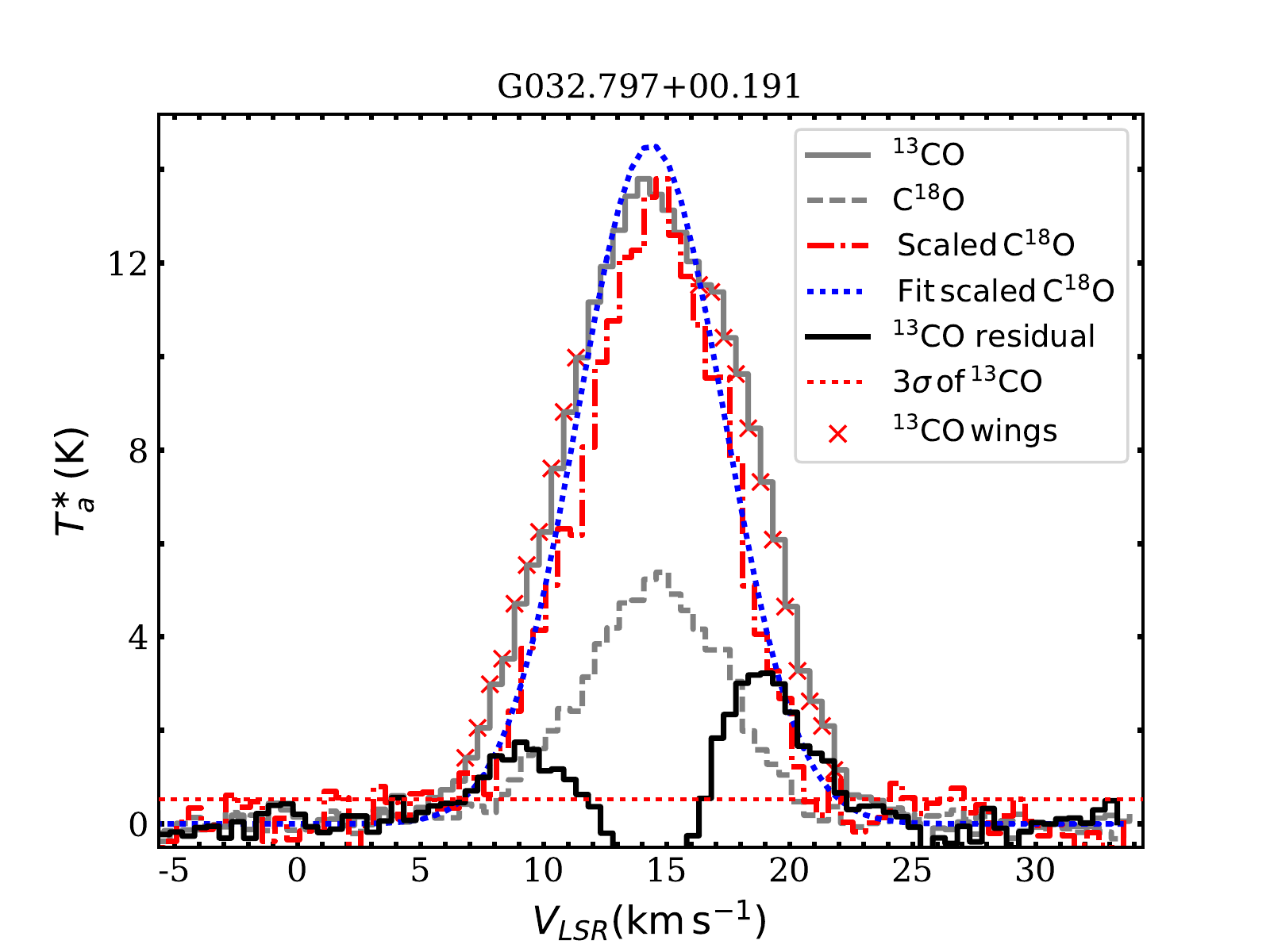}&\includegraphics[width = 0.50\textwidth]{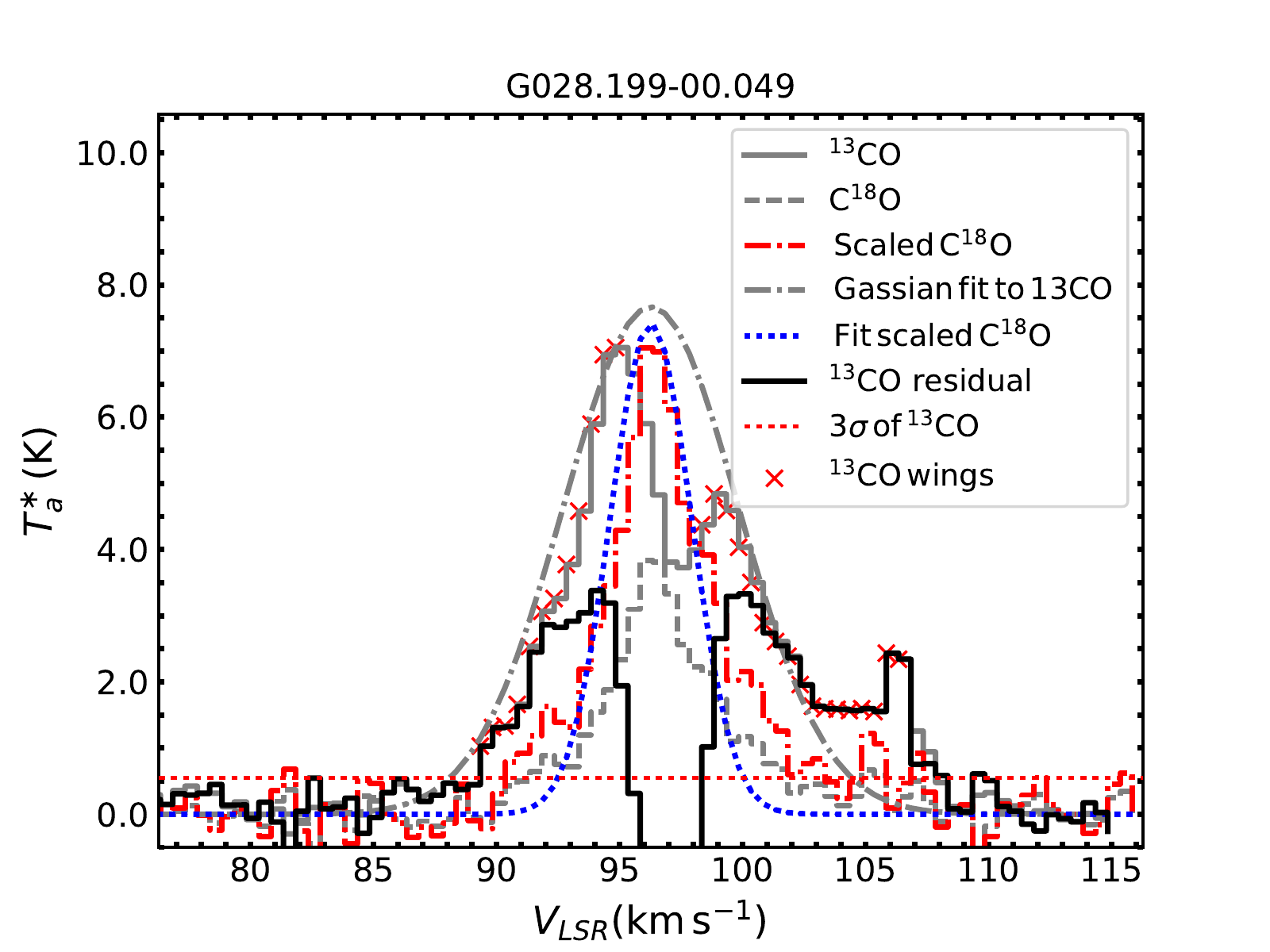}\\  
      \end{tabular}
 \caption{Left panel: an example of outflow wing selection by using spectra of the $\rm ^{13}CO$  (gray solid line) and $\rm C^{18}O$  (gray dashed line) for the ATLASGAL clump G032.797$+$00.191. Blue wings and red wings identification process: (a) scaling the $\rm C^{18}O$ spectrum to $\rm ^{13}CO$ peak, shown by the red dash-dotted line; (b), fitting a Gaussian to the scaled $\rm C^{18}O$, shown by the blue dotted line; (c), obtaining the $\rm ^{13}CO$ residuals spectra (black solid line) by subtracting the Gaussian fit to scaled $\rm C^{18}O$ (red dash-dotted) from $\rm ^{13}CO$ (gray solid line); (d) blue wings ($\rm 6.8-11.8\,km\,s^{-1}$) and red wings ($\rm 16.3-21.8\,km\,s^{-1}$), shown as red cross symbols, can be determined from where the$\rm ^{13}CO$ residuals are larger than the 3$\sigma$ line. Here $\sigma$ is the noise level of the emission-free spectrum. Right panel: example of outflow wing selection toward ATLASGAL clump G028.199$-$00.049 in which the $\rm ^{13}CO$ profile shows clear evidence of self-absorption. First, a Gaussian fit to the shoulders of the $\rm ^{13}CO$ spectra (gray dash-dotted line) and the fitted Gaussian is used as the true peak temperature of the  $\rm ^{13}CO$. This indicates the expected actual peak for the scaled $\rm C^{18}O$ spectra. Then, following the above procedures (a)$-$(d) blue wings ($\rm 89.3-95.8\,km\,s^{-1}$) and red wings ($\rm 98.3-107.3\,km\,s^{-1}$) are thus determined for G028.199$-$00.049. For more details, please see Figures 2 and 3 in \citet{deVilliers2014MNRAS444}.}
\label{wingselection}
\end{figure*}

We illustrate the basic steps in the procedure in Figure\,\ref{wingselection}. Starting from the observed spectra of $^{13}$CO ( gray solid line) and C$^{18}$O ( gray dashed line) obtained at the peak position of the ATLASGAL clump, the basic procedures to identify outflow wings are (a) scaling the $\rm{C^{18}O}$ lines to the peak temperature of $^{13}$CO, shown by the in red dash-dotted line; (b) fitting a Gaussian to the scaled the $\rm{C^{18}O}$ spectra, 
shown as the blue dotted line; (c) obtaining the $\rm ^{13}CO$ residuals spectra ( black solid line), by subtracting the scaled Gaussian fit $\rm{C^{18}O}$ ( red dash-dotted line) from the $\rm{^{13}CO}$ ( gray solid line); and (d) identifying the blue and red line wings (red cross symbols) where the $^{13}$CO residual is larger than 3$\sigma$, where $\sigma$ is the noise level of the emission-free spectrum. The line wings are defined by the velocity where the $\rm ^{13}CO$ profile is broader than the scaled Gaussian $\rm C^{18}O$ profile (core-only emission). In order to avoid subtracting any emission from high-velocity structures that may be included in the scaled $\rm C^{18}O$ profile, a Gaussian was fitted to the scaled $\rm C^{18}O$, by gradually removing points from the outer high-velocity edges until the $\rm C^{18}O$ spectra could be fitted, as suggested by \citet{vanderWalt2007AA464} and \citet{deVilliers2014MNRAS444}. Following the above procedures, blue wings ($\rm 6.8-11.8\,km\,s^{-1}$) and red wings ($\rm 16.3-21.8\,km\,s^{-1}$) for the emission spectra of the ATLASGAL clump were determined via custom-written scripts using Astropy, a Python package for Astronomy \citep{Astropy2013AA33A}; (see Figure\,\ref{wingselection}(a) for an example toward the ATLASGAL clump G032.797$+$00.191). 
%%%

For those $\rm ^{13}CO$ profiles showing clear evidence of self-absorption 
(e.g. G028.199$-$00.049 as shown in Figure\,\ref{wingselection}(b)), the method was adjusted as follows. First, a Gaussian fit to the shoulders of its $\rm ^{13}CO$ profile ( gray dash-dotted line in Figure \ref{wingselection}(b)) and the fitted Gaussian peak is used as the true peak temperature of the $\rm ^{13}CO$. This gives an indication of the expected peak to the scaled $\rm C^{18}O$ profile. Then, following the procedures (a)$-$(d), blue wings ($\rm 89.3-95.8\,km\,s^{-1}$) and red wings ($\rm 98.3-107.3\,km\,s^{-1}$) are thus determined for these clumps. For more details, please see Figures 2 and 3 in \citet{deVilliers2014MNRAS444}. 
%%%

This method of searching for outflows is affected by uncertainties due to confusion (the observed sources lie along the Galactic plane where most of the molecular material resides), spectral noise (in the case of weak sources), and outflow geometry (which determines the width of the wings in the profile; \citep{Codella2004AA615C}). A consequence of these limitations is that we might miss some outflows, but given the homogeneity of the present sample and the large number of the observed objects, these results should be representative of the general population and therefore provide an accurate {  picture} of the commonality of outflows and their properties.  
%%%

In total, we find that 225 out of 325 clumps are associated with high-velocity structures based on the method outlined above. The source velocities and blueshifted and redshifted velocity ranges are given in Table\,\ref{outflow_wings}. Ten of  the 225 sources show single red/blue high-velocity wings and the remaining 215 show both blue and red wings. 
% for a small portion of the outflows identified, with the total 225 outflows listed in Appendix Table\,\ref{total_outflow_wings}

\begin{figure*}
\centering
%\begin{tabular}{cc}
\includegraphics[width = 0.49\textwidth, trim=30 0 50 0, clip]{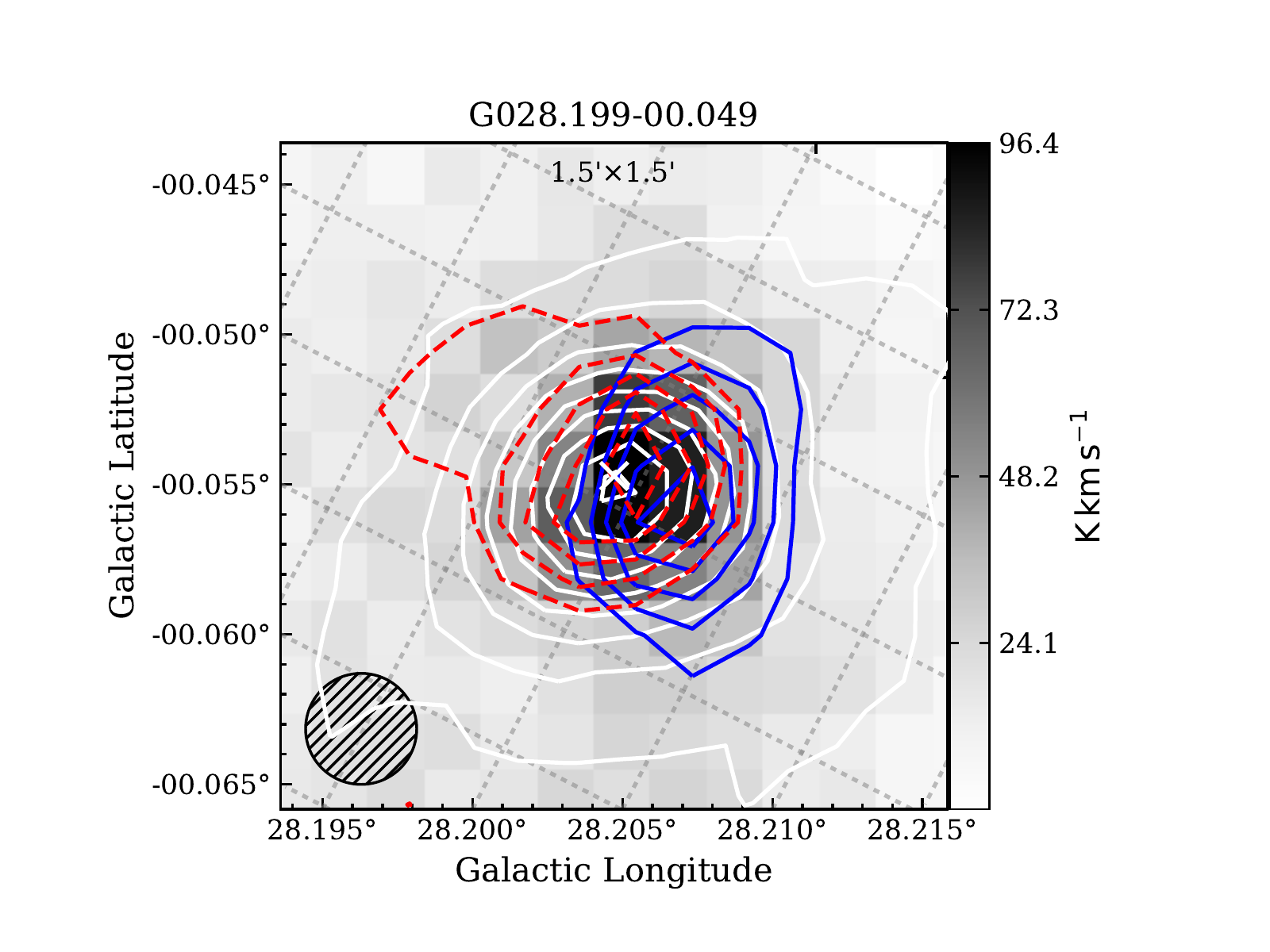}
\includegraphics[width = 0.49\textwidth, trim=30 0 50 0, clip]{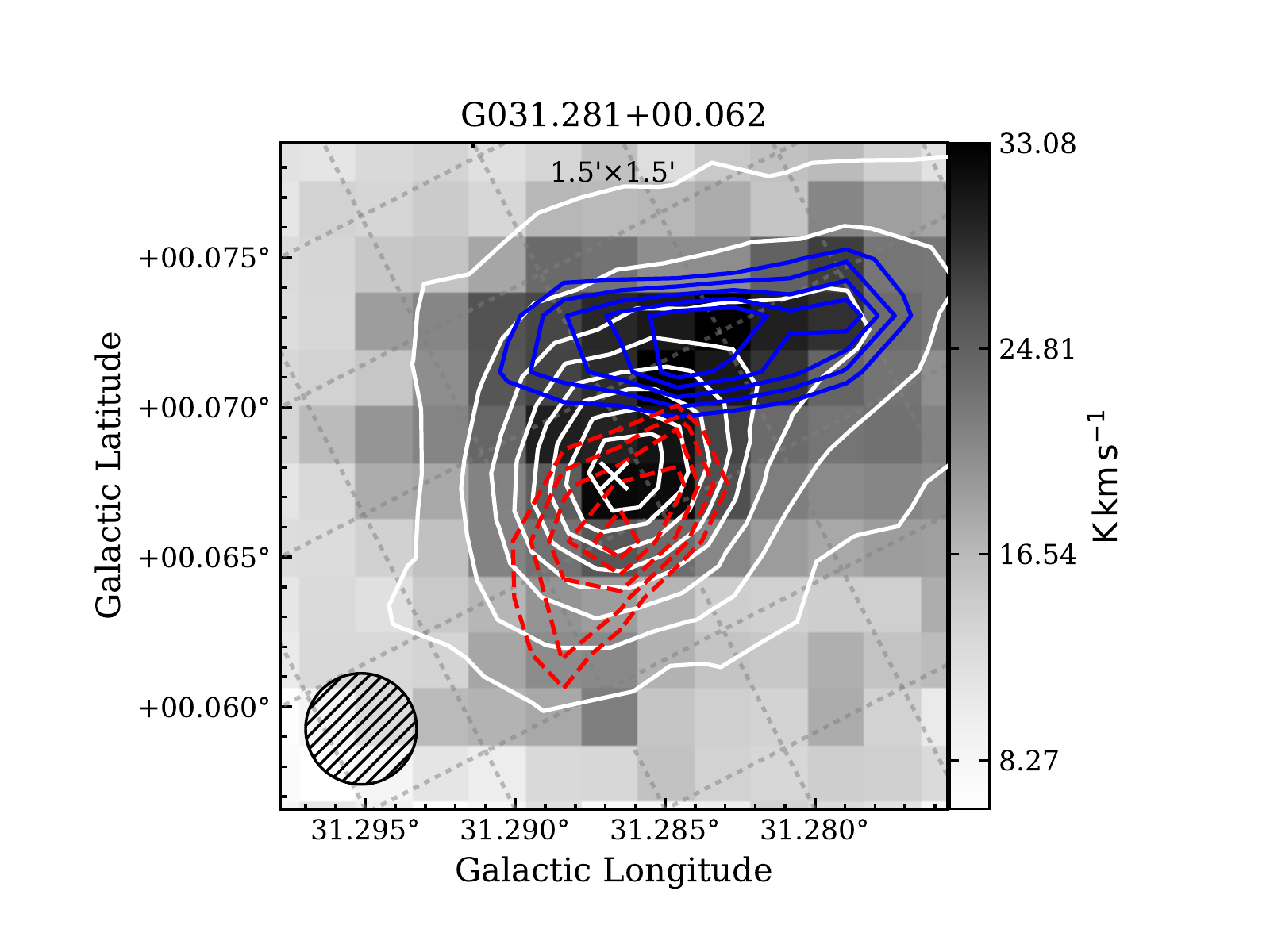}\\  
%\end{tabular} %
 \caption{Examples of the outflow mapping: the intensity integrated image ($1\arcmin .5 \times1\arcmin.5$) of the blue and red wings centered on the symbol of white cross at the ATLASGAL clump G028.199$-$00.049 (left-hand panel) and G031.281$+$00.062 (right-hand panel). Gray-scale images show the $\rm ^{13}CO$ integrated emission with blue wings (blue solid line) and red wings (red dashed lines). These wing emissions are integrated with velocity ranges of $\rm 101.2-104.2\,km\,s^{-1}\,(blue\,wings)$ and $\rm 111.2-112.7\,km\,s^{-1}\,(red\,wings)$ for G031.281$+$00.062 and $\rm 89.3-95.8\,km\,s^{-1}(blue\,wings)$ and $\rm 98.3-107.3\,km\,s^{-1}\,(red\,wings)$ for G028.199$-$00.049. Red and blue contours have been divided by 5/6 levels with a starting value $3\sigma$ or 30\% of maximum intensity and and ending value 95\% of maximum intensity with units of $\rm K\,km\,s^{-1}$. White contours show the  870$\mu m$ ATLASGAL emission with levels determined by a dynamic-range power-law fitting scheme \citep{Thompson2006AA}. The beam of CHIMPS (15\arcsec) is shown by the hatched black circle presented in the lower left of each image. }
\label{outflowmap}
\end{figure*}

Next, we created $^{13}$CO integrated-intensity images of each wing, integrated over the velocity ranges determined in the previous step. This is so that we can spatially separate the wing emission into distinct red and blue outflow lobes and subsequently calculate the physical properties of the outflows using the methods presented in Section\,\ref{outflow_props}. We show two examples in Figure \ref{outflowmap}, where solid blue and dashed red contours representing blue and red outflow lobes are overlaid onto the $\rm{^{13}CO}$ integrated-intensity image ( in gray scale), and the 870\,$\mu m$ emission from ATLASGAL is shown as white contours. The ATLASGAL emission is optically thin and traces the bulk of the dense gas, revealing the column density distribution and the clump centroid. 
%%%

%%%%
%%%
%%%%

As some massive clumps are located in clusters, their outflow properties may have been contaminated by similar high-velocity components from different clumps \citep{Shepherd1996ApJ472S}, and their red and/or blue outflow lobes may be mixed with other high-velocity components from nearby sources in the field of view. We thus exclude 48 clumps where it is difficult to identify their red and blue lobe area, as the contours of the outflow lobes are mixed with a complex environment. In addition, 12 sources show blue and red wings but their integrated emission is too weak to show two outflow lobes on their $\rm{^{13}CO}$ integrated images. 
In summary, we have obtained outflow maps in 155 of our 215 sources, which display well-defined blue and red lobes. Excluding two sources without distances (\citealt{Urquhart2017arXiv392U}), we are left with a final sample of 153 massive clumps with mapped outflows suitable for further analysis. Outflow wing spectra of 225 clumps and $\rm{^{13}CO}$ integrated images of 155 clumps are shown online as supporting information. 
%%%

%%%%%%%%%%%%%%%%%%
\section{Results}
Here we present the results of our outflow search and determine the physical properties of the identified outflow sample. 

%%%%%%%%%%%%%%%%%%
\begin{table*}
\centering
\caption[]{ \it \rm $\rm{^{13}CO}$ Outflow Properties of All Blue and Red Lobes for 153 ATLASGAL Clumps : Blue/Red Lobe Length $ l_{b/r}[\rm pc]$, 
 Mass $ M_{\rm b}(\rm blue)$,  $ M_{\rm r}(\rm red)$, $\rm M_{\rm out}(M_{\rm out} = M_{\rm b} + M_{\rm r}) [M_\odot]$, Momentum $\rm p [10\,M_\odot\,\rm km\,s^{-1}]$, Energy $\rm E [10^{39}\,J]$, Dynamic Time $ t_{\rm d}[10^4\,\rm yr]$, 
 Mass Entrainment Rates $ \dot{M}_{\rm out} [10^{-4}\,M_\odot/\rm yr]$, Mechanical force $ F_{\rm CO} [10^{-3}\,M_\odot\,\rm km\,s^{-1}/yr]$, 
 and Mechanical Luminosity $ L_{\rm CO} [L_\odot]$. Only eight sources are listed here, with full version at Appendix Table\,\ref{total_outflow_phyparam}. }
\begin{tabular}{llllllllllll}
\hline
\hline
ATLASGAL            &$ l_b$ & $ l_r $ & $ M_{\rm b}$ & $ M_{\rm r}$ & $ M_{\rm out}$ & $p$ & E & $ t_{\rm d}$ & $\dot{ M}_{\rm out}$ &  $   F_{\rm CO}$ & $   L_{\rm CO}$  \\
 \footnotesize CSC Gname            &  ($\rm pc$) & ($\rm pc) $ &\scriptsize ($ M_\odot$) &\scriptsize ($ M_\odot$) &\scriptsize ($ M_\odot$) &\scriptsize  ($ 10\,M_\odot\,km\,s^{-1}$) 
 &\scriptsize  ($\rm 10^{39}\,J$) &\scriptsize ($\rm 10^4\,yr$) &\scriptsize ($ 10^{-4}M_\odot/yr$) &\scriptsize ($ 10^{-3}M_\odot\,km\,s^{-1}/yr$) &\scriptsize ($ L_\odot$)  \\
\hline
  G027.784$+$00.057 &  1.1   &  0.6   &  39.4    &  5.4     &  44.8      &  {  20.8}        &  {  2.4}      &  14.8     &   2.9                  &   {  1.2}                &   {  1.2}     \\
  G027.903$-$00.012 &  0.8   &  1.0   &  18.8    &  18.8    &  37.6      &  {  14}         &  {  0.8}      &  24.5     &   1.5                  &   {  0.6}                &  {  0.28}     \\
  G027.919$-$00.031 &  0.5   &  0.5   &  3.0     &  9.5     &  12.4      &  {  4.6}         &  {  0.16}     &  16.0     &   0.7                  &   {  0.2}                &   {  0.08}    \\
  G027.936$+$00.206 &  0.2   &  0.2   &  1.9     &  3.2     &  5.1       &  {  3.8}         &  {  0.4}      &  3.1      &   1.6                  &   { 1.2}                &   {  0.8}     \\
  G027.978$+$00.077 &  0.5   &  1.0   &  7.4     &  13.3    &  20.7      &  {  12.6}         &  { 1.2}      &  16.1     &   1.2                  &   {  0.8}                &   {  0.4}     \\
  G028.148$-$00.004 &  0.5   &  0.6   &  5.4     &  8.5     &  13.9      &  {  5.0}         &  {  0.32}     &  14.6     &   0.9                  &   {  0.4}                &   {  0.16}    \\
  G028.151$+$00.171 &  0.6   &  1.2   &  6.0     &  2.7     &  8.7       &  {  4.0}         &  {  0.28}     &  25.5     &   0.3                  &   {  0.14}               &   {  0.08}    \\
  G028.199$-$00.049 &  0.8   &  1.5   &  83.5    &  86.0    &  169.5     &  {  176.0}        &  {  38.8}      &  9.7      &   16.8                 &  {  16.6}                 &   {  30.8}     \\ 
\hline
\hline
\end{tabular}
\label{outflow_phyparam}
\end{table*}    
%%%%%%%%%%%%%%%%%table%%

\setlength{\tabcolsep}{0pt}
\begin{table}
\scriptsize
\centering
\caption {\rm   Detection rate vs. bins range of $ M_{\rm clump}(M_{\odot})$, $ L_{\rm bol}(L_{\odot})$, $ L_{\rm bol}/M_{\rm clump}(L_{\odot}/M_{\odot})$, and $ N_{\rm H_2}(\rm cm^{2})$}
\begin{tabular}{cl| cl |cl |cl}
\hline
\hline
\multicolumn{2}{c|}{$ logM_{\rm clump}$} &       \multicolumn{2}{c|}{$ logL_{\rm bol}$}  &      \multicolumn{2}{c|}{$ log[L_{\rm bol}/M_{\rm clump}]$} &      \multicolumn{2}{c}{$ logN_{\rm H_2}$} \\
\hline

[-0.30,2.31] & $61\pm7\%$ & [0.0,2.89] & $52\pm8\%$ & [-1.00,0.20] & $52\pm8\%$ & [21.76,22.06] & $61\pm7\%$ \\

[2.31,2.60] & $69\pm7\%$ & [2.89,3.30] & $53\pm7\%$ & [0.20,0.62] & $49\pm7\%$ & [22.06,22.22] & $49\pm7\%$ \\

[2.60,2.78] & $56\pm7\%$ & [3.30,3.70] & $62\pm7\%$ & [0.62,0.78] & $71\pm7\%$ & [22.22,22.32] & $56\pm7\%$ \\

[2.78,3.05] & $57\pm7\%$ & [3.70,3.96] & $70\pm7\%$ & [0.78,0.98] & $69\pm7\%$ & [22.32,22.42] & $55\pm8\%$ \\

[3.05,3.30]  & $76\pm6\%$ & [3.96,4.25] & $73\pm7\%$ & [0.98,1.26] & $82\pm6\%$ & [22.42,22.58] & $76\pm6\%$ \\

[3.30,3.59]  & $69\pm7\%$ & [4.25,4.77] & $80\pm6\%$ & [1.26,1.51] & $78\pm6\%$ & [22.58,22.84] & $89\pm5\%$ \\

[3.59,5.04]  & $96\pm3\%$ & [4.77,6.91] & $91\pm4\%$ & [1.51,2.65] & $82\pm6\%$ & [22.84,23.92] & $96\pm3\%$ \\
\hline
\end{tabular}
\label{tab_detection_rate}
\end{table}    

\subsection{Detection Statistics of Outflows}
\label{sect:detection}
%%%figure%%%
\begin{figure}
\centering
\includegraphics[width = 0.49\textwidth]{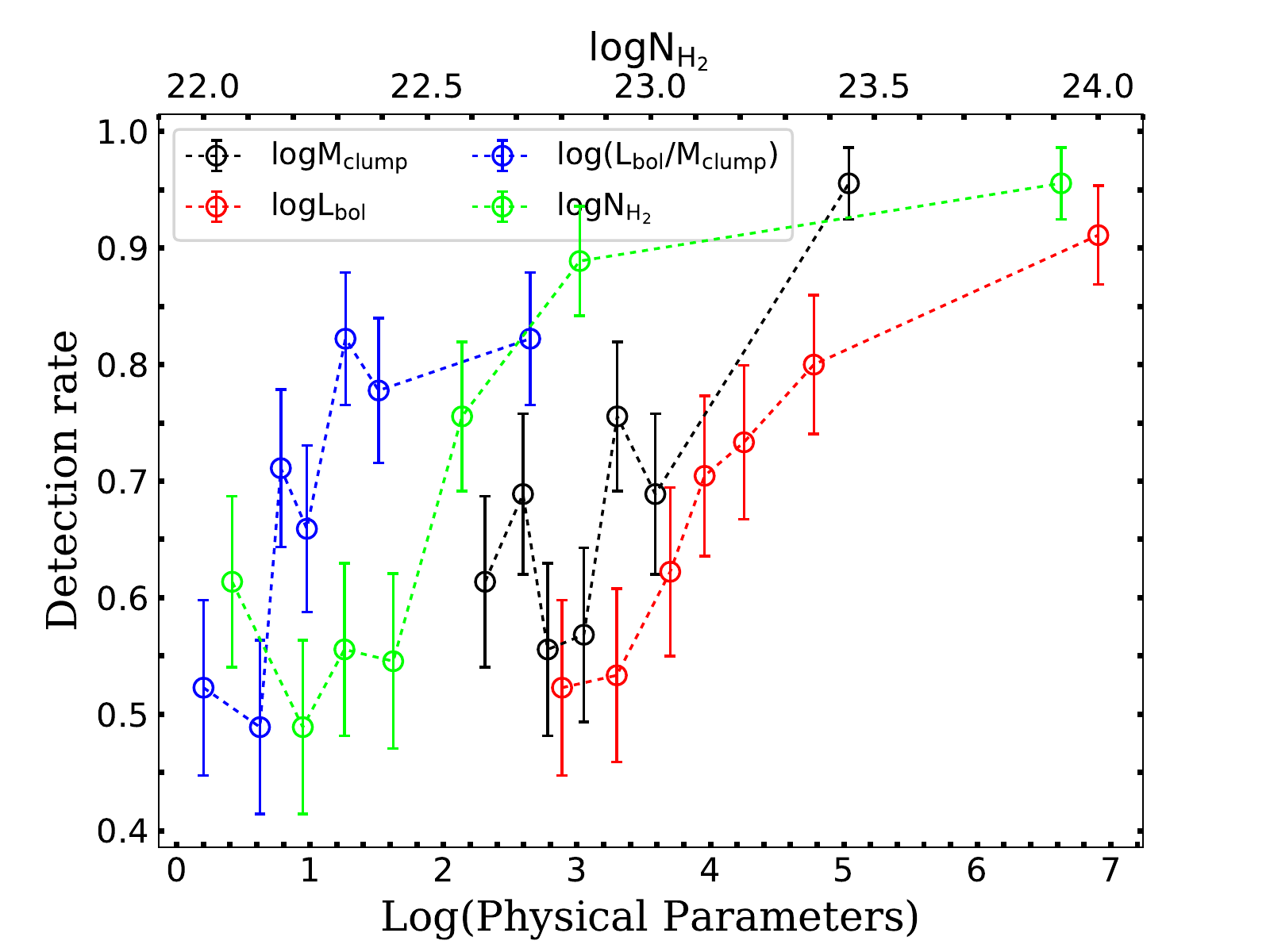}\\  
 \caption{Detection rate as a function of clump mass $\rm M_{clump}\,(M_{\odot})$, 
 bolometric luminosity of central objects $\rm L_{bol}\,(L_{\odot})$, 
 luminosity-to-mass ratio $\rm L_{bol}/M_{clump}\,(L_{\odot}/M_{\odot})$, 
 and the peak $\rm H_2$ column density of clumps $\rm N_{H_2}\,(cm^{-2})$ in logarithmic scales. The values on the x-axis for these parameters correspond to the bin values from the second to the end value, while $ logN_{\rm H_2}$ shows bin values on the top x-axis from the second to the end. The bin values and detection rates are presented in Table \ref{tab_detection_rate}} 
\label{detectionrate_vs_clumpparam}
\end{figure}

Among the 325 clumps in our outflow search sample, 225 of them were found to show high-velocity line wings, resulting in a detection rate of $69\%\pm3\%$ for the whole sample. 
Within the 225 sources that show high-velocity line wings, 10 clumps have a single blue/red wings, and the remaining 215 have both blue and red wings.  The monopolar features of the 10 clumps might be affected by uncertainties due to confusion, spectral noise, and outflow geometry \citep{Codella2004AA615C}.   The detection frequency of bipolar outflows is subsequently reduced to $\sim66\pm3\%$. 

 This detection rate is comparable to that of \citet[66\%]{Maud2015MNRAS645M} and \citet[57\%]{Zhang2001ApJ552L,Zhang2005ApJ625}, which may be due to the similarity of the evolutionary stages of our sample with those of \cite{Maud2015MNRAS645M}$-$ i.e., compact \hii\ regions or MYSOs$-$ and of luminosity with those of \cite{Zhang2001ApJ552L,Zhang2005ApJ625}, i.e., $\rm 10^{2}L_{\odot}\sim 10^{5}L_{\odot}$. This detection rate is slightly larger than that of \citet[39\%$-$50\%]{Codella2004AA615C} partly because they include a number of sources at a later stage of \uchii\ regions when outflows tend to disappear \citep{Codella2004AA615C}. 
Our detection rate is slightly smaller than that of some previous results \citep[e.g.,][]{Shepherd1996ApJ457S,Beuther2002AA383B,Lopez_Sepulcre2009AA499L,deVilliers2014MNRAS444}, {  likely due to the fact that} they are targeted observations toward markers of MSF.

%This result is comparable to the detection rates determined by some previous studies \citep[e.g.,][]{Shepherd1996ApJ457S,Zhang2001ApJ552L,Beuther2002AA383B,Codella2004AA615C,Zhang2005ApJ625,Lopez_Sepulcre2009AA499L,deVilliers2014MNRAS444,Maud2015MNRAS645M}. 

 ATLASGAL clumps were classified into four types of evolutionary sequence based on their infrared-to-radio SED by \citet{koenig2017} and \citet{Urquhart2017arXiv392U}, including the youngest quiescent phase (i.e., a starless or prestellar phase with $\rm 70\mu m$ weak), protostellar (i.e., clumps with mid-infrared $\rm 24\mu m$ weak but far-infrared bright), YSO-forming clumps (YSO clumps; i.e., mid-infrared $\rm 24\mu m$ bright clumps), and MSF clumps (i.e., mid-infrared $\rm 24\mu m$ bright clumps with a MSF tracer). Among our outflow search sample of 
325 clumps, with the exception of six clumps that have not yet
been classified, there are 125 MSF, 171 YSO, 19 protostellar, and four quiescent clumps. We detect outflow line wings toward  102 (102/125; $82\%\pm3\%$) MSF clumps, 105 (105/171; $61\%\pm4\%$) YSO clumps, 10 protostellar clumps (10/19; $53\%\pm11\%$), and 2 quiescent clumps (2/4; $50\%\pm25\%$) respectively.

Looking at the MSF subsample in more detail, there are 56 clumps associated with
\uchii\ regions from \citet{Urquhart2013MNRAS}, 52 of which are found to have high-velocity line wings ($93\%\pm3\%$). Four clumps are associated with four hypercompact (HC) \HII~regions \citep{Sewilo2004ApJ,Keto2008ApJ672,Sewilo2011ApJS,Zhang2014ApJ}, of which 100\% show high-velocity line wings. Seventy clumps are associated with maser (water or methanol) emissions \citep{Codella2004AA615C,Urquhart2014MNRAS1555U,deVilliers2014MNRAS444,Urquhart2017arXiv392U} and  
60 of the 70 maser-associated clumps ($86\%\pm4\%$) are associated with high-velocity line wings, which is consistent with the detection rate for maser-associated sources in \citet{Codella2004AA615C} and \citet{deVilliers2014MNRAS444}. These high detection rates confirm that outflows are a common feature in the early stages of MSF, as suggested in many previous studies 
\citep{Shepherd1996ApJ472S,Kurtz2000prpl299K,Beuther2002AA383B,Molinari2002ApJ570M,Wu2004AA426W}.

 Among the 325 total clumps and 225 
 outflow-associated clumps, 314 and 216, respectively, have measured distances and hence physical parameters from \citet{Urquhart2017arXiv392U}. We are therefore able to examine the detection rate as a function of clump mass ($ M_{\rm clump}$), bolometric luminosity of central objects ($ L_{\rm bol}$), luminosity-to-mass ratio $\rm (L_{bol}/M_{clump}$), and the peak $\rm H_2$ column density of clumps ($ N_{\rm H_2}$); these are shown in Figure \ref{detectionrate_vs_clumpparam}. For each parameter, we divide the clumps into seven bins covering the minimum to maximum values in Table\,\ref{tab_summary_param} and then determine the detection fraction for each bin {(  see Table \ref{tab_detection_rate})}. The results are plotted in Figure \ref{detectionrate_vs_clumpparam} showing that the detection rate increases from $\sim 50\%$ to $\sim 90\%$ as the clump evolves, which reveals an obvious trend in that more massive, luminous, dense, and evolved sources show  a much higher outflow detection fraction. 
 
 Our overall detection rate of $69\%\pm3\%$ for the whole sample is probably a lower limit due to the sensitivity of CHIMPS and the inclusion of less massive clumps that may not be capable of forming massive stars {  (e.g., roughly 8\% of the clumps in this sample have masses $\rm M_{clump}<100\,M_{\odot}$ with the fraction of low-mass clumps higher than in other studies \citep[e.g.,][]{Beuther2002AA383B,deVilliers2014MNRAS444})}. This explains why the overall detection rate determined in this work is lower than previously reported in the literature \citep{Shepherd1996ApJ457S,Beuther2002AA383B,Kim2006ApJ643,Lopez_Sepulcre2009AA499L,deVilliers2014MNRAS444}, as the previous literature samples  were very specifically targeted toward markers of MSF, and outflows are said to be ubiquitous properties of MSF.  Our unbiased survey reveals strong selection functions in the outflow detection fraction in luminosity, clump mass, column density, and luminosity-to-mass ratio. At later evolutionary stages of the central objects in the clumps, the detection rates of outflows in our sample can be as high as 90\% when $\rm L_{bol}/M_{clump}>10\,(L_{\odot}/M_{\odot})$, $\rm L_{bol}>2.7\times10^{4}\,L_{\odot}$, $\rm M_{clump}>3.9\times10^{3}\,M_{\odot}$, and $\rm N_{H_2} > 3.8\times10^{22}\,cm^{-2}$. 
In particular, the {  rise} in the fraction of detected outflows with peak $\rm H_2$ column density at $\rm logN_{H_2} > 22.2\,cm^{-2}$ (or $\rm   \sim 250\,M_{\odot}\,pc^{-2}$ ) is larger than the concept of a surface density threshold for efficient star formation of $\rm \sim 120\,M_{\odot}\,pc^{-2}$, as found by \citet{Lada2010ApJ687L} and \citet{Heiderman2010ApJ1019H}.

\subsection{Determination of Outflow Parameters}
\label{outflow_props}

The physical properties of outflows  provide useful information on the energy and mass exchange process and have been derived by many previous works \citep[e.g.][]{Cabrit1990ApJ530C,Beuther2002AA383B,deVilliers2014MNRAS444}. Following the strategy outlined by \citet{deVilliers2014MNRAS444}, we make the following assumptions. (1) The $\rm{J = 3-2}$ transition temperature of $\rm{^{13}CO}$ is $\rm T\rm{_{trans}}=31.8$ K \citep{Minchin1993AA595M} and excitation temperature $\rm T_{ex} = 35\,K$ \citep[e.g.][]{Shepherd1996ApJ472S,Henning2000AA211H,Beuther2002AA383B}. 
(2) The $\rm{^{13}CO}$ line wings are optically thin. 
The column density of $\rm{^{13}CO}$ may thus be written as \citep{Curtis2010MNRAS401}
\begin{equation}
N\rm{\left(^{13}CO\right)} = 5 \times 10^{12} 
T\rm{_{ex}} \exp \left( \frac{T\rm{_{trans}}}{T\rm{_{ex}}} \right) 
\int{T\rm{_{mb}} dv} \rm\,{(cm^{-2})}, 
\end{equation}

\noindent where $\int{T_{mb}} dv$ is calculated from the mean temperature of $\rm{^{13}CO}$ within the outflow lobe area defined by the lowest contours, dividing it by the main beam-correction factor of 0.72 from CHIMPS \citep{Rigby2016MNRAS456R}. The abundance ratios of $\rm{[CO]/[H_2]=10^{-4}}$ \citep{Frerking1982ApJ590F} 
and $\rm{[^{12}CO]/[^{13}CO]}$= $\rm 7.5D_{\rm{gal}}+7.6$, where $\rm D_{gal}$ is galactocentric distance in kpc \citep{Wilson1994ARAA191W}, are used to convert to the $\rm H_2$ column density. The column density $\rm N(H_{2})$ is, therefore, given by

\begin{equation}
\rm N(H_{2})= (7.5D_{gal}+7.6)\times 10^{4}N(^{13}CO).  
\end{equation}
The $\rm N(H_{2})$ column densities of the blue and red lobes ($\rm N_{b/r}$) are then used to calculate the mass of each lobe ($\rm M_{b/r}$), and then obtain the total outflow mass $ M_{\rm out}$, 
\begin{equation}
	M\rm{_{out}} = M_{r} + M_{b} = \left( N\rm{_{b}} \times A\rm{_{b}}  + N\rm{_{r}} \times A\rm{_{r}} \right)m\rm{_{H_2}},
	\label{eq:mass}
\end{equation}
where $\rm A\rm{_{b/r}}$ is the surface area of each lobe and $\rm m_{H_2}$ is the mass of a hydrogen molecule. The surface area of each lobe is calculated using the same threshold used to calculate $\rm T_{mb}$, defined by the lowest contours. 

For each pixel in the defined outflow lobe area, 
we calculate the outflow momentum and energy per velocity channel ($\Delta v$), by using the channel velocity relative to the systemic velocity ($ v_i$) and gas mass ($\rm M_{i}$) corresponding to the emission in that channel. The outflow momentum and energy can thus be obtained by summing their corresponding value over all velocity channels,  
\begin{equation}
p = \sum_{A_b} \left[ \sum_{i=v_b} M_{b_i} v_i \right]\Delta v + \sum_{A_r} \left[ \sum_{i=v_r} M_{r_i} v_i \right]\Delta v
\label{eq:momentum_me}
\end{equation}
\begin{equation}
E = \frac{1}{2} \sum_{A_b} \left[ \sum_{i=v_b} M_{b_i} v^{2}_{i} \right]\Delta v  + \frac{1}{2} \sum_{A_r} \left[ \sum_{i=v_r} M_{r_i} v^{2}_{i} \right]\Delta v.
\label{eq:Energy_me}
\end{equation}

The maximum characteristic length $ l_{max}$ refers to the maximum length of each outflow lobe $ l_{b/r}$ that is measured from the clump centroid to each extreme of each lobe. Therefore, we can estimate the dynamic time scale $\rm t_{d}$, the mass rate of the molecular outflow $\dot{ M}_{\rm out}$, the mechanical force $F_{\rm CO}$ and the mechanical luminosity $L_{ \rm CO}$ using the following equations: 

\begin{equation}
t_d = \frac{l_{max}}{\left(\rm V_{maxb} + \rm V_{  maxr} \right) /2}.
\label{eq:timescale}
\end{equation}
\begin{eqnarray}
\dot{M}\rm{_{out}} & = & \frac{\rm M\rm{_{out}}}{t} \\
F_{CO} & = & \frac{p}{t} \\
L_{CO} & = & \frac{E}{t},
\label{eq:Mechanics}
\end{eqnarray}
%%%
where $\rm V_{maxb}$ and $\rm  V_{maxr}$ are the maximum blue and red velocities relative to the peak $\rm{C^{18}O}$ velocity (see Table\,\ref{outflow_wings}). Please see \citet{deVilliers2014MNRAS444} for further details. 
We adopt an average inclination angle of $\theta=57.3^{\circ}$  to correct the results for the unknown angle between the flow axis and the line of sight \citep{Beuther2002AA383B,Zhang2005ApJ625}. The inclination-corrected physical properties of outflows with mapped blue and red lobes are listed in Table~\ref{outflow_phyparam} for a small portion, and the properties of total 153 outflows are shown in Appendix Table\,\ref{total_outflow_phyparam}. 

In Table\,\ref{tab_summary_param} we give a summary of the maximum, minimum, median, and standard deviation of the distribution of the clump properties with and without outflows, as well as  the outflow properties of the 153 clumps with mapped outflow lobes. The outflows from our survey have a range of physical properties similar to those of previously studied massive outflows \citep[e.g.,][]{Beuther2002AA383B,Wu2004AA426W,Zhang2005ApJ625,deVilliers2014MNRAS444}, and are more than 2 orders of magnitude more massive and more energetic than low-mass outflows \citep[e.g.,][]{Bontemps1996AA311,Wu2004AA426W,Arce2007prplconf245A,Bjerkeli2013AA8B}. The  mean outflow mass-loss rates imply a mean accretion rate of $\rm\sim10^{-4}\,M_\odot\,yr^{-1}$ \citep{Beuther2002AA383B,deVilliers2014MNRAS444}, which agrees with the accretion rates  predicted by theoretical models of MSF \citep[e.g.,][]{Bonnell2001MNRAS785B,Krumholz2007ApJ959K}. 

%%table%%%%
\setlength{\tabcolsep}{4pt}
\begin{table}
%\scriptsize
\caption {\rm Summary of the Physical Parameters of Clumps and Outflows. In Columns 2$-$5, we give the minimum, maximum, $\rm mean\pm standard\,deviation$, and median values of these  parameters for each subsample. The physical parameters of clump are measured by \cite{Urquhart2017arXiv392U}.}
\begin{tabular}{lllll}
\hline
\hline
Parameter  &  $x_{min}$ & $x_{max}$ & $x_{mean}\pm x_{std}$ & $x_{med}$\\
\hline
\multicolumn{5}{c}{919 ATLASGAL Clumps in CHIMPS}     \\
\hline
$\rm log(M_{clump}/M_{\odot})$ &-0.30 &5.04&$2.84\pm0.62$&2.88\\
$\rm log(L_{bol}/L_{bol})$ &0.00&6.91&$3.19\pm0.99$&3.15\\
$\rm log[L_{bol}/M_{clump}(\scriptsize L_{\odot}/M_{\odot})]$ & -1.77&2.65&$0.35\pm0.79$& 0.39\\
$\rm log(N_{H_2}/cm^{2})$ & 21.76&23.92&$22.35\pm0.29$&22.30\\
\hline
\multicolumn{5}{c}{325 Clumps with Good Data}     \\
\hline
$\rm log(M_{clump}/M_{\odot})$ &-0.30 &5.04&$2.93\pm0.64$&2.93\\
$\rm log(L_{bol}/L_{bol})$ &0.00&6.91&$3.82\pm0.96$&3.82\\
$\rm log[L_{bol}/M_{clump}(\scriptsize L_{\odot}/M_{\odot})]$ &-1.00&2.65&$0.89\pm0.62$& 0.90\\
$\rm log(N_{H_2}/cm^{2})$ &21.76&23.92&$22.45\pm0.36$&22.36\\
\hline
\multicolumn{5}{c}{225 Clumps with Outflows}     \\
\hline
$\rm log(M_{clump}/M_{\odot})$ &1.50 &4.5&$3.00\pm0.61$&3.05\\
$\rm log(L_{bol}/L_{bol})$ &1.64&6.21&$3.99\pm0.90$&3.96\\
$\rm log[L_{bol}/M_{clump}(\scriptsize L_{\odot}/M_{\odot})]$ &-0.97&2.65&$0.99\pm0.61$& 0.99\\
$\rm log(N_{H_2}/cm^{2})$ &21.82&23.92&$22.51\pm0.37$&22.47\\
\hline
\multicolumn{5}{c}{100 Clumps without Outflows}     \\
\hline
$\rm log(M_{clump}/M_{\odot})$ &-0.30 &5.04&$2.77\pm0.66$&2.77\\
$\rm log(L_{bol}/L_{bol})$ &0.00&6.91&$3.44\pm0.98$&3.42 \\
$\rm log[L_{bol}/M_{clump}(\scriptsize L_{\odot}/M_{\odot})]$ & -1.00&2.37&$0.67\pm0.61$& 0.74\\
$\rm log(N_{H_2}/cm^{2})$ &21.76&23.89&$22.28\pm0.27$&22.25\\
\hline
\multicolumn{5}{c}{Outflow Properties for 153 Clumps with Further Analysis}     \\
\hline
$\rm M_{out}/M_{\odot}$ &1.36&2065.26 &$121.16\pm250.82$ & 45.89\\
$\rm \ell_{max}/pc$ &0.20&3.02&$1.10\pm0.57$&0.99             \\
$\rm t_{d}\,(10^{5}\,yr)$ &0.25&8.90&$1.78\pm1.30$&1.51\\
$\dot{\rm M}_{\rm out}\,(\scriptsize \rm 10^{-4}M_{\odot}/yr)$ &0.08 &172.34&$9.26\pm21.11$&2.72 \\
$\rm p\,(\scriptsize 10\,M_{\odot}\,km\,s^{-1})$ &{  0.54}&{  2964.65} &{$  124.76\pm359.60$} & {  23.39}\\
$\rm E\,(\scriptsize 10^{39}\,J)$ & {  0.02} & {  786.51} & {$  20.45\pm79.91$} & {  2.00} \\
$\rm L_{CO}\,(L_{\odot})$ & {  0.01} & {  502.88} & { $  14.71\pm53.28$} & {  0.89} \\
$\rm F_{CO}\,(\scriptsize 10^{-3}\,M_{\odot}\,km\,s^{-1}\,yr^{-1})$ & {  0.03} & {  225.26} & {$  9.98\pm28.65$} & {  1.32} \\
\hline
\hline
\end{tabular}
\label{tab_summary_param}
\end{table} 
\setlength{\tabcolsep}{6pt}
%%%%%end table%%%
  
 Typically, the uncertainties on derived outflow physical properties are a factor of $\sim3$ on outflow mass $ M_{\rm out}$, a factor of $\sim10$ on mechanical force $F_{\rm CO}$, and a factor of $\sim30$ on mechanical luminosity $L_{ \rm CO}$, in previous studies \citep[e.g.,][]{Cabrit1990ApJ530C,Shepherd1996ApJ472S,Beuther2002AA383B,Wu2004AA426W,deVilliers2014MNRAS444}. These are mainly due to uncertainties in distance, $\rm ^{12}CO/H_2$, $\rm T_{ex}$,  and inclination angles \citep{Cabrit1990ApJ530C}. The uncertainty in kinematic distance described by \citet{Urquhart2017arXiv392U} could also have a large influence on these parameters. 
%{  The typical errors of the calculated outflow parameters are mainly from the uncertainties in many assumptions used in determining them (such as $\rm ^{12}CO/H_2$, $\rm T_{ex}$, and inclination angles) \citep[e.g.,][]{Cabrit1990ApJ530C,deVilliers2014MNRAS444}. An additional uncertainty is caused by distance ($\sim6\%$, \cite{Urquhart2017arXiv392U}) could also have a large influence on these parameters.} 
However, many of these uncertainties are systematic and so are unlikely to have a significant effect on the overall distribution and correlations between individual quantities. Therefore,  the homogeneity of our sample and the large number of the observed objects should ensure any results drawn from our statistical analysis are robust. 

\section{Discussion}

\subsection{Outflow Activity as a Function of MYSO Evolutionary State}

The outflow properties presented in Section\,\ref{sect:detection} allow us to investigate at which stage outflows ``switch on'' and how outflow properties change with respect to different evolutionary phases. Interestingly, two clumps in the youngest quiescent stage, i.e., $\rm 70\mu m$ weak \citep{Urquhart2017arXiv392U}, show outflow wings, which suggests some clumps that are in a quiescent stage are associated with outflow activity and therefore may be in a very early protostellar stage. This is supported by \cite{Feng2016ApJ100F} and \cite{Tan2016ApJ3T}, who have reported bipolar outflow toward a high-mass protostar associated with a 70$\rm \mu m$ dark source. This makes these two $70\mu m$ weak clumps interesting candidates to investigate outflow activity in the earliest stages of a protostar's evolution in more detail. 
%In addition, \cite{koenig2017} suggest that star formation may already be underway in some of these $\rm 70 \mu m$ weak clumps and \cite{Traficante2017MNRAS3882T} reported some $\rm 70 \mu m$ weak sources associated with molecular outflows. 

There is a clear trend for increasing detection frequency of outflows along the four evolutionary sequences, i.e., from the youngest quiescent ($50\%\pm25\%$) to protostellar ($53\%\pm11\%$), to YSO ($61\%\pm4\%$), and then to MSF clump ($82\%\pm3\%$). This suggests that outflow activity becomes much more common as clumps evolve. 
%A detailed investigation of the outflow detection rate toward clumps at different evolutionary stage, found that outflow activity becomes much more common with increasing age of clumps, as detection rate increases from the youngest quiescent clump ($50\pm25\%$) to protostellar ($53\pm11\%$), to YSO ($61\pm4\%$), and then to MSF clump ($82\pm3\%$). 
%This provides evidence that outflow activity start to develop in the earliest stages and outflow signature could be more observable as the clumps evolve. 
In a detailed study of the subgroup MSF clump (i.e., a mid-infrared $\rm 24\mu m$ bright clump associated with a MSF tracer), higher detection rates occurred for subclass of \hchii\ regions-associated clumps (100\%), \uchii\ regions-associated clumps ($93\%\pm3\%$), and masers-associated clumps ($86\%\pm4\%$). For masers-associated clumps, the detection rate is 100\% (i.e., 11/11) for water maser-associated clumps and $86\pm3\%$ (i.e., 53/63) for methanol maser-associated clumps, and 100\% (i.e., 4/4) for water-methanol maser-associated clumps. For maser-associated \uchii\ regions, the detection rate is 100\% (i.e., 27/27) and reduces to $86\%\pm4\%$ (i.e., 25/29) for non-maser-associated \uchii\ regions. Therefore, in the MSF clump group, the detection rate can be very high, $\sim$ 90\%$-$100\% for pre-\uchii\ (e.g., \hchii\ regions \citep{Kurtz2005IAUS111K}, water and/or methanol masers \citep{Codella2004AA615C,koenig2017}), and the early 
\uchii\ region phase (e.g., maser-associated \uchii\ regions; \citep{Codella2004AA615C}). 
Then, outflow detection frequency is likely to decrease as the \uchii\ region evolves, which is 
also supported by the decreasing outflow activity at the end of the \uchii\ region stage reported by \citet{Codella2004AA615C}. 
 
%As \hchii\ region is younger than \uchii\ region \citep{Kurtz2005IAUS111K}, and  masers (water and methanol) \citep{Codella2004AA615C} are thought to be developed before \uchii\ region stage, and \uchii\ regions in our sample are very young \citep{koenig2017}, which suggest that outflow detection rates of clumps at pre-\uchii\ or early \uchii\ phase could be very high $\sim 90\% $ by up to 100\%. Then, outflow detection frequency decrease as \uchii\ region evolves. tflow wings, i.e., G043.166$+$00.01 and G043.178$-$00.01

In summary, the outflow detection rate is increasing as the clumps evolve in this young sample (see Figure\,\ref{detectionrate_vs_clumpparam}), and appears to  peak (100\%) at the pre-\uchii\ region or early \uchii\ region stage. However, there are a few clumps at a later stage of evolution  with large values of $ L/M$ that are associated with a complex star formation region in the Galactic plane (e.g., G043.166$+$00.01 in W49A), that show no evidence for outflow wings. The nondetection of outflows toward these sources may be due to the complexity of the CO emission \citep[e.g.,][]{Zhang2001ApJ552L}, interactions of the sources within the clumps below our resolution \citep[e.g.,][]{Codella2004AA615C}, or external winds/shocks \citep[e.g.,][]{Maud2015MNRAS645M}.
However, these nonoutflow sources with high $\rm L/M$ show extended emission or are part of extended emission at 1.4 GHz MAGPIS survey \citep{Helfand2006AJ}. The high $\rm L/M$ may also indicate that the \hii\ region has started to disrupt their environment and that the central YSOs are no longer accreting.

%{\color{purple} The high L/M may also indicate that the HII region has started to disrupt their environment and are no longer accreting. Take a look at the radio emission towards these sources and see if it's extended.}
%{\color{red}reply: Yes, they all show extended emission or be part of extended emission at 1.4GHz MAGPIS survey.}} {\color{purple} Then the high L/M is because they have have ceased accretion and are now disrupting their natal cloud and not due to confusion.}

%% Mark edited to here. 

\subsection{Comparison between Clumps with and without Outflows}

%%%figure%%%
\begin{figure*}
\centering
\begin{tabular}{cc}
\includegraphics[width = 0.5\textwidth]{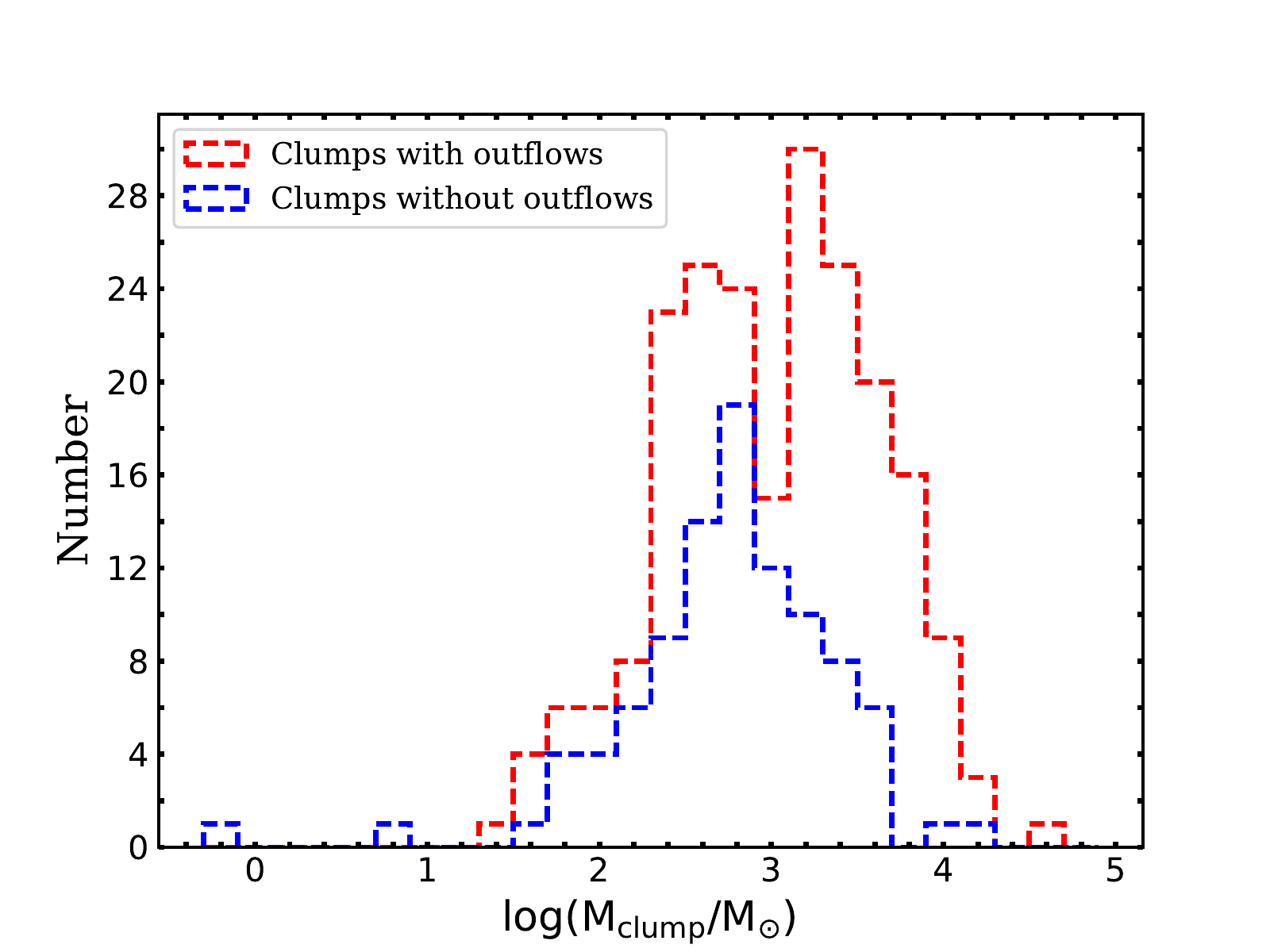}& 
\includegraphics[width = 0.5\textwidth]{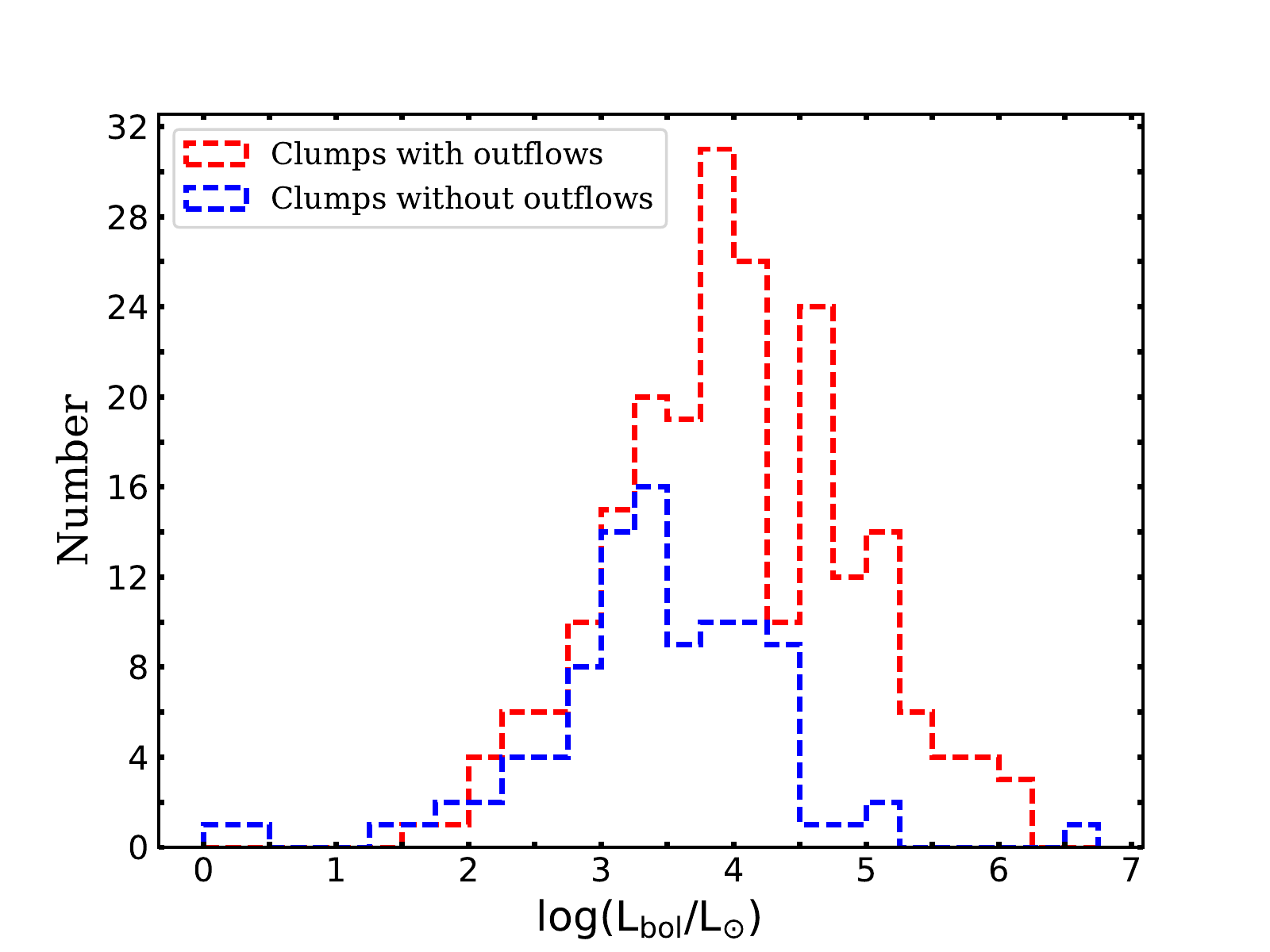}\\ 
\includegraphics[width = 0.5\textwidth]{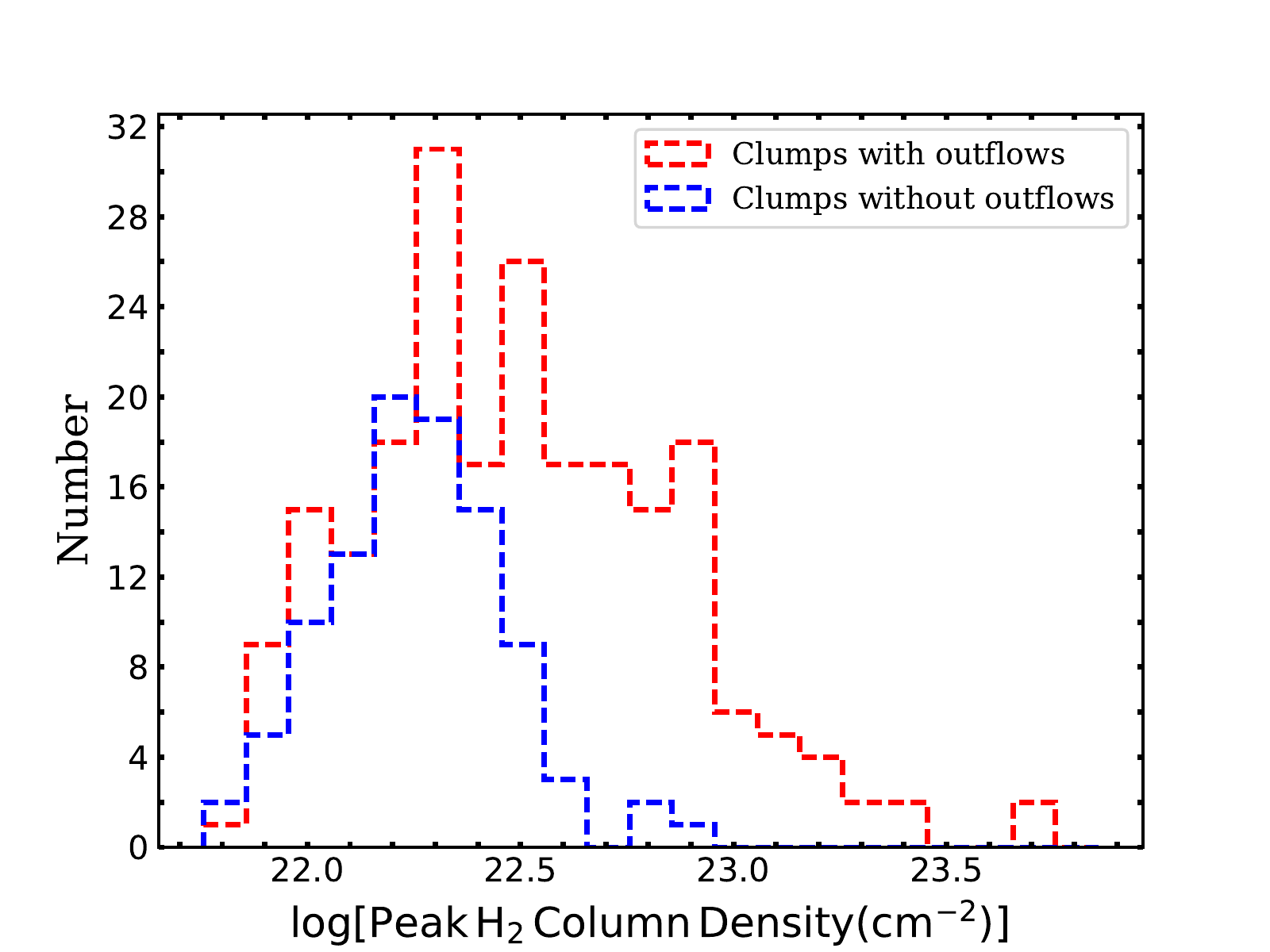}& 
\includegraphics[width = 0.5\textwidth]{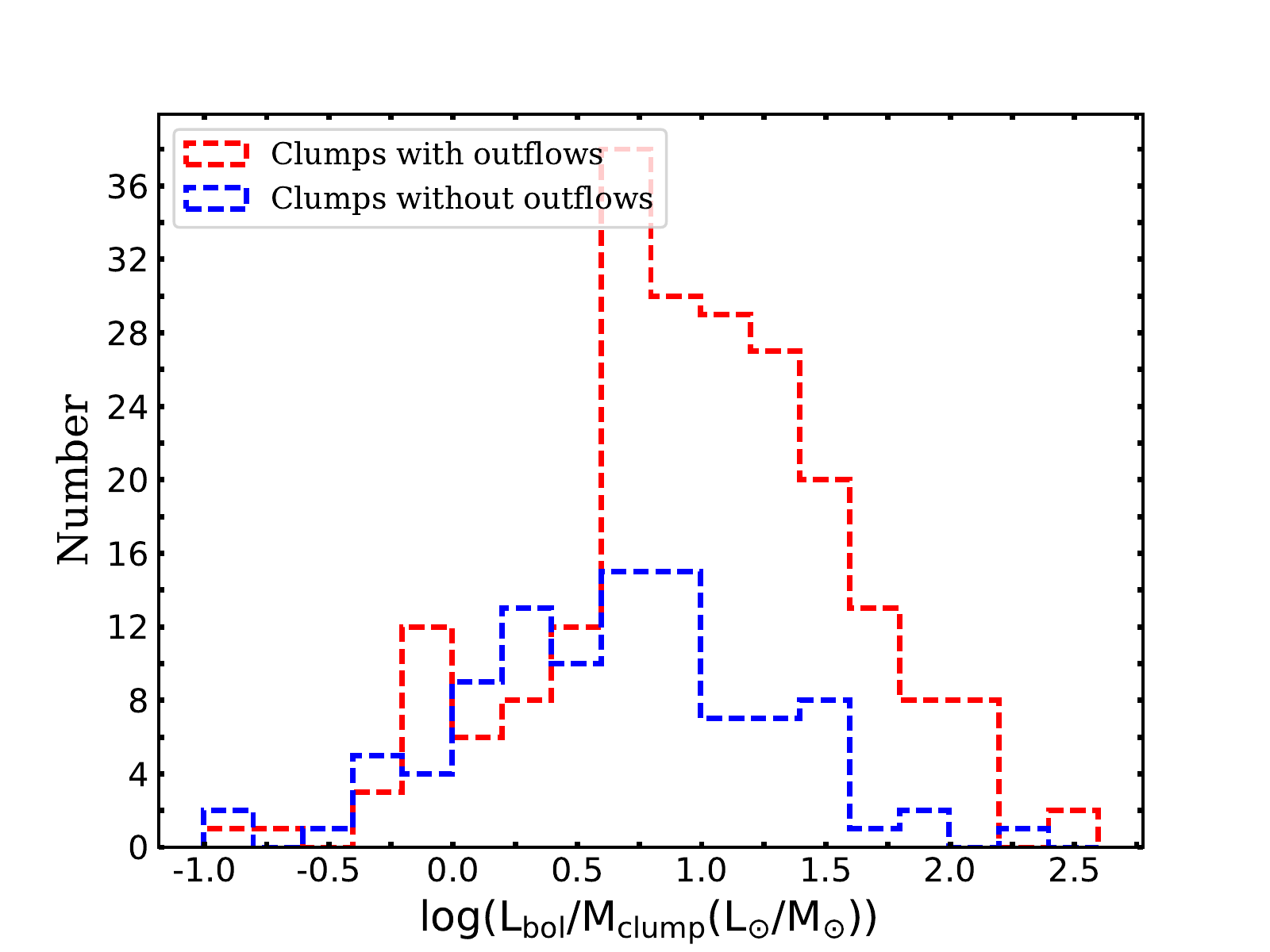}\\ 
\end{tabular}
\caption{Top left to bottom right: logarithmic distributions of the clump mass ($ M_{\rm clump}/M_{\odot}$), bolometric luminosity ($ L_{\rm bol}/L_{\odot}$), the peak $\rm H_2$ column density ($ N_{\rm H_2}/\rm cm^{2}$), and luminosity-to-mass ratio ($ L_{\rm bol}/M_{\rm clump}(L_{\odot}/M_{\odot})$), for the 225 outflow clumps (red dashed lines) compared to the 100 clumps without outflows (blue dashed lines). The bin sizes are 0.2, 0.25, 0.1, and 0.2\,dex from top left to bottom right. The $ L/M$ ratio is a well-identified indicator of clump evolution with larger values for more evolved clumps, and the peak $\rm H_2$ column density shows a very strong positive correlation with the fraction of clumps associated with MSF \citep{Urquhart2017arXiv392U}.}
\label{distribution_param}
\end{figure*}
%\citet{Urquhart2017arXiv392U} reported that the fraction of clumps associated with massive stars reach 100\% for the peak $\rm H_2$ column density above $\rm 10^{23} cm^{-2}$, which is supported by our finding that clumps active outflow at 100\% for $\rm log(N_{H_2}/cm^{2}) >23$  (see bottom-left figure).
%%%figure%%%

Our search for high-velocity line wings in the outflow search sample allows us to divide clumps into two subsamples: those that are associated with outflows and those that are not. In Fig.\,\ref{distribution_param} we present histograms that compare the distribution of the physical properties of the clumps associated with outflows (red) and unassociated clumps (blue). The average properties for the two samples are summarized in Table\,\ref{tab_summary_param}. It is clear from these Figure \ref{distribution_param} that the clumps associated with outflows are significantly more massive, have higher column densities, and host more luminous and evolved objects. K-S tests confirm that these two samples are significantly different from each other ($p$-values $\ll$ 0.001).  This implies that clumps with  more luminous central sources are much more likely to be associated with outflows than clumps hosting lower-luminosity central sources. This is consistent with the study by \citet{Urquhart2014MNRAS1555U} who found that the more massive and dense clumps are more likely to be associated with MYSOs and \hii\ regions and therefore more likely to be associated with outflows.

\subsection{Comparison of Outflow Parameters to Properties of Their Corresponding Clumps}
\label{section_correlation}
Our large homogeneous and uniformly selected sample allows us to examine the correlation between the physical properties of the outflows and the properties of their corresponding clumps. As most of the derived physical properties depend on the distance to the clump, we use a nonparametric measure of the statistical dependence to measure their correlation and allow for the effects of distance being a common variable between parameters. We use Spearman's rank correlation coefficient ($\rho$) to control the effect of distance-dependent parameters \citep{Kim2006_1068}. The results of these correlations are listed in Table\,\ref{outflowparam_vs_clumparam}. The relations between outflow properties and their natal clump mass, bolometric luminosity, and luminosity-to-mass ratio are shown in Figures\,\ref{fig:Mout_vs_clumpparam}$-$\ref{fig:Lco_vs_clumpparam}.

%%%table%%%
\setlength{\tabcolsep}{8pt}
\begin{table*}
\caption {\rm Outflow Parameters versus Clump Properties. \\{  We use non-parametric Spearman's rank correlation test to determine the level of correlation between these distance-dependent parameters when take distance as the control variable. The p-value gives the significance of all correlations and is lower than 0.0013 for a significant correlation. If a significant correlation is found, we fit the data in log-log space using a linear least-squares fit method.} }
\begin{tabular}{| l | ccc | l |}
\hline
Relations  &  \multicolumn{3}{c | }{Spearman's Rank Correlation}  &  Linear least-squares fits \\
\hline
& $\rho$  & p-value & control variable &       log-log space\\
\hline
$\rm M_{out}\, vs\,\rm M_{clump}$ &  0.35 & $\rm \ll0.001$& Dist.         &\scriptsize $\rm   log(M_{out}/M_\odot )= (0.6\pm0.06)log(M_{clump}/M_\odot)\,-(0.2\pm0.20)$       \\
\hline
$\rm M_{out}\, vs\,\rm L_{bol}$ &  0.66 & $\rm \ll0.001$& Dist.         &\scriptsize$\rm   log(M_{out}/M_\odot )= (0.5\pm0.03)log(L_{bol}/L_\odot)\,-(0.3\pm0.13)$       \\
\hline
$\rm M_{out}\, vs\,\rm L_{bol}/M_{clump}$ &  0.59 & $\rm \ll0.001$& Dist.         &\scriptsize$\rm   log(M_{out}/M_\odot )= (0.6\pm0.07)log(L_{bol}/M_{clump}(L_\odot/M_{\odot}))\,+(1.1\pm0.08)$       \\
\hline
$\dot{\rm  M}_{\rm out}\,\rm vs\,\rm M_{clump}$ &  0.47 & $\rm \ll0.001$& Dist.         &\scriptsize $ {\rm   log}(\dot{\rm   M}_{\rm   out}/\rm   M_\odot yr^{-1})= (0.6\pm0.07)log(M_{clump}/M_\odot)\,-(5.3\pm0.20)$       \\
\hline
$\dot{\rm  M}_{\rm out}\,\rm vs\,\rm L_{bol}$ &  0.80 & $\rm \ll0.001$& Dist.         & \scriptsize$  {\rm   log}(\dot{\rm   M}_{\rm   out}/\rm    M_\odot yr^{-1})= (0.6\pm0.03)  log(L_{bol}/L_\odot)\,-(5.7\pm0.13)$       \\
\hline
$\dot{\rm M}_{\rm out}\,\rm vs\,\rm L_{bol}/M_{clump}$ &  0.68 & $\rm \ll0.001$& Dist.         & \scriptsize$  {\rm   log}(\dot{\rm   M}_{\rm out}/\rm   M_\odot yr^{-1})= (0.7\pm0.06)  log(L_{bol}/M_{clump}(L_\odot/M_{\odot}))\,-(4.2\pm0.07)$       \\
\hline
$\rm F_{CO}\, vs\,\rm M_{clump}$   &  0.51 & $\rm \ll0.001$& Dist.         &\scriptsize $  \rm   log(F_{CO}/M_\odot\,km\,s^{-1}yr^{-1})= (0.8\pm0.09)log(M_{clump}/M_\odot)\,-(5.1\pm0.30)$       \\
\hline
$\rm F_{CO}\, vs\,\rm L_{bol}$   &  0.79 & $\rm \ll0.001$ & Dist.         & \scriptsize$\rm    log(F_{CO}/M_\odot\,km\,s^{-1}yr^{-1})= (0.7\pm0.04)log(L_{bol}/L_\odot)\,-(5.5\pm0.17)$       \\
\hline
$\rm F_{CO}\, vs\,\rm L_{bol}/M_{clump}$   &  0.65 & $\rm \ll0.001$ & Dist.         & \scriptsize$  \rm   log(F_{CO}/M_\odot\,km\,s^{-1}yr^{-1})= (0.9\pm0.08)log(L_{bol}/M_{clump}(L_\odot/M_{\odot}))\,-(3.6\pm0.10)$       \\
\hline
$\rm L_{CO} \, vs\,\rm M_{clump} $ &  0.54 & $\rm \ll0.001$ & Dist.         &\scriptsize$  \rm   log(L_{CO}/L_\odot)= (1.0\pm0.10)log(M_{clump}/M_\odot)\,-(2.8\pm0.3)$ \\
\hline
$\rm L_{CO} \, vs\,\rm L_{bol} $ &  0.79 & $\rm \ll0.001$ & Dist.         &\scriptsize$  \rm   log(L_{CO}/L_\odot)= (0.8\pm0.05)log(L_{bol}/L_\odot)\,-(3.2\pm0.2)$ \\
\hline
$\rm L_{CO} \, vs\,\rm L_{bol}/M_{clump} $ &  0.62 & $\rm \ll0.001$ & Dist.         &\scriptsize$  \rm   log(L_{CO}/L_\odot)= (1.1\pm0.1)log(L_{bol}/M_{clump}(L_\odot/M_{\odot}))\,-(1.0\pm0.1)$ \\
\hline
\end{tabular}
\label{outflowparam_vs_clumparam}
\end{table*}    
%%%%table%%%

\subsubsection{$ M_{\rm out}$ versus $ M_{\rm clump}$, $ L_{\rm bol}$, 
and $ L_{\rm bol}/M_{\rm clump}$ }

%%%%figure%%%
\begin{figure}
%\centering
\begin{tabular}{c}
\includegraphics[width = 0.5\textwidth]{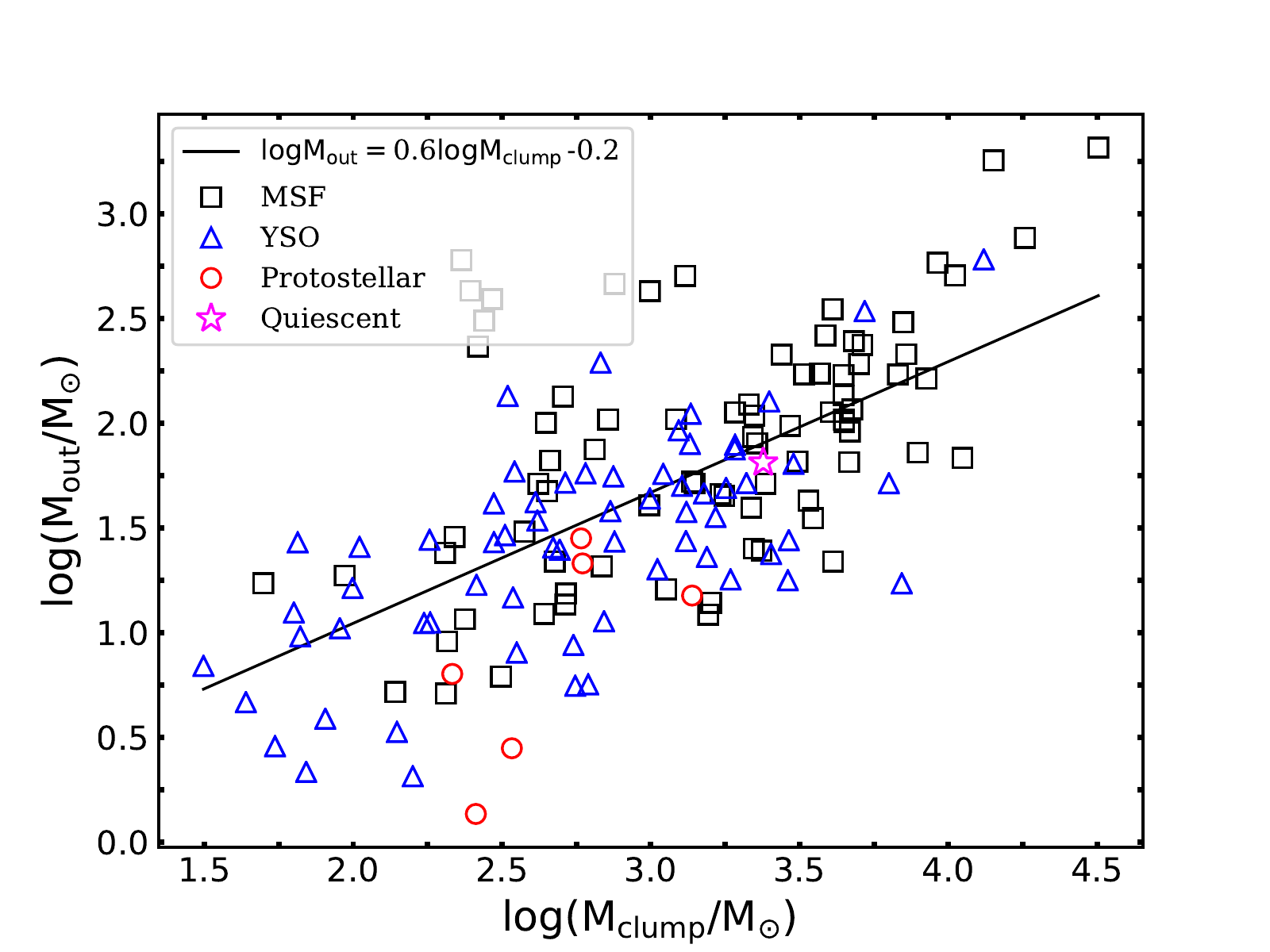}\\
\includegraphics[width = 0.5\textwidth]{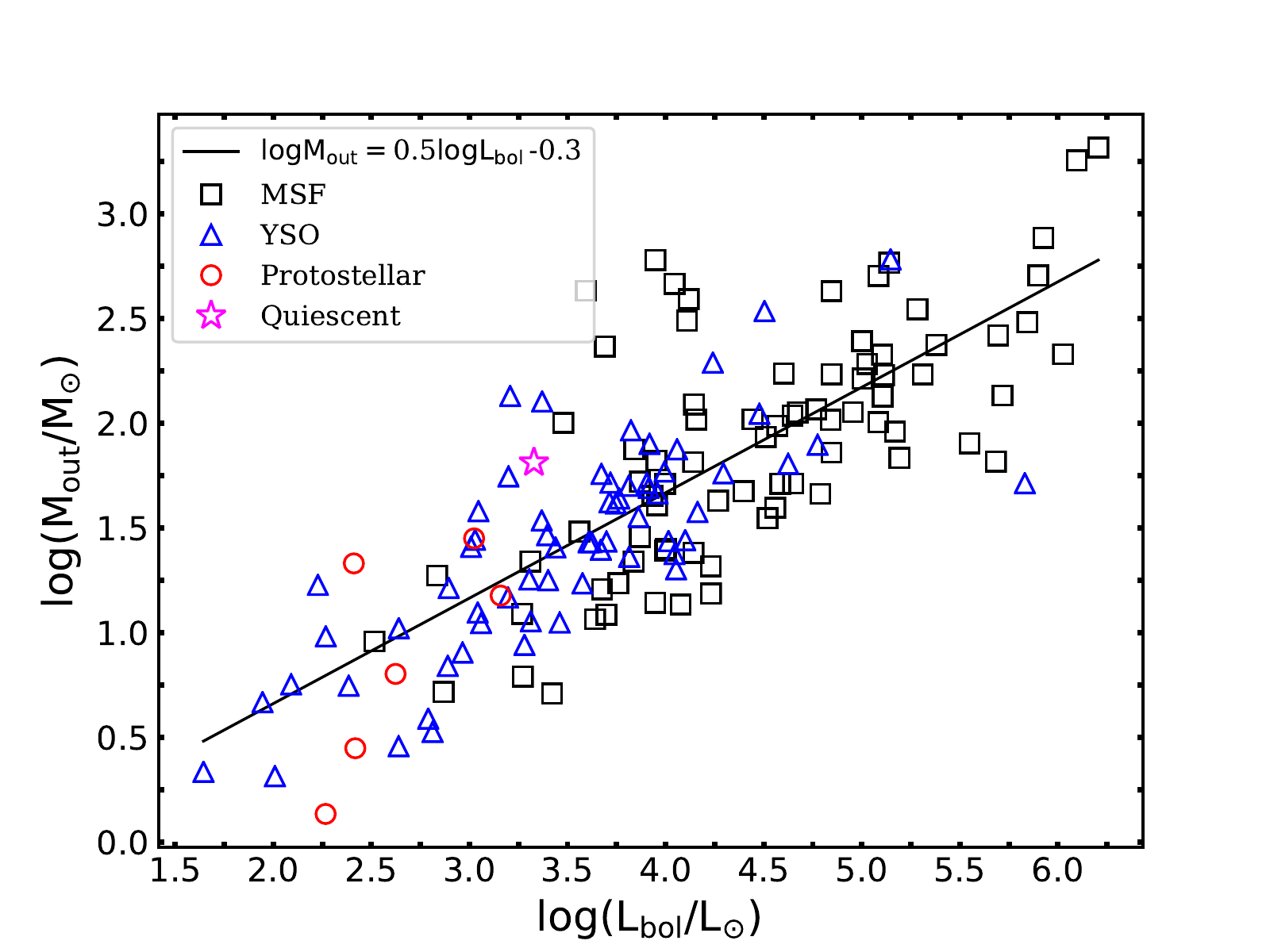}\\
\includegraphics[width = 0.5\textwidth]{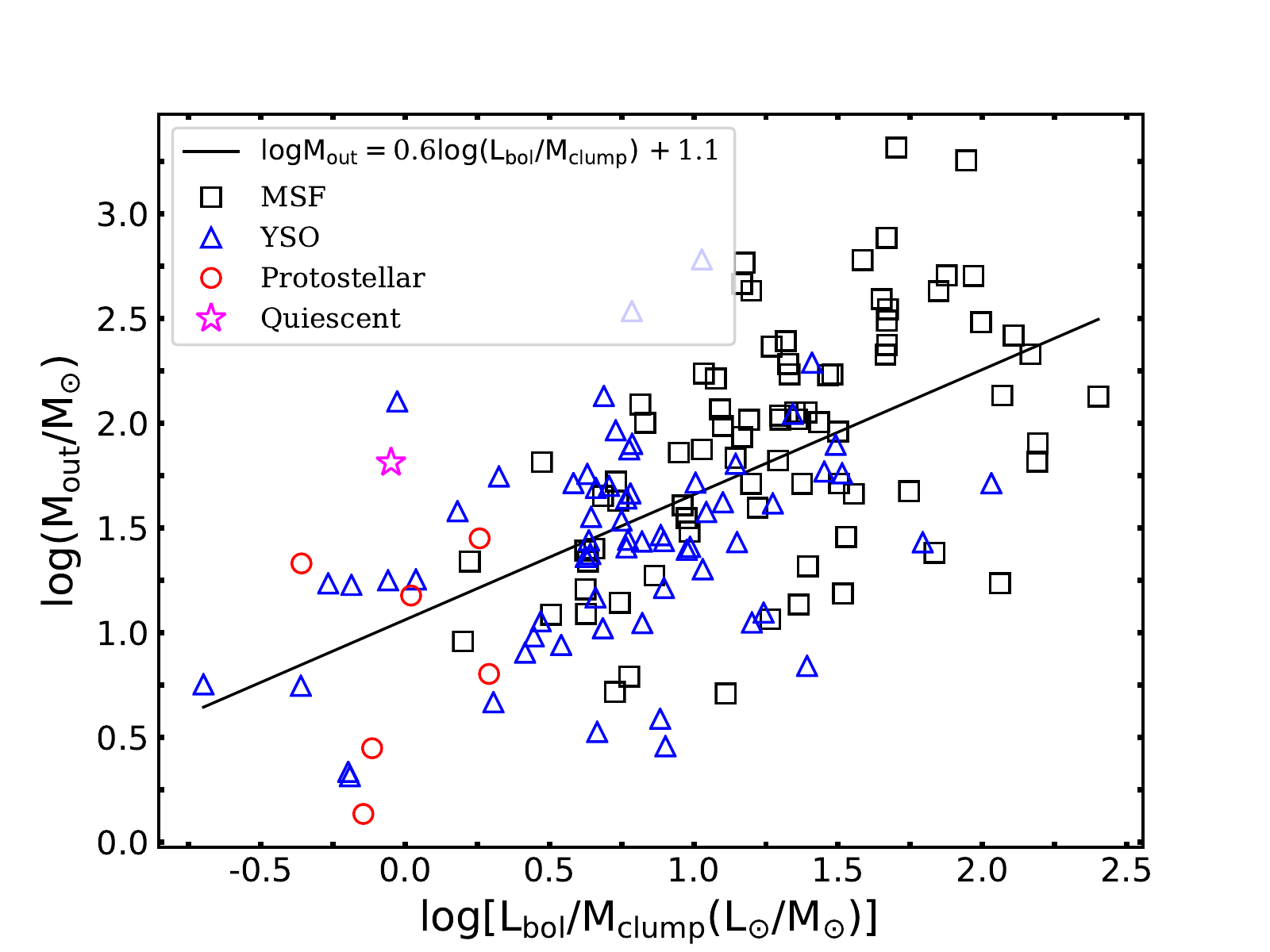}\\  
\end{tabular}
\caption{Top panel: outflow mass versus clump mass. 
  Middle panel: outflow mass  versus bolometric luminosity.  
  Bottom panel:  outflow mass versus luminosity-to-mass ratio. 
 The black squares, blue triangles, red circles, and magenta stars refer to MSF, YSO, protostellar, and quiescent clumps. The solid line in each plot is the least-squares linear fit line in logarithmic scale. }
\label{fig:Mout_vs_clumpparam}
\end{figure}

%%%%figure%%%

The mass of the outflow is a fundamental parameter, and {  we plot the relation between outflow mass $ M_{\rm out}$ and clump properties $ M_{\rm clump}$, $ L_{\rm bol}$, and $L_{\rm bol}/M_{\rm clump}$  } in the top, middle and bottom panels of Figure \ref{fig:Mout_vs_clumpparam}. The correlation coefficients and results of linear fits to the data are presented in Table \ref{outflowparam_vs_clumparam}. A similar relation ($\rm M_{out}\propto M_{clump}^{0.8}$) was reported by \citet[][]{deVilliers2014MNRAS444} for 44 methanol maser-associated objects using the same method as this work. \citet{Beuther2002AA383B} reported a correlation of $\rm M_{out}\sim 0.1M_{clump}^{0.8}$ for 21 high-mass star-forming regions. \citet{Lopez_Sepulcre2009AA499L} gave a correlation of $\rm M_{out}=0.3M_{clump}^{0.8}$ for 11 very luminous objects, with their clump masses derived from $\rm C^{18}O$ and millimeter-wave dust emission. \citet{Sanchez_Monge2013AA557A} found a similar relation as \citet{Lopez_Sepulcre2009AA499L} for 14 high-mass star-forming regions, with outflow masses derived from $\rm SiO$ and clump masses from infrared SED fits.  The correlation derived from our sample of 153 massive clumps  ($\rm M_{out}\propto M_{clump}^{0.6\pm0.06}$) is similar to these previous results, while the marginally shallower index most likely results from a larger range of clump masses and wider spread of  evolutionary stages in this work. The ratio of $\rm M_{out}/M_{clump}$ has a median value of 0.05 for the sample, and 92\% of the sample has the ratio in $0.005\sim0.32$. Thus, approximately 5\% of the core gas is entrained in the molecular outflow, which is similar to the mean entrainment ratio of 4\% in \citet{Beuther2002AA383B}. 

The correlation between $ M_{\rm out}$ and $ L_{\rm bol}$ ($\rho$ = 0.66) suggests that the two parameters are physically related. The fit to the logs of these parameters gives a slope of $0.5\pm0.03$, which is similar to the slope reported by \citet{Wu2004AA426W}  
for a sample of high-mass and low-mass sources ($0.56\pm0.02$) spanning a wide range for $ L_{\rm bol}$ between $\rm 10^{-1}\,L_\odot$ and $\rm 10^{6}\,L_\odot$. \citet{Lopez_Sepulcre2009AA499L} also find a similar relation  toward a sample of O-type YSOs. The agreement between all of these studies suggests that the correlation is applicable over a broad range of luminosities {  (i.e., $\rm   10^{-1}\,L_\odot \sim 10^{6.5}\,L_\odot$)} from low-mass objects to massive objects and that the outflow driving mechanism is likely to be similar for all luminosities. 

In addition, the relation between $ M_{\rm out}$ and $L_{\rm bol}/M_{\rm clump}$  indicates that the outflow mass increases as the embedded protostar in the clump evolves.  In Figure\,\ref{fig:Mout_vs_clumpparam}, the largest amount of entrainment mass comes from the most evolved MSF clumps. While there is no clear evolutionary trend of outflow mass for the four stages, this is probably because the properties of the four proposed evolutionary stages are likely to overlap with each other (see \citet{koenig2017} and \citet{Urquhart2017arXiv392U}).  The partial correlation coefficient, $\rho=0.59$, is larger than found for the $ M_{\rm out}$ and $ M_{\rm clump}$ ($\rho=0.35$), but smaller than found for $ M_{\rm out}$ and $ L_{\rm bol}$ ($\rho=0.66$).

\subsubsection{${\dot{ M}}_{\rm out}$ versus $ M_{\rm clump}$, $ L_{\rm bol}$ and $ L_{\rm bol}/M_{\rm clump}$}

\begin{figure}
 %\centering
    \begin{tabular}{c}
   \includegraphics[width = 0.5\textwidth]{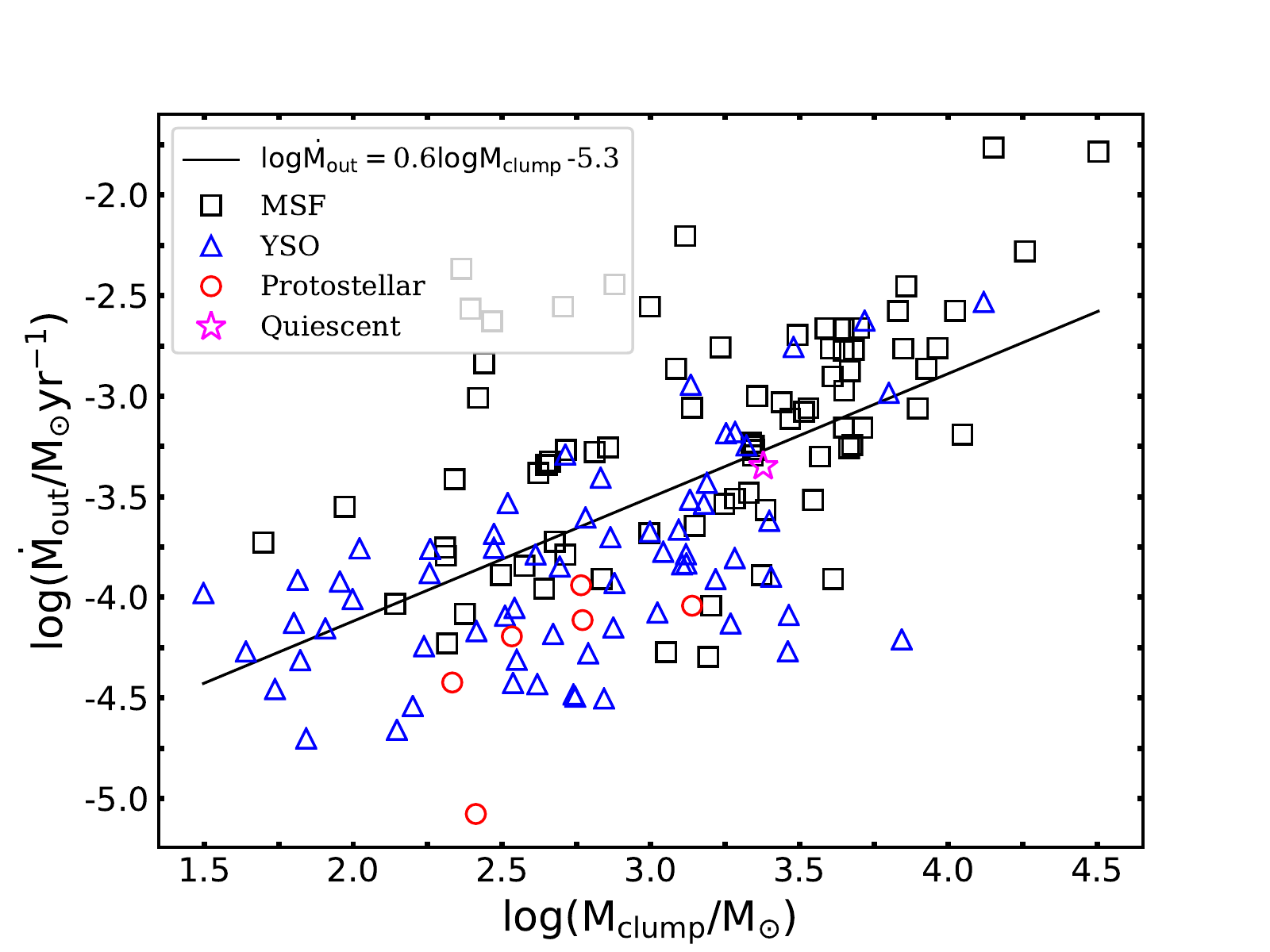}\\
   \includegraphics[width = 0.5\textwidth]{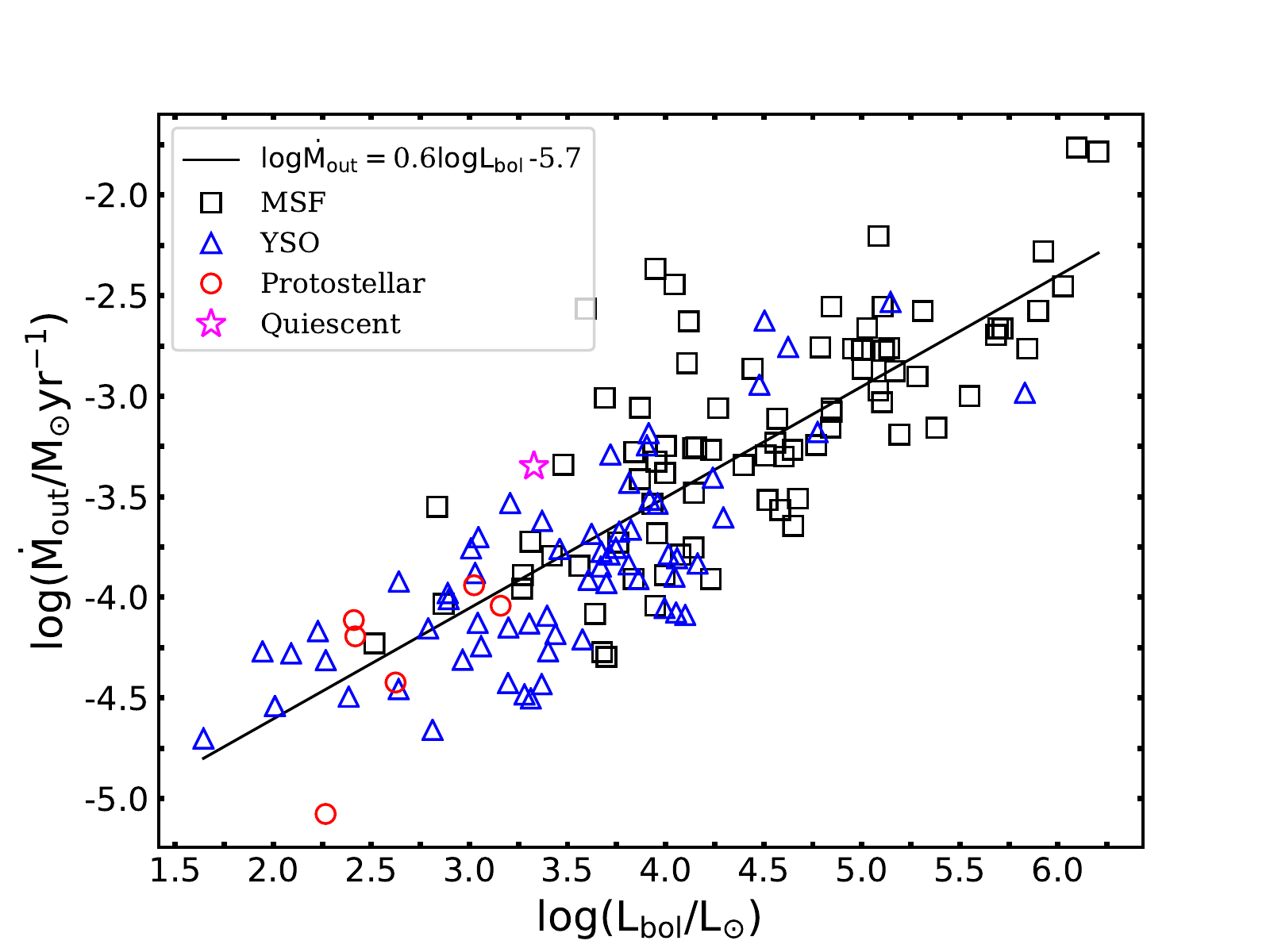}\\
   \includegraphics[width = 0.5\textwidth]{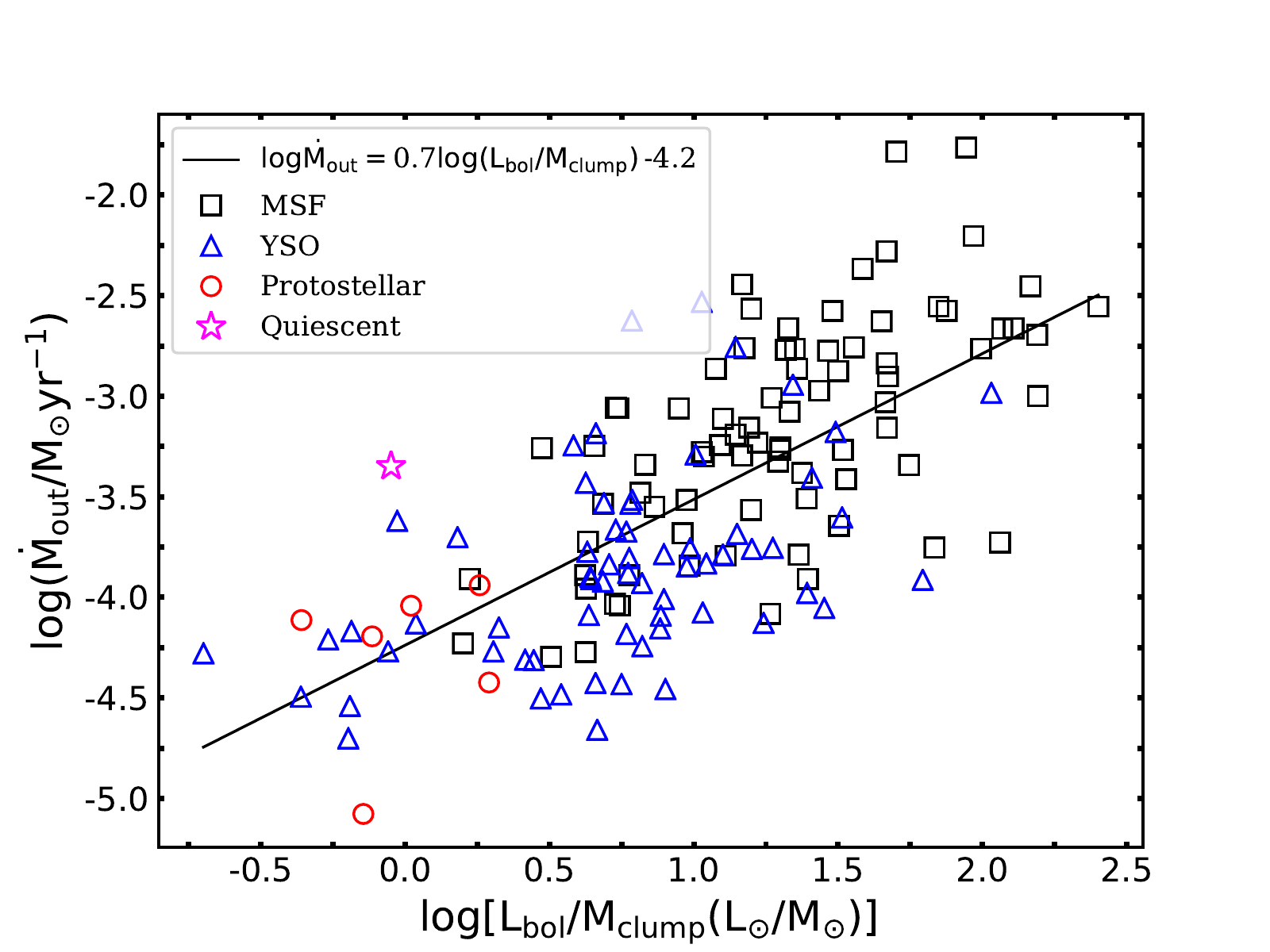}\\  
   \end{tabular}
 \caption{ Top panel: outflow mass-loss rate versus clump masses. Middle panel: outflow mass-loss rate  versus bolometric luminosity.  
 Bottom panel:  outflow mass-loss rate versus luminosity-to-mass ratio. 
 The symbols represent the same source types as in Figure \ref{fig:Mout_vs_clumpparam}. 
 The solid line in each plot is the least-squares linear fit line in logarithmic scale. }
\label{fig:Moutrate_vs_clumpparam}
\end{figure}

We present the relationships  of $\dot{ M}_{\rm out}$ as a function of $ M_{\rm clump}$, $ L_{\rm bol}$, and $L_{\rm bol}/M_{\rm clump}$   in the top, middle and bottom panels of Figure~\ref{fig:Moutrate_vs_clumpparam} and the correlation coefficients and results of linear fits to the data in Table \ref{outflowparam_vs_clumparam}.
The tight correlation between outflow mass-loss rate and clump mass suggests that higher-mass clumps host protostellar objects that have higher outflow activity, which agrees with previous results \citep{deVilliers2014MNRAS444} within the uncertainties. Furthermore, it is possible to give a rough estimation for the average accretion rate (${\dot{\rm M}}_{\rm accr}$) from the mean outflow mass-loss rate as ${\dot{\rm M}}_{\rm accr}\sim{\dot{\rm M}}_{\rm out}/6$, by following the same strategy as in \citet{Beuther2002AA383B} and \citet{deVilliers2014MNRAS444}, which are based on star formation models \citep[e.g.,][]{Tomisaka1998ApJ502L,Shu1999ASIC193S}. The mean outflow mass-loss rate is $ {\dot{\rm M}}_{\rm out} = 9.2\times 10^{-4}\,M_\odot\,yr^{-1}$ in our sample, and the approximate mean accretion rate is ${\dot{\rm M}}_{\rm accr}\sim{\dot{\rm M}}_{\rm out}/6 \sim 1.5 \times10^{-4}\,\rm M_{\odot}\,yr^{-1}$, which is the same order of magnitude as the $\rm \sim 10^{-4}\,M_{\odot}\,yr^{-1}$ found by previous studies of luminous YSOs and \hii\ regions \citep[e.g.,][]{Beuther2002AA383B,Zhang2005ApJ625,Kim2006ApJ643,Lopez_Sepulcre2009AA499L,deVilliers2014MNRAS444}. 

The correlation of mass entrainment rate ($\dot{ M}_{\rm out}$) and bolometric luminosity ($\rm L_\odot$) has been discussed in a number of previous studies  \citep{Cabrit1992AA274C,Shepherd1996ApJ472S,Henning2000AA211H,Beuther2002AA383B,Lopez_Sepulcre2009AA499L}, all of which have reported a similar relation that higher-luminosity objects are associated with higher outflow mass entrainment rates.  From this relation, \cite{Shepherd1996ApJ472S} suggested that massive stars are responsible for the observed outflow power. \citet{Beuther2002AA383B} proposed that the mass entrainment rate does not depend strongly on the luminosity for sources $\rm L_{bol}>10^{3}\,L_\odot$.
However, \citet{Henning2000AA211H} suggested that a correlation between the mass entrainment rate and luminosity for low-, intermediate-, and high-luminosity objects. Our study confirms that a tight positive correlation exists between outflow mass-loss rates and luminosity for objects of all luminosities. 

Furthermore, the entrainment rates ($\rm \dot{M}_{\rm out}$) are also related to the luminosity-to-mass ratio ($L_{\rm bol}/M_{\rm clump}$ ) of the clump, which suggests that a higher entrainment rate is associated with more evolved protostars with larger values of $L_{\rm bol}/M_{\rm clump}$ . This indicates that the accretion rate increases with the evolution of star formation in the clump, providing strong support for theoretical models that predict that accretion rates  increase as a function of time \citep[e.g.,][]{Bernasconi1996AA829B,Norberg2000AA1025N,Behrend2001AA190B,Haemmerl2013AA112H}. The partial correlation coefficient $\rho=0.68$ is larger than $\rho=0.47$ for $\dot{ M}_{\rm out}$ and $ M_{\rm clump}$, but smaller than $\rho=0.80$ for $\dot{ M}_{\rm out}$ and $ L_{\rm bol}$. 

%%%figure%%%
\subsubsection{$F_{\rm CO}$ versus $ M_{\rm clump}$, $ L_{\rm bol}$ and 
$L_{\rm bol}/M_{\rm clump}$ } 

%%%figure%%%
\begin{figure}
%\centering
\begin{tabular}{c}
\includegraphics[width = 0.5\textwidth]{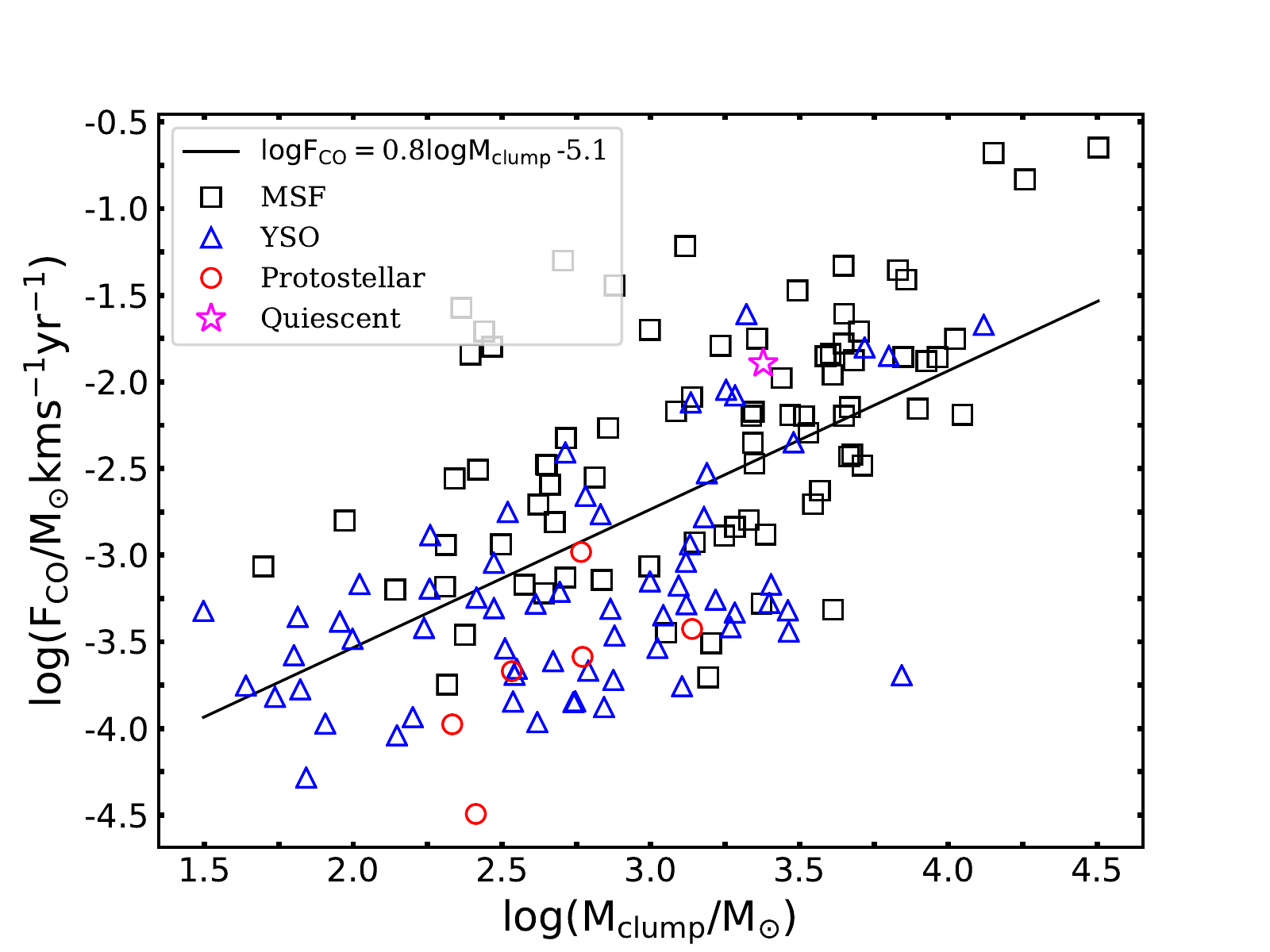} \\
\includegraphics[width = 0.5\textwidth]{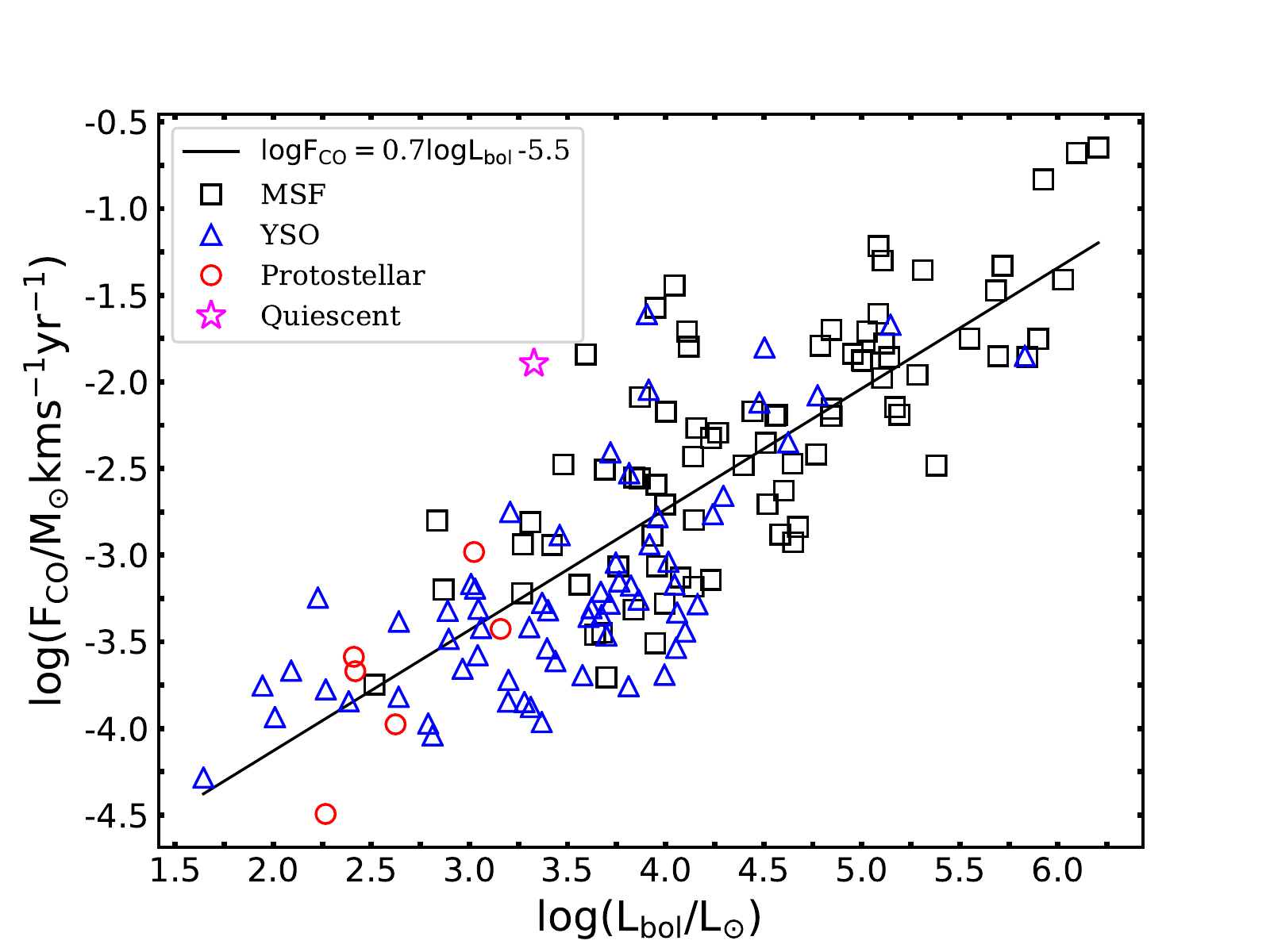}\\
\includegraphics[width = 0.5\textwidth]{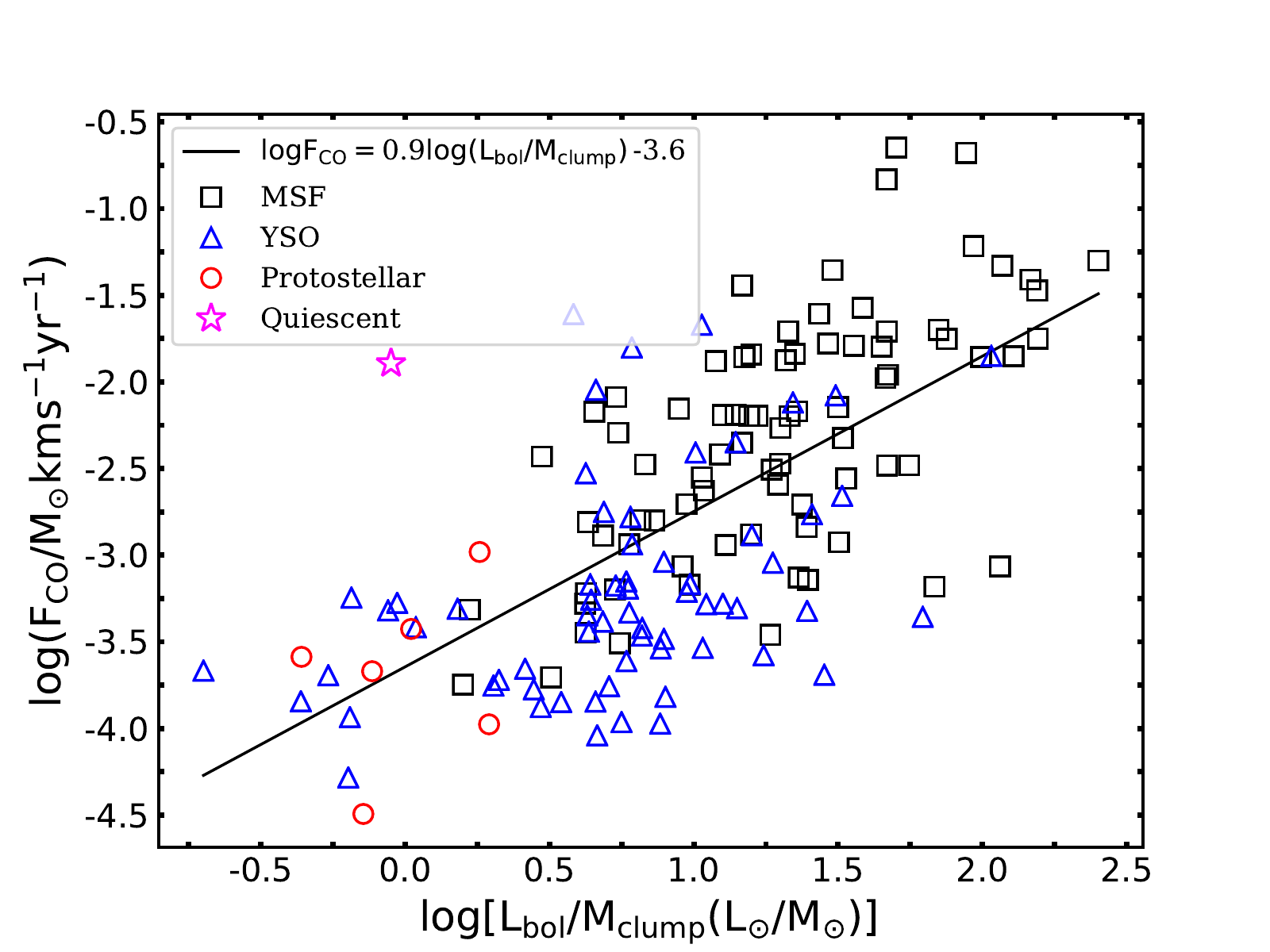}\\  
\end{tabular}
\caption{ Top panel:  outflow mechanical force versus clump mass. 
  Middle panel: outflow mechanical force  versus bolometric luminosity.  
  Bottom panel: outflow mechanical force versus luminosity-to-mass ratio. 
 The symbols represent the same source types as in Figure \ref{fig:Mout_vs_clumpparam}. 
 The solid line in each plot is the least-squares linear fit line in logarithmic scale. }
\label{fig:Fco_vs_clumpparam}
\end{figure}
 %%%figure%%%
 
 We present the outflow mechanism force $F_{\rm CO}$ as a function of clump properties of $ M_{\rm clump}$, $ L_{\rm bol}$, and $L_{\rm bol}/M_{\rm clump}$  in the top, middle and bottom panels of Figure~\ref{fig:Fco_vs_clumpparam} and the correlation coefficients and results of linear fits to the data in Table \ref{outflowparam_vs_clumparam}.
 
The mechanical force of an outflow ($F_{\rm CO}$, also known as the outflow momentum flux) is the ratio of the momentum to the dynamical age of the outflow and can be used as a measure of the outflow strength and the rate at which momentum is injected into the clump by the outflow \citep{Bachiller1999ASIC227B,Downes2007AA873D}. 
Many central studies have reported that outflow force is positively correlated with  clump (or core) mass and luminosity \citep[e.g.][]{Cabrit1992AA274C,Bontemps1996AA311,Shepherd1996ApJ472S,Wu2004AA426W,Zhang2005ApJ625}. We find similar {  positive} correlations in our sample in Figure \ref{fig:Fco_vs_clumpparam}. 

% which is usually taken to mean that there is a common driving mechanism of outflows for low and high mass star formation
%which provides further confirmation that low-mass and high-mass star formation may share a similar mechanism for outflows

Interestingly, we also find a positive correlation between the outflow force and the luminosity-to-mass ratio of the clump (see Figure \ref{fig:lum_vs_Fco} and \ref{fig:Fco_vs_clumpparam}), which suggests that as the star formation evolves within the clump, the outflow force increases.   In our sample (see figure \ref{fig:Fco_vs_clumpparam}), the most powerful outflows originate within the most evolved MSF clumps, whereas the first three evolutionary stages (e.g., quiescent, protostellar, and YSO) are associated with less powerful outflows. This is in contradiction to  studies of low-mass star formation, which show a decrease  in the outflow force  between the Class 0 and I stages \citep{Bontemps1996AA311, Curtis2010MNRAS408}.  We investigate this point in more detail in Section \ref{sect:implications}. The partial correlation coefficient $\rho=0.65$ is larger than $\rho=0.51$ for $F_{\rm CO}$ versus $ M_{\rm clump}$, but smaller than $\rho=0.79$ for $F_{\rm CO}$ versus $ L_{\rm bol}$. 

\subsubsection{$L_{ \rm CO}$ versus $ M_{\rm clump}$, $ L_{\rm bol}$ and $L_{\rm bol}/M_{\rm clump}$ } 
%%%figure%%%

\begin{figure}
%\centering
\begin{tabular}{c}
\includegraphics[width = 0.5\textwidth]{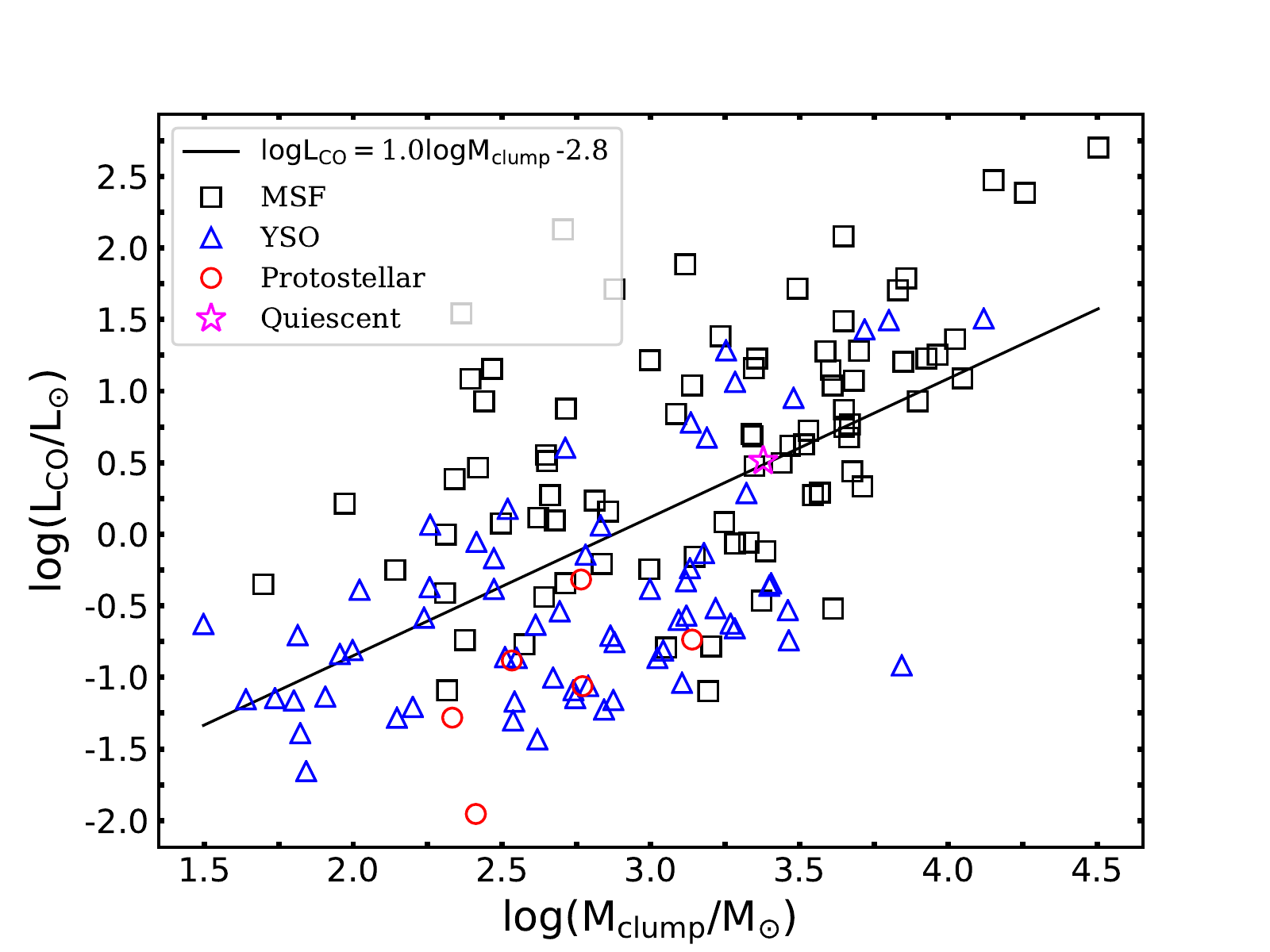}\\ 
\includegraphics[width = 0.5\textwidth]{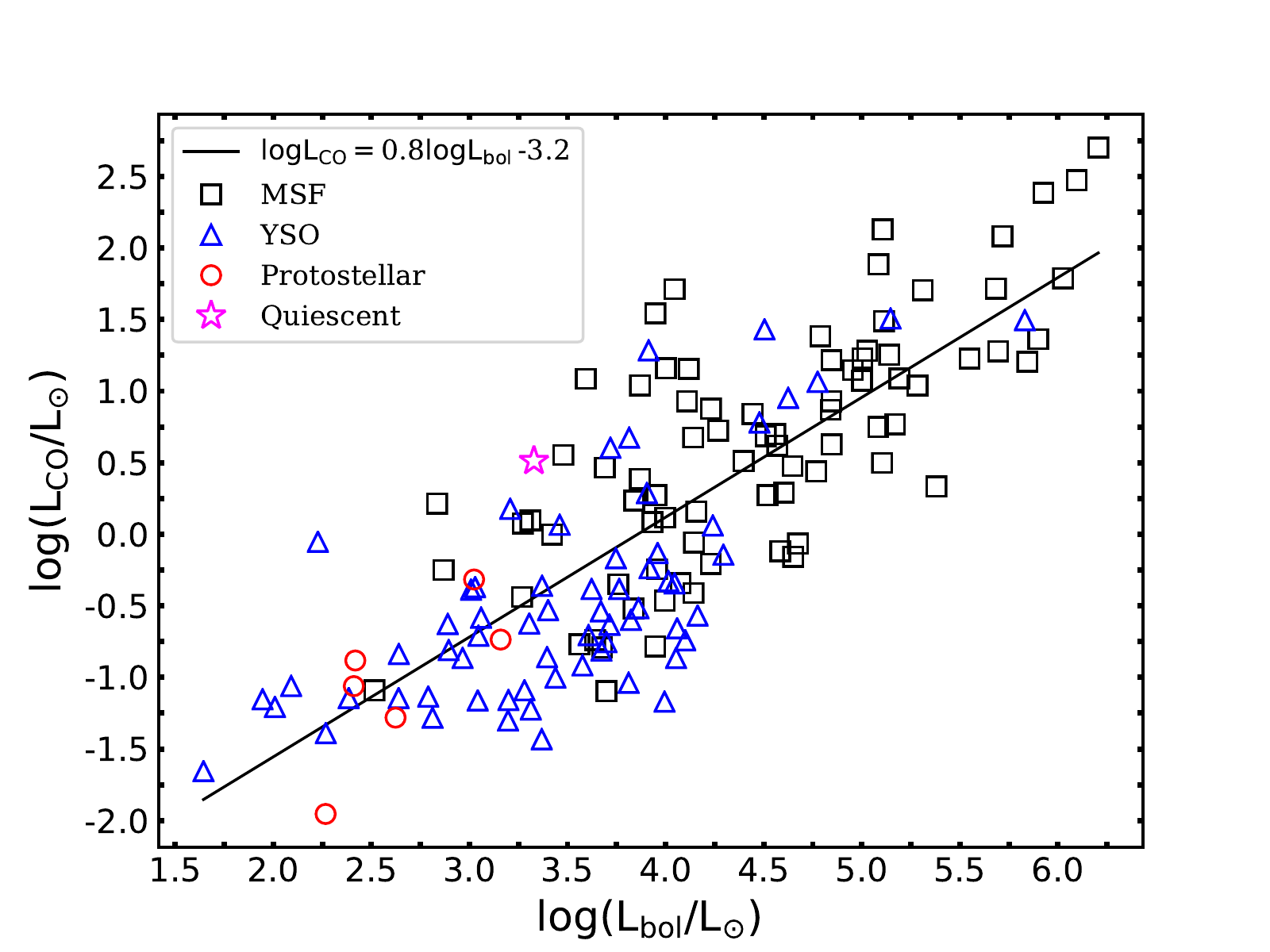}\\
\includegraphics[width = 0.5\textwidth]{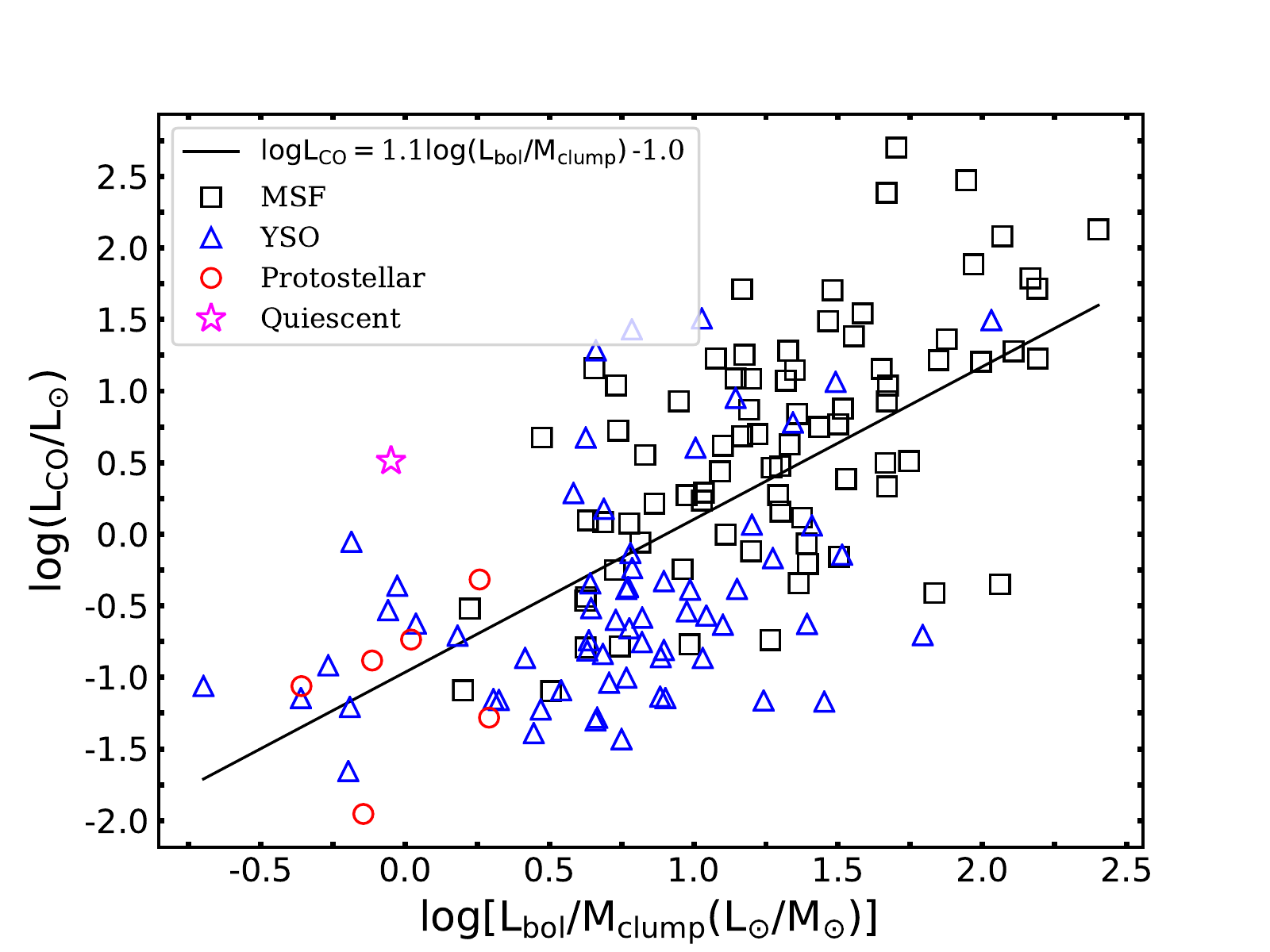}\\  
\end{tabular}
\caption{  Top panel: outflow mechanical luminosity versus clump mass. 
   Middle panel: outflow mechanical luminosity  versus bolometric luminosity. Bottom panel: outflow mechanical luminosity versus luminosity-to-mass ratio. The markers represent the same source types as in Figure \ref{fig:Mout_vs_clumpparam}. 
 The solid line in each plot is the least-squares linear fit line in logarithmic scale. }
\label{fig:Lco_vs_clumpparam}
\end{figure}
%%%figure%%%
%

We present the relations between the  outflow mechanism luminosity $L_{ \rm CO}$ and the outflow properties of $ M_{\rm clump}$, $ L_{\rm bol}$, and $L_{\rm bol}/M_{\rm clump}$  in the top, middle and bottom panels of Figure \ref{fig:Lco_vs_clumpparam} and  the correlation coefficients and results of linear fits to the data in Table \ref{outflowparam_vs_clumparam}.
Tight relations exist between outflow mechanical luminosity $L_{ \rm CO}$ and clump mass, bolometric luminosity $ L_{\rm bol}$, and luminosity-to-mass ratio $L_{\rm bol}/M_{\rm clump}$ . The relation between $L_{ \rm CO}$ and $ L_{\rm bol}$ ($\rm L_{CO}\propto L_{bol}^{0.8}$) is similar to the reported correlations of $\rm L_{CO}\propto L_{bol}^{0.8}$ for embedded YSOs in \citet{Cabrit1992AA274C} and slightly larger than $\rm L_{CO}\propto L_{bol}^{0.6}$ for both low-mass and high-mass groups in \citet{Wu2004AA426W}. 
The average of value of $\rm L_{CO}/L_{bol}$ is  $\sim 3\times10^{-4}$. 

Similar to outflow force, the mechanical luminosity is also related to the luminosity-to-mass ratio of the clump, suggesting that clumps with more evolved star formation are associated with more powerful outflows. {  In Figure\,\ref{fig:Lco_vs_clumpparam}, the most luminous outflow comes from the most evolved MSF clumps, and the first three evolutionary stages (e.g., quiescent, protostellar, and YSO) are associated with less luminous outflows.} The partial correlation coefficient $\rho=0.62$ is larger than $\rho=0.54$ for $L_{ \rm CO}$ and $ M_{\rm clump}$, but smaller than $\rho=0.79$ for $L_{ \rm CO}$ and $ L_{\rm bol}$. There is a close correlation between the clump luminosity and the mechanical luminosity.

\subsubsection{Implications of the Clump-Outflow Correlations}
\label{sect:implications}

As suggested by \citet{McKee2003ApJ850M}, the accretion rate during star-formation is proportional to the surface density of the clumps $\rm \Sigma^{0.75}$. This indicates that the most massive and dense clumps harbor stars with higher accretion rates than those forming in lower-mass clumps. 
\citet{Urquhart2013MNRAS} found that the bolometric luminosities of a sample of MSF clumps were tightly correlated with the Lyman continuum flux emitted from their embedded \hii\ regions and therefore demonstrated that the vast majority of the observed luminosity could be directly attributed to the most massive stars in the clumps. Furthermore, \citet{Urquhart2013MNRAS} also found a tight correlation between the clump mass and the mass of the most massive stars that showed that the most massive stars are preferentially found toward the center of the most massive clumps in the highest column density regions. The correlations between outflow and clump properties in the above section suggest that higher clump masses and more luminous and evolved central sources are associated with much more powerful outflows, together with higher entrainment masses, larger entrainment mass rates, stronger outflow force, and higher outflow mechanical luminosity. 
Furthermore, the luminosity of the clumps shows that the strongest relation with outflow properties as its correlation coefficient is the highest. This provides support that the outflow may be dominated by the most luminous and massive source within clumps, as the luminosity of the clump is in turn largely provided by that of the most massive protostar or YSO within the clump \citep{Urquhart2013MNRAS,Urquhart2014MNRAS1555U}. It is difficult to resolve the contributions from a single massive protostar or YSO in clumps \citep{Urquhart2013MNRAS}; however, the tight relation between the most massive and luminous clumps associated with most powerful outflows from our investigation is statistically reliable.

The mean accretion rates $\rm \sim 10^{-4}\,M_\odot\,yr^{-1}$ estimated by our sample are large enough to overcome the strong radiation from massive protostars, which supports the expected accretion rates in theoretical models of MSF \citep[e.g.,][]{Bonnell2001MNRAS785B,Krumholz2007ApJ959K}.

We saw a positive correlation between outflow force and mechanical luminosity with {  clump} luminosity and luminosity-to-mass {  ratio}, which at first sight indicates that the outflows increase in force and luminosity as the star formation evolves. However, the slopes between outflow properties and clumps parameters are rather shallow ($<1$) in log-log space (see Table\,\ref{outflowparam_vs_clumparam}), which suggests that the outflow properties evolve more slowly than do $ L_{\rm bol}$, $ M_{\rm clump}$ and $L_{\rm bol}/M_{\rm clump}$ . This may indicate a decrease in the mass accretion rate (and resulting mass outflow rate) whilst the luminosity of the central YSOs continues to increase. Alternatively this may be caused by a decrease in the amount of entrained material as the outflow cavities become more developed. Another possibility is that while larger clumps are associated with more massive and luminous sources and these drive more powerful outflows perhaps less of the total luminosity is emanating from the star driving the outflow. Finally, almost all of our outflow sample are comprised of mid-infrared bright clumps with $\rm L_{bol}/M_{clump}\sim\,10\,(L_{\odot}/M_{\odot})$, which indicates that they are all likely to be at a similar evolutionary stage close to the end of the main accretion phase \citep[see Figure 24 in][]{Urquhart2017arXiv392U} and which may explain the tight correlations between parameters.

\subsection{Comparison with Low-mass Outflows}

%%%figure%%%
The results of our least-squares fitting between outflow parameters and clump properties in this work are consistent with the relations seen in low-mass outflows \citep{Cabrit1992AA274C,Wu2004AA426W}. Here we present a comparison between the outflow force for low luminosities and high luminosities to illustrate their connection.

\begin{figure}
%\centering
\includegraphics[width = 0.50\textwidth]{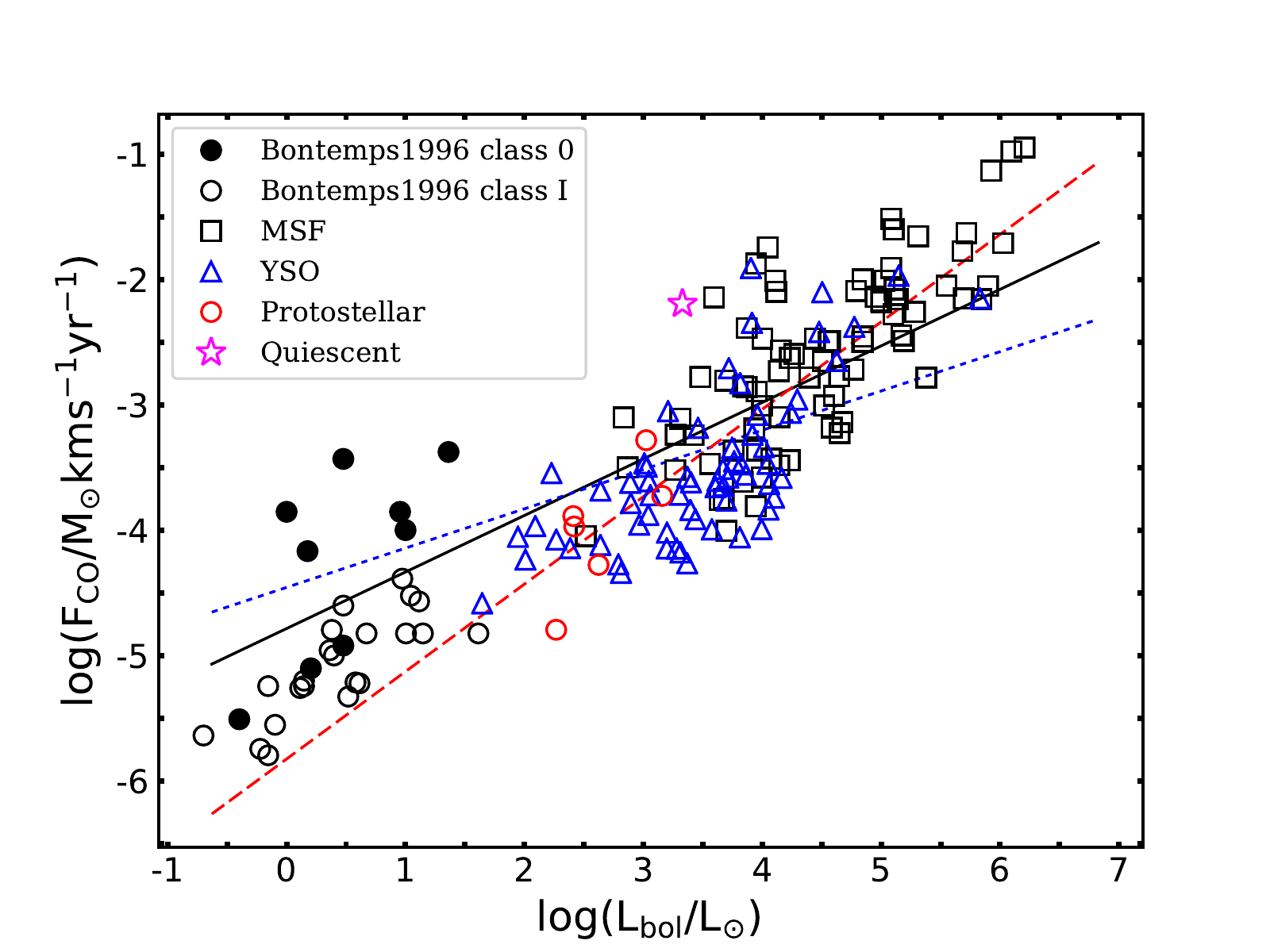}   
\caption{  Outflow force $ F_{\rm CO}=$$\dot{\rm P}$ versus 
 the bolometric luminosity $ L_{\rm bol}$ of central sources. The symbols represent the same source types as in Figure \ref{fig:Mout_vs_clumpparam}. To compare with low-mass outflow force, the filled and open circles, respectively, indicate the Class 0 and Class I from \citet{Bontemps1996AA311}. The black solid line   ($\rm logF_{CO}=0.5logL_{bol}-4.7$)} represents the best fit in $\rm log-log$ space between $F_{\rm CO}$ and $ L_{\rm bol}$ for both low-mass and high-mass outflows. The red dashed line { ($\rm logF_{CO}=0.7logL_{bol}-5.5$) is the best fit for all massive outflows in this study and extrapolated to lower luminosities. The blue dotted line ($\rm logF_{CO}=0.3logL_{bol}-4.5$) indicates the best fit for low-mass outflows from \citet{Bontemps1996AA311}, extended to higher luminosities. }
 
\label{fig:lum_vs_Fco}
\end{figure}

Figure \ref{fig:lum_vs_Fco} plots the outflow force $ \rm F_{CO}$ 
against luminosity for MSF, YSO, protostellar, 
and quiescent clumps in the sample of 153 mapped outflows, together with outflows 
associated with Class 0 and Class I protostars/YSOs from \citet{Bontemps1996AA311}. The outflow mechanical force values have been inclination-corrected using an average angle of $57\degr.3$ for this work, and this has also been applied to the results of \citet{Bontemps1996AA311}. In Figure \ref{fig:lum_vs_Fco}, we see a continuous relationship between outflow force and luminosity that holds over 7 orders of magnitude.  This supports the hypothesis that a similar outflow mechanism may operate for both low-mass and high-mass star formation. However, when low-luminosity and high-luminosity sources are fitted separately, we find a slight difference between low and high luminosity samples, which implies that the existence of a common outflow mechanism is not as clear-cut as Figure \ref{fig:lum_vs_Fco} suggests. This small difference was also found by \citet{Maud2015MNRAS645M}. Nevertheless, we \citep[and][]{Maud2015MNRAS645M} cannot exclude the possibility of  systematic error between the outflow force of low-luminosity and high-luminosity samples given that they lie at very different distances and were observed using different techniques. The main cause of the different slope in the outflow force-luminosity relation is the Class 0 sources observed by \citet{Bontemps1996AA311}. A larger and more consistently analyzed sample of low-luminosity sources, perhaps from the JCMT Gould Belt survey \citep{Ward-Thompson2007}, is required to investigate any potential systematic bias. However, this work lies beyond the scope of our study. 

In conclusion, we find that outflows are ubiquitous phenomena for both high-mass and low-mass groups with a potentially similar driving mechanism.

\subsection{The Evolution of the Outflows and Clump Turbulence} 

Outflow feedback can help address two main questions: (a) do outflows inject enough momentum to maintain turbulence, and (b) can outflows properly couple to clump gas to drive turbulent motions \citep{Frank2014prplconf451F}. 
It remains a challenge to quantify the cumulative impact of  the outflow-driven turbulence on molecular clouds. One method to measure the effect that outflows have on their parent clumps is to compare the total outflow energy and the turbulent kinematic energy. The turbulent kinematic energy can be estimated by $\rm E_{turb}=(3/16\,ln2)M_{core}\times {FWHM}^2$ {  \citep{Arce2001ApJ132A}}, if the thermal motions are a negligible contribution to the full width half maximum (FWHM) of $\rm C^{18}O\,J=3-2 $  \citep{Arce2001ApJ132A,Maud2015MNRAS645M}. Our large sample of clumps and outflows with well-determined physical properties allows us to statistically investigate the correlation between outflow energy and turbulence energy at different evolutionary stages of central sources, as well as examine the impact that outflows have on their natal clumps. 

\begin{figure}
%\centering
\includegraphics[width = 0.50\textwidth]{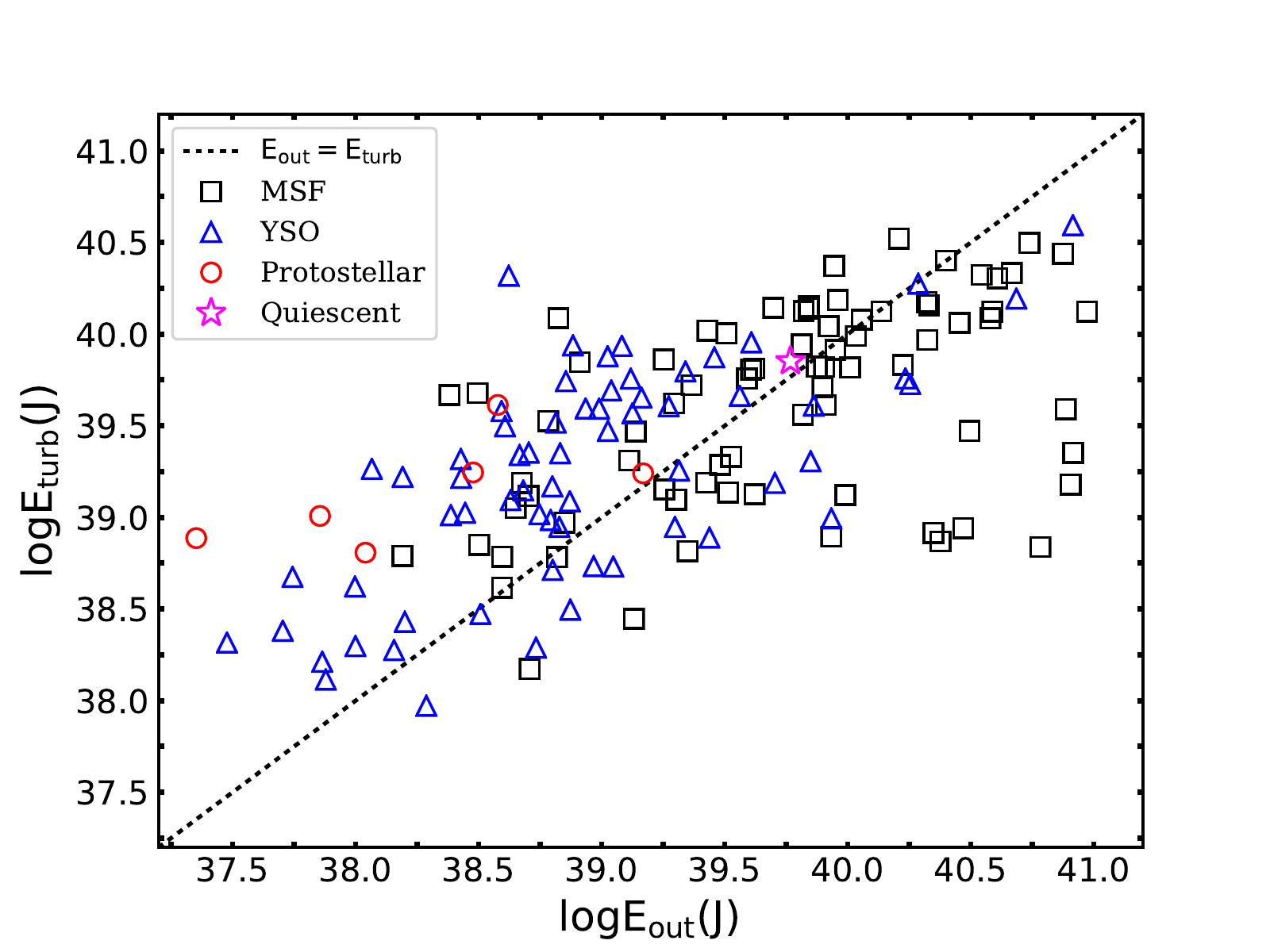}\\    
\caption{  Outflow energy of the 153 outflow clumps against turbulence energy. The symbols represent the same source types as in Figure \ref{fig:Mout_vs_clumpparam} and the dotted line shows $\rm E_{out}=E_{turb}$. This may indicate that outflow energy is comparable to turbulence energy for the majority clump. } %with mean $\rm %E_{out}/E_{turb}\sim0.83$, however, outflow energy is insufficient to maintain turbulence in the clump.and have enough power to drive turbulence in the local environment.

\label{fig:Eout_vs_Eturb}
\end{figure}

In Figure\,\ref{fig:Eout_vs_Eturb}, we see that the outflow energy ($\rm E_{out}$) appears to be comparable to the turbulent energy ($\rm E_{turb}$ ) for the 153 clumps with mapped outflows in our sample.   The mean value of $\rm E_{out}/E_{turb}\sim3.3$, with a spread of $0.02\sim88$. This suggests that the outflows associated with most clumps have enough energy to maintain turbulence. \citet{Cunningham2009ApJ816C} proposed that jet-driven outflows can provide an efficient form of dynamical feedback and act to maintain turbulence in the molecular cloud. 
However, some authors suggested that outflows have enough power to drive turbulence in the local environment \citep[e.g.,][]{Arce2010ApJ1170A,Mottram2012MNRAS420M,Maud2015MNRAS645M} but do not contribute significantly to the turbulence of the clouds \citep[e.g.,][]{Arce2010ApJ1170A,Maud2015MNRAS645M,Plunkett2015ApJ22P}.

 Looking at each subgroup in more detail in Figure\,\ref{fig:Eout_vs_Eturb}, we can see that all clumps in the first two evolutionary stages (i.e., quiescent and protostellar) lie above the line of equality of $\rm E_{turb}$ and $\rm E_{out}$ (i.e., $\rm E_{turb}> E_{out}$). 
 This is consistent with \citet{Graves2010MNRAS1412G}, who found the total outflow energy to be smaller than the turbulent kinetic energy of the cloud.
For the three more evolved stages, the mean turbulence kinematic energy is consistent with each other within the uncertainties, i.e., protostellar clumps ($\rm E_{turb}\sim1.4\times10^{39}\,J$), YSO clumps ($\rm E_{turb}\sim3.9\times10^{39}\,J$) and MSF clumps ($\rm E_{turb}\sim9.8\times10^{39}\,J$), while the mean outflow energy increases as the clump evolves, i.e., protostellar clumps ($\rm E_{out}\sim3.9\times10^{38}\,J$), YSO clumps ($\rm E_{out}\sim4.0\times10^{39}\,J$), and MSF clumps ($\rm E_{out}\sim3.5\times10^{40}\,J$). 
  Thus, the mean ratio of $\rm E_{out}/E_{turb}$ increases from $\sim0.3$ to $\sim1$ and then to $\sim3.6$ as the clump evolves from the protostellar to MSF stage. 
 This may imply that no matter whether the outflows have not  (e.g., protostellar stage) or have (e.g., YSO or MSF phase) enough energy to fully drive the turbulence, outflow energy does not significantly contribute to the turbulence energy of the parent clump as the protostar evolves. This is consistent with simulation preformed by \citet{Krumholz2012ApJ71K} that showed that outflow-driven feedback has a smaller impact on MSF regions.

%%figure%%%
\begin{figure*}
\centering
\begin{tabular}{cc}
\includegraphics[width = 0.50\textwidth]{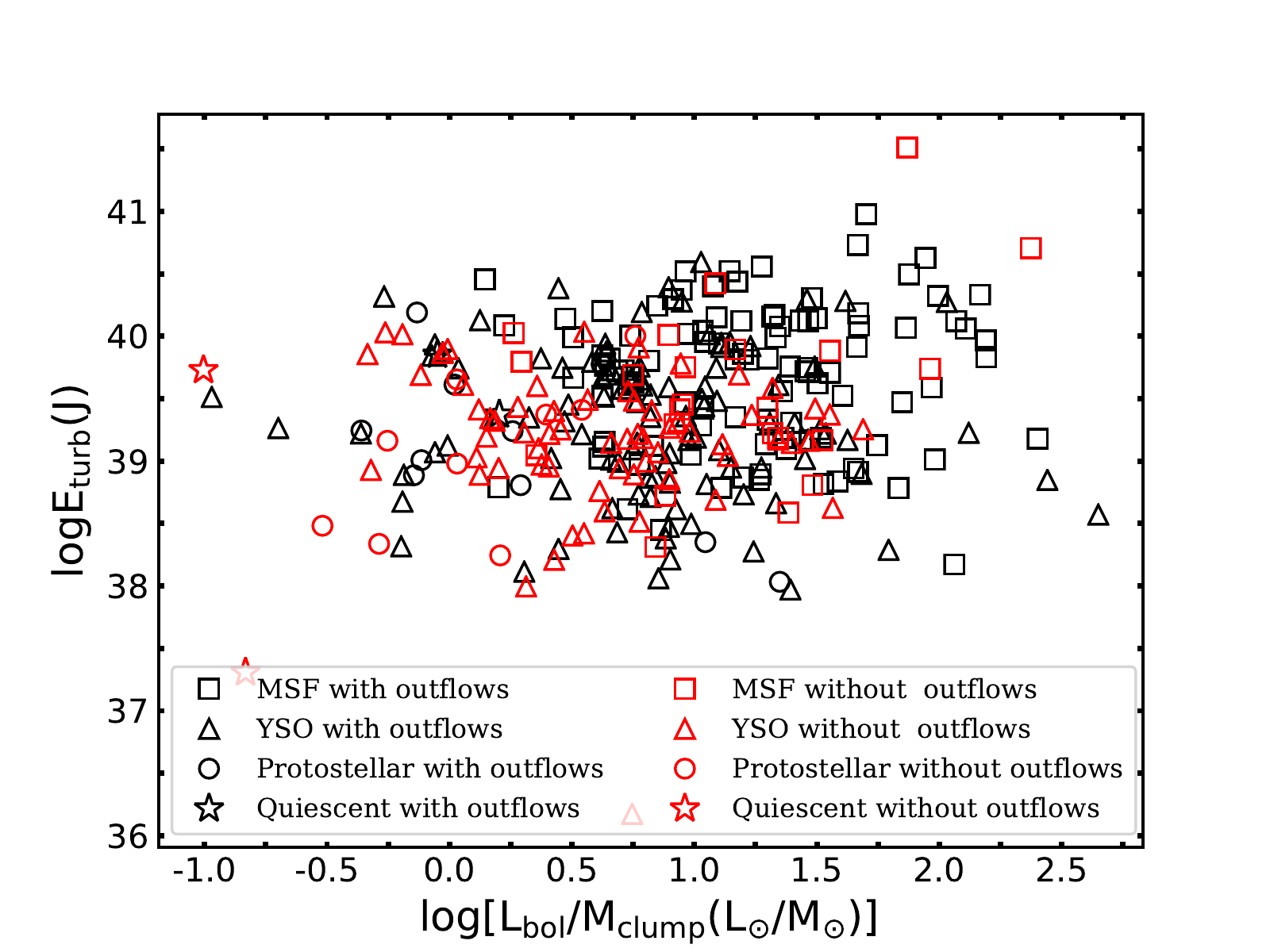}& \includegraphics[width = 0.50\textwidth]{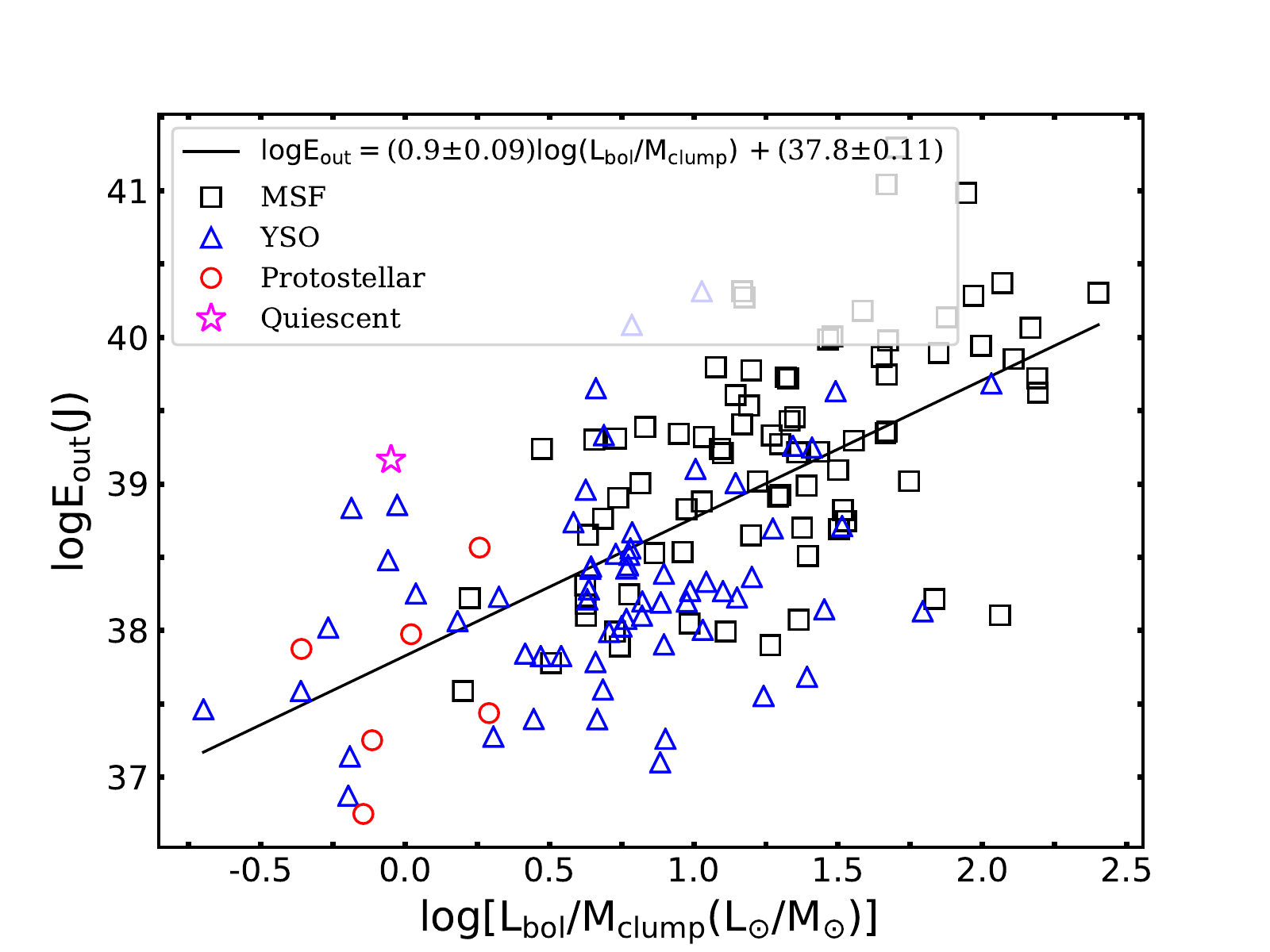}\\   %
\end{tabular}
\caption{Left panel: turbulence energy ($\rm E_{turb}$) of the 314 clumps against the luminosity-to-mass ratio of central objects. The symbols refer to the same source type as in Figure~\ref{fig:Mout_vs_clumpparam}, with all black markers representing clumps with detected outflows and all red markers indicating clumps with no outflows. It is shown that the clumps without outflows appear to have the range of turbulence energy similar to the clumps with no outflows. More interestingly, the level of turbulence is not significantly affected by the evolution of central sources in the clumps.   Right panel: the plot of outflow energy ($\rm E_{out}$) versus the luminosity-to-mass ratio of central sources for the 153 clumps with mapped outflows. This suggests that higher outflow energies are associated with more evolved objects (i.e., larger value of $L_{\rm bol}/M_{\rm clump}$ ).
  }
\label{fig:lrm_vs_Eturb}
\end{figure*}
%%%figure%%%

In the left panel of Figure\,\ref{fig:lrm_vs_Eturb},  we show the turbulent energy ($\rm E_{turb}$) versus the luminosity-to-mass ratio of the clump ($L_{\rm bol}/M_{\rm clump}$ ) for the 314 clumps in our sample with measured clump properties and C$^{18}$O detections. The average value of $\rm E_{turb}$ is $\rm \sim7.0\times10^{39}\,J$ for clumps that show outflows, which is consistent with the value of $\rm \sim6.8\times10^{39}\,J$ for clumps that do not contain outflows, within uncertainties. In addition, there is no difference for the range of turbulent energy values between clumps with outflows and clumps without outflows, which implies that star-forming clumps have a similar level of turbulence as quiescent clumps. This is consistent with studies mentioned in \citet{Hennebelle2012AARv55H} for clouds with and without star-forming activity showing similar velocity dispersion \citep{Kawamura2009ApJs184K} and presenting comparable levels of turbulence \citep{Williams1994ApJ693W}. 
Furthermore, it is clear that there is no obvious correlation between the $\rm E_{turb}$ and $L_{\rm bol}/M_{\rm clump}$ , with Spearman's rank coefficient $\rho=0.08$ and $p$-value=0.33, which suggests that the level of turbulence in the clump is not significantly affected by the evolution of the central object. This is consistent with the analysis of $\rm NH_3$ line widths of  $\sim\,8000$ dense clumps as a function of evolution of embedded protostars ($L_{\rm bol}/M_{\rm clump}$ ) in \citet{Urquhart2017arXiv392U}. In addition, the right panel of Figure \ref{fig:lrm_vs_Eturb} shows that the outflow energy is strongly correlated with the evolution of central objects, with Spearman's rank coefficient $\rho=0.6$ and $p$-value$\ll0.001$, indicating that higher outflow energies are associated with more evolved objects (i.e., larger value of $L_{\rm bol}/M_{\rm clump}$ ).

All these findings imply that the outflow energy from embedded protostars should increase as they evolve; in addition, this outflow energy is large enough to maintain the turbulence in the clump (see Figure\,\ref{fig:Eout_vs_Eturb}). However, outflow energy does not significantly contribute to the energy of the turbulence in the clump as the protostar evolves (see left panel of Figure\,\ref{fig:lrm_vs_Eturb}).  The level of turbulence is similar for clumps associated with outflows and not associated with outflows at four evolutionary stages, which suggests that the origin of the turbulence occurs before the star formation begins.  This is consistent with several other studies \citep[e.g.,][]{Ossenkopf2002AA307O,Brunt2009AA883B,Padoan2009ApJ707L} that suggest that turbulence is mostly driven by large-scale mechanisms that originate outside the cloud (e.g.~supernovae). 

\section{Summary and Conclusions}

We have carried out an unbiased outflow survey toward 919 ATLASGAL clumps located in the CHIMPS survey, 325 of which have $\rm{^{13}CO}$ and $\rm C^{18}O$ data that are suitable for outflow identification. The physical properties of the 325 clumps are shown in Table\,\ref{clump_properties}. We detect high-velocity outflow wings toward 225 clumps by inspecting the line wings in the one-dimensional $\rm{^{13}CO}$ spectra extracted at the centroid of each clump (see Table\,\ref{outflow_wings} for details). We investigate these wings further by mapping the $\rm{^{13}CO}$  integrated intensity corresponding to each wing. We are able to estimate the outflow properties for 153 clumps that are found to have well-defined bipolar outflows and reliable distances. These properties are given in Table\,\ref{outflow_phyparam}. The overall physical properties of the clumps are summarized in Table\,\ref{tab_summary_param}.
We show that the outflows discovered here are more than 2 orders of magnitude more massive and energetic than the outflows associated with low-mass objects. We compare outflow properties with clump characteristics, discuss how the properties of this large homogeneous sample change as a function of evolution, and examine their impact on the turbulence of their natal clumps.

The main results are summarized as follows.  
\begin{enumerate}
\item 
 Of the 325 massive clumps, 225 show high-velocity line wings indicative of outflows, implying a $69\%\pm3\%$ detection frequency of CO outflows. Among the 225 sources, 10 clumps have single blue/red wings and the rest 215 show both blue and red wings. The detection frequency bipolar outflows is $66\%\pm3\%$,  while we find significantly higher outflow detection rates in \uchii\ regions (52/56, $93\%\pm3\%$), maser-associated sources (60/70, $86\%\pm4\%$), and \hchii\  regions (4/4, 100\%) in our sample. 
\item 
The 225 clumps with detected outflows have significantly higher $ M_{\rm clump}$, $ L_{\rm bol}$, and $L_{\rm bol}/M_{\rm clump}$  and higher $ N_{\rm H_2}$ compared to 100 clumps with no outflows. K-S tests for these parameters suggest that the two samples are from different populations. 
\item 
The detection rate of outflows increases with increasing of $ M_{\rm clump}$, $ L_{\rm bol}$, $L_{\rm bol}/M_{\rm clump}$  and $ N_{\rm H_2}$, which can be as high as 90\% when $\rm M_{clump}>3.9\times10^{3}\,M_{\odot}$, $\rm L_{bol}>2.7\times10^{4}\,L_{\odot}$, $\rm L_{bol}/M_{clump}>10\,(L_{\odot}/M_{\odot})$, and $\rm N_{H_2}>3.8\times10^{22}\,cm^{-2}$. 
The detection rates as a function of  $ N_{\rm H_2}$ are entirely consistent with the gas surface threshold density for efficient star formation suggested by \citet{Lada2010ApJ687L} and \citet{Heiderman2010ApJ1019H}. 

\item 
Outflow activity begins to switch on at the youngest quiescent stage (i.e., 70$\mu m$ weak) in this young sample. The detection frequency of the outflow is increasing as the clumps evolve from quiescent ($50\%\pm25\%$), to protostellar ($53\%\pm11\%$), to YSO ($61\%\pm4\%$), and then to MSF clump ($82\%\pm2\%$). The detection of outflow activity appears to peak (i.e., 100\%) at the pre-\uchii\ (e.g., \hchii\ regions, water and/or methanol masers) or early \uchii\ region phase (e.g., maser associated \uchii\ regions).

\item 
Outflow properties ($ M_{\rm out}$, $\rm \dot{M}_{out}$, $F_{\rm CO}$ and $L_{ \rm CO}$) are tightly correlated with the $ M_{\rm clump}$, $ L_{\rm bol}$, and $L_{\rm bol}/M_{\rm clump}$  of the clump when the effect of distance is controlled. The strongest relation between the outflow parameters and the clump luminosity may indicate that outflows are dominated by the energy of the most luminous source in the clump. These correlations are consistent with studies of both low-mass and high-mass samples, which leads us to conclude that they share a similar mechanism for outflows, although there exists the potential for systematic bias between low- and high-mass samples.

\item 
The mean outflow mass entrainment rate is $\rm 9.2\times10^{-4}\,M_\odot\,yr^{-1}$, suggesting a mean accretion rate of $\rm \sim10^{-4}\,M_\odot\,yr^{-1}$. This is the same order found in high-mass star formation regions \citep[e.g.,][]{Beuther2002AA383B,Kim2006ApJ643,deVilliers2014MNRAS444}, 
and is in agreement with the accretion rates predicted theoretical models of MSF \citep[e.g.,][]{Bonnell2001MNRAS785B,Krumholz2007ApJ959K}. 
Moreover, our results are also consistent with an increasing accretion rate as a function of time, which is an expected consequence of the number of theoretical models \citep[e.g.,][]{Bernasconi1996AA829B,Norberg2000AA1025N,Behrend2001AA190B,Haemmerl2013AA112H}. 
 
\item 
The outflow energy is comparable to the turbulent energy of the cloud {  with mean $\rm E_{out}/E_{turb}\sim3.3$.} While the outflow energy increases with increasing of $L_{\rm bol}/M_{\rm clump}$ , i.e., with the evolution of the central protostar, the turbulent energy does not. 
We find no obvious correlation between $\rm E_{turb}$ and $L_{\rm bol}/M_{\rm clump}$ . Thus, the outflow does not contribute significantly to clump turbulence as the clump evolves. This suggests that core turbulence might exist before star formation begins, which is consistent with  turbulence being mostly driven by large-scale mechanisms \citep[e.g.,][]{Ossenkopf2002AA307O,Brunt2009AA883B,Padoan2009ApJ707L}.

\end{enumerate}

These results  may suggest that outflow energies are dominated by the most massive and luminous protostars in the clumps. However, it is a challenge to resolve the contributions from single massive protostars in clumps. High angular resolution observations are needed to resolve individual outflows within the clumps. The large and homogeneously selected sample that we present here should form the basis for subsequent interferometric observations with ALMA and NOEMA.  

In addition, we have demonstrated the potential of wide-field Galactic plane surveys to discover a relatively unbiased selection of outflows. We look forward to the expansion of our study using the forthcoming CHIMPS2 survey, which will expand the area covered by CHIMPS to the remaining section of the first Galactic quadrant and potentially double the number of outflows discovered here.

\section*{Acknowledgements}
We would like to thank the anonymous referee for the helpful comments. We acknowledge support from the NSFC~(11603039, 11473038) from China's Ministry of Science and Technology. M.\,A.\,T. acknowledges support from the UK Science $\&$ Technology Facilities Council via grant ST/M001008/1. 
A.\,Y.\,Y. would like to thank the UK Science and Technology Facilities Council and the China Scholarship Council for grant funding through the China-UK SKA joint PhD program. 

This document was produced using the Overleaf web application, which can be found at www.overleaf.com.

\bibliographystyle{mn2e_new}
\bibliographystyle{apj}
\bibliography{ref}

\label{lastpage}
\appendix

\iffalse
\section{Appendix Figures}

\begin{figure*}
 \centering
     \begin{tabular}{cc}
   \includegraphics[width = 0.50\textwidth]{G029p954-00p016co_outflows.pdf}&\includegraphics[width = 0.50\textwidth]{G029p954-00p016_coimgw.pdf}\\  
      \end{tabular}
 \caption{Outflow wing selection: using spectra of the $\rm ^{13}CO$  (gray solid line) and $\rm C^{18}O$  (gray dashed line) for the ATLASGAL clumps. Blue and red wings identification process: (a) scaling the $\rm C^{18}O$ spectrum to $\rm ^{13}CO$ peak, shown as a red dash-dotted line; (b), fitting a Gaussian to the scaled $\rm C^{18}O$, shown by a blue dotted line; (c), obtaining the $\rm ^{13}CO$ residual spectra (black solid line) by subtracting the Gaussian fit to the scaled $\rm C^{18}O$ (red dash-dotted) from $\rm ^{13}CO$ (gray solid line); (d) blue and red wings can be determined from where the$\rm ^{13}CO$ residuals are larger than the 3$\sigma$ line, where $\sigma$ is the noise level of the emission-free spectrum. The outflow wing selection toward ATLASGAL clumps shows clear evidence of self-absorption as follows. First, a Gaussian fit to the shoulders of $\rm ^{13}CO$ spectra and the fitted Gaussian is used as the true peak temperature of the  $\rm ^{13}CO$.  Then,  the above procedures (a)$-$(d) are followed. }
\label{appendix_fig_wings_inten}
\end{figure*}
\fi

\section{Tables}
\label{sect:table}

\begin{table*}
\setlength{\tabcolsep}{2pt}
\centering
 \scriptsize
\caption[]{{ \it \rm Clump properties of all 325 ATLASGAL clumps$^{\star}$ to search for outflows: clumps Galactic name and coordinates, 
 integrated flux density at 870$\mu m$ ($\rm F_{int}$), heliocentric distance (Dist.), peak $\rm H_2$ column density ($ N_{\rm H_2}$), 
 bolometric luminosity ($ L_{\rm bol}$), and clump mass ($ M_{\rm clump}$). {  These physical values are from \citep{Urquhart2017arXiv392U}.}  }}
\begin{tabular}{llllllll|llllllllll}
\hline
\hline
ATLASGAL          & $\ell$  & $b$ & $\rm F_{int}$ & Dist. &  $ N_{\rm H_2}$   & $\rm logL_{bol}$& $\rm logM_{clump}$  & ATLASGAL          & $\ell$  & $b$ & $\rm F_{int}$ & Dist. &  $\rm logN_{H_2}$   & $\rm logL_{bol}$& $\rm logM_{clump}$ \\
CSC Gname            & ($\rm \degr$)  & ($\rm \degr$) & ($\rm Jy$) & (kpc) & ($\rm cm^{-2}$) & ($\rm logL_\odot$)&($\rm logM_{\odot}$) & CSC Gname            & ($\rm \degr$)  & ($\rm \degr$) & ($\rm Jy$) & (kpc) & ($\rm cm^{-2}$) & ($\rm L_\odot$)&($\rm M_{\odot}$) \\
\hline
  G027.784$+$00.057 &     27.784    &      0.057     &      9.11     &      5.9          &      22.578      &      3.9         &     3.2          &  G031.281$+$00.062 &     31.281    &      0.062     &      40.97    &      5.2          &      23.105      &      4.8         &     3.6          \\
  G027.796$-$00.277 &     27.796    &      $-$0.277  &      4.48     &      2.9          &      22.36       &      3.1         &     2.2          &  G031.386$-$00.269 &     31.386    &      $-$0.269  &      5.57     &      5.2          &      22.046      &      4.2         &     2.7          \\
  G027.883$+$00.204 &     27.883    &      0.204     &      9.16     &      8.3          &      22.19       &      3.3         &     3.6          &  G031.394$+$00.306 &     31.394    &      0.306     &      25.56    &      5.2          &      22.512      &      4.3         &     3.5          \\
  G027.903$-$00.012 &     27.903    &      $-$0.012  &      8.36     &      6.1          &      22.437      &      4.2         &     3.1          &  G031.396$-$00.257 &     31.396    &      $-$0.257  &      12.95    &      5.2          &      22.865      &      4.8         &     3.0          \\
  G027.919$-$00.031 &     27.919    &      $-$0.031  &      2.11     &      3.0          &      21.866      &      3.0         &     1.8          &  G031.412$+$00.307 &     31.412    &      0.307     &      61.68    &      5.2          &      23.696      &      4.8         &     3.9          \\
  G027.923$+$00.196 &     27.923    &      0.196     &      7.02     &      8.3          &      22.125      &      3.4         &     3.4          &  G031.498$+$00.177 &     31.498    &      0.177     &      1.89     &      5.2          &      22.259      &      2.9         &     2.5          \\
  G027.936$+$00.206 &     27.936    &      0.206     &      7.48     &      2.7          &      22.416      &      3.4         &     2.3          &  G031.542$-$00.039 &     31.542    &      $-$0.039  &      2.03     &      2.1          &      22.211      &      1.6         &     1.8          \\
  G027.978$+$00.077 &     27.978    &      0.077     &      9.49     &      4.5          &      22.381      &      4.2         &     2.8          &  G031.568$+$00.092 &     31.568    &      0.092     &      2.88     &      5.2          &      22.299      &      2.4         &     2.8          \\
  G028.013$+$00.342 &     28.013    &      0.342     &      1.76     &      8.3          &      21.872      &      3.5         &     2.7          &  G031.581$+$00.077 &     31.581    &      0.077     &      12.6     &      5.2          &      22.894      &      4.4         &     3.1          \\
  G028.033$-$00.064 &     28.033    &      $-$0.064  &      1.98     &      6.1          &      22.133      &      3.0         &     2.6          &  G031.594$-$00.192 &     31.594    &      $-$0.192  &      3.46     &      2.1          &      22.22       &      2.7         &     1.8          \\
  G028.144$+$00.321 &     28.144    &      0.321     &      5.76     &      $-$          &      22.276      &      $-$         &     $-$          &  G031.596$-$00.336 &     31.596    &      $-$0.336  &      1.38     &      5.2          &      22.026      &      2.8         &     2.3          \\
  G028.148$-$00.004 &     28.148    &      $-$0.004  &      9.19     &      6.1          &      22.628      &      3.9         &     3.2          &  G031.644$-$00.266 &     31.644    &      $-$0.266  &      1.66     &      2.1          &      22.169      &      1.9         &     1.6          \\
  G028.151$+$00.171 &     28.151    &      0.171     &      4.56     &      4.8          &      22.313      &      3.3         &     2.7          &  G031.668$+$00.241 &     31.668    &      0.241     &      2.33     &      5.2          &      22.267      &      3.3         &     2.5          \\
  G028.199$-$00.049 &     28.199    &      $-$0.049  &      35.49    &      6.1          &      23.169      &      5.1         &     3.6          &  G031.676$+$00.244 &     31.676    &      0.244     &      2.02     &      5.2          &      22.222      &      3.2         &     2.5          \\
  G028.231$+$00.041 &     28.231    &      0.041     &      19.78    &      6.1          &      $-$         &      $-$         &     $-$          &  G031.734$-$00.182 &     31.734    &      $-$0.182  &      1.54     &      5.2          &      21.993      &      2.9         &     2.3          \\
  G028.233$+$00.002 &     28.233    &      0.002     &      3.07     &      6.1          &      22.143      &      3.2         &     2.7          &  G032.019$+$00.064 &     32.019    &      0.064     &      19.64    &      5.2          &      22.942      &      3.8         &     3.7          \\
  G028.234$+$00.062 &     28.234    &      0.062     &      5.33     &      6.1          &      22.398      &      3.2         &     3.1          &  G032.044$+$00.059 &     32.044    &      0.059     &      41.74    &      5.2          &      23.151      &      4.6         &     3.8          \\
  G028.244$+$00.012 &     28.244    &      0.012     &      12.11    &      6.1          &      22.175      &      4.7         &     3.1          &  G032.051$-$00.089 &     32.051    &      $-$0.089  &      1.64     &      9.9          &      21.95       &      3.7         &     2.8          \\
  G028.288$-$00.362 &     28.288    &      $-$0.362  &      25.73    &      11.6         &      22.577      &      5.9         &     4.0          &  G032.117$+$00.091 &     32.117    &      0.091     &      14.15    &      5.2          &      22.556      &      4.5         &     3.1          \\
  G028.291$+$00.007 &     28.291    &      0.007     &      4.4      &      3.0          &      22.3        &      3.3         &     2.2          &  G032.149$+$00.134 &     32.149    &      0.134     &      25.01    &      5.2          &      22.893      &      4.6         &     3.5          \\
  G028.294$-$00.192 &     28.294    &      $-$0.192  &      1.37     &      10.4         &      21.981      &      3.5         &     2.9          &  G032.424$+$00.081 &     32.424    &      0.081     &      6.27     &      11.2         &      22.203      &      4.1         &     3.6          \\
  G028.301$-$00.382 &     28.301    &      $-$0.382  &      17.05    &      9.7          &      22.357      &      5.4         &     3.7          &  G032.456$+$00.387 &     32.456    &      0.387     &      1.99     &      3.0          &      22.267      &      2.6         &     2.0          \\
  G028.316$-$00.032 &     28.316    &      $-$0.032  &      10.62    &      6.1          &      22.288      &      3.0         &     3.4          &  G032.471$+$00.204 &     32.471    &      0.204     &      9.87     &      3.0          &      22.524      &      3.5         &     2.6          \\
  G028.321$-$00.009 &     28.321    &      $-$0.009  &      12.7     &      6.1          &      22.617      &      4.0         &     3.4          &  G032.604$-$00.256 &     32.604    &      $-$0.256  &      3.14     &      5.2          &      21.995      &      3.4         &     2.5          \\
  G028.388$+$00.451 &     28.388    &      0.451     &      1.52     &      4.6          &      21.896      &      3.1         &     2.1          &  G032.706$-$00.061 &     32.706    &      $-$0.061  &      4.28     &      6.5          &      22.505      &      3.6         &     3.0          \\
  G028.398$+$00.081 &     28.398    &      0.081     &      26.07    &      4.5          &      23.166      &      4.0         &     3.5          &  G032.739$+$00.192 &     32.739    &      0.192     &      8.28     &      13.0         &      22.515      &      4.1         &     4.0          \\
  G028.438$+$00.036 &     28.438    &      0.036     &      1.05     &      4.5          &      21.905      &      3.6         &     1.8          &  G032.744$-$00.076 &     32.744    &      $-$0.076  &      14.26    &      11.7         &      23.023      &      5.0         &     3.9          \\
  G028.469$-$00.282 &     28.469    &      $-$0.282  &      1.73     &      11.6         &      22.07       &      3.8         &     3.1          &  G032.797$+$00.191 &     32.797    &      0.191     &      31.65    &      13.0         &      23.171      &      6.1         &     4.2          \\
  G028.579$-$00.341 &     28.579    &      $-$0.341  &      7.37     &      4.7          &      22.313      &      3.8         &     2.9          &  G032.821$-$00.331 &     32.821    &      $-$0.331  &      5.79     &      5.1          &      22.609      &      4.2         &     2.7          \\
  G028.596$-$00.361 &     28.596    &      $-$0.361  &      12.48    &      4.7          &      22.179      &      4.4         &     2.9          &  G032.990$+$00.034 &     32.99     &      0.034     &      10.18    &      9.2          &      22.681      &      4.8         &     3.5          \\
  G028.601$-$00.377 &     28.601    &      $-$0.377  &      9.73     &      4.7          &      22.064      &      4.1         &     2.9          &  G033.133$-$00.092 &     33.133    &      $-$0.092  &      15.27    &      9.4          &      22.869      &      5.0         &     3.7          \\
  G028.608$+$00.019 &     28.608    &      0.019     &      20.21    &      7.4          &      22.683      &      5.0         &     3.6          &  G033.134$-$00.021 &     33.134    &      $-$0.021  &      2.22     &      6.5          &      22.231      &      3.1         &     2.8          \\
  G028.608$-$00.027 &     28.608    &      $-$0.027  &      2.48     &      2.8          &      22.017      &      2.8         &     1.9          &  G033.203$+$00.019 &     33.203    &      0.019     &      9.75     &      6.5          &      22.153      &      3.7         &     3.3          \\
  G028.649$+$00.027 &     28.649    &      0.027     &      13.53    &      7.4          &      22.758      &      4.5         &     3.5          &  G033.206$-$00.009 &     33.206    &      $-$0.009  &      11.79    &      6.5          &      22.289      &      4.7         &     3.2          \\
  G028.681$+$00.032 &     28.681    &      0.032     &      5.34     &      7.4          &      22.186      &      3.5         &     3.1          &  G033.264$+$00.067 &     33.264    &      0.067     &      7.45     &      6.5          &      22.507      &      3.9         &     3.2          \\
  G028.701$+$00.404 &     28.701    &      0.404     &      2.73     &      4.7          &      22.39       &      2.9         &     2.6          &  G033.288$-$00.019 &     33.288    &      $-$0.019  &      2.75     &      6.5          &      22.22       &      3.7         &     2.7          \\
  G028.707$-$00.294 &     28.707    &      $-$0.294  &      14.36    &      4.7          &      22.629      &      3.7         &     3.3          &  G033.338$+$00.164 &     33.338    &      0.164     &      4.11     &      5.2          &      22.26       &      3.0         &     2.9          \\
  G028.787$+$00.237 &     28.787    &      0.237     &      6.02     &      7.4          &      22.524      &      4.4         &     3.1          &  G033.388$+$00.199 &     33.388    &      0.199     &      2.62     &      5.2          &      22.282      &      3.6         &     2.6          \\
  G028.802$-$00.022 &     28.802    &      $-$0.022  &      6.37     &      7.4          &      22.65       &      3.4         &     3.5          &  G033.389$+$00.167 &     33.389    &      0.167     &      2.14     &      13.1         &      22.491      &      4.1         &     3.3          \\
  G028.812$+$00.169 &     28.812    &      0.169     &      12.91    &      7.4          &      22.576      &      4.7         &     3.4          &  G033.393$+$00.011 &     33.393    &      0.011     &      19.02    &      6.5          &      22.795      &      4.1         &     3.7          \\
  G028.831$-$00.252 &     28.831    &      $-$0.252  &      23.85    &      4.7          &      22.941      &      4.6         &     3.3          &  G033.416$-$00.002 &     33.416    &      $-$0.002  &      15.89    &      5.4          &      22.484      &      4.1         &     3.3          \\
  G028.861$+$00.066 &     28.861    &      0.066     &      23.13    &      7.4          &      22.786      &      5.1         &     3.7          &  G033.418$+$00.032 &     33.418    &      0.032     &      2.52     &      6.5          &      22.187      &      3.7         &     2.6          \\
  G028.881$-$00.021 &     28.881    &      $-$0.021  &      3.5      &      7.4          &      22.541      &      3.8         &     3.0          &  G033.494$-$00.014 &     33.494    &      $-$0.014  &      2.59     &      6.5          &      22.339      &      3.8         &     2.7          \\
  G028.974$+$00.081 &     28.974    &      0.081     &      3.3      &      10.4         &      21.821      &      3.8         &     3.1          &  G033.643$-$00.227 &     33.643    &      $-$0.227  &      3.78     &      6.5          &      22.362      &      4.1         &     2.7          \\
  G029.002$+$00.067 &     29.002    &      0.067     &      1.92     &      10.4         &      21.955      &      4.2         &     2.8          &  G033.651$-$00.026 &     33.651    &      $-$0.026  &      11.84    &      6.5          &      22.574      &      4.1         &     3.5          \\
  G029.016$-$00.177 &     29.016    &      $-$0.177  &      3.11     &      5.8          &      22.244      &      3.2         &     2.8          &  G033.656$-$00.019 &     33.656    &      $-$0.019  &      3.1      &      6.5          &      22.368      &      4.1         &     2.7          \\
  G029.119$+$00.087 &     29.119    &      0.087     &      4.54     &      5.6          &      22.023      &      3.7         &     2.8          &  G033.739$-$00.021 &     33.739    &      $-$0.021  &      12.56    &      6.5          &      22.997      &      3.6         &     3.7          \\
  G029.126$-$00.146 &     29.126    &      $-$0.146  &      3.74     &      $-$          &      21.87       &      $-$         &     $-$          &  G033.809$-$00.159 &     33.809    &      $-$0.159  &      2.49     &      3.2          &      21.981      &      2.9         &     2.0          \\
  G029.241$+$00.251 &     29.241    &      0.251     &      3.95     &      4.0          &      22.086      &      3.2         &     2.4          &  G033.811$-$00.187 &     33.811    &      $-$0.187  &      6.46     &      10.8         &      22.299      &      5.2         &     3.3          \\
  G029.276$-$00.129 &     29.276    &      $-$0.129  &      3.57     &      4.1          &      22.216      &      2.9         &     2.5          &  G033.914$+$00.109 &     33.914    &      0.109     &      31.26    &      6.5          &      23.048      &      5.2         &     3.7          \\
  G029.362$-$00.316 &     29.362    &      $-$0.316  &      1.44     &      5.6          &      21.756      &      3.7         &     2.2          &  G034.096$+$00.017 &     34.096    &      0.017     &      10.46    &      1.6          &      22.626      &      2.9         &     2.1          \\
  G029.396$-$00.094 &     29.396    &      $-$0.094  &      5.47     &      7.7          &      22.841      &      3.4         &     3.4          &  G034.133$+$00.471 &     34.133    &      0.471     &      5.61     &      11.6         &      22.375      &      5.0         &     3.4          \\
  G029.464$+$00.009 &     29.464    &      0.009     &      1.44     &      8.7          &      21.966      &      3.4         &     2.7          &  G034.169$+$00.089 &     34.169    &      0.089     &      2.09     &      1.6          &      22.182      &      1.8         &     1.5          \\
  G029.476$-$00.179 &     29.476    &      $-$0.179  &      5.35     &      7.7          &      22.368      &      4.0         &     3.2          &  G034.221$+$00.164 &     34.221    &      0.164     &      7.39     &      1.6          &      22.184      &      2.9         &     1.9          \\
  G029.779$-$00.261 &     29.779    &      $-$0.261  &      2.36     &      5.2          &      22.445      &      2.4         &     2.7          &  G034.229$+$00.134 &     34.229    &      0.134     &      60.99    &      1.6          &      $-$         &      $-$         &     $-$          \\
  G029.841$-$00.034 &     29.841    &      $-$0.034  &      2.72     &      5.2          &      22.381      &      2.9         &     2.7          &  G034.241$+$00.107 &     34.241    &      0.107     &      3.73     &      1.6          &      22.314      &      2.4         &     1.6          \\
  G029.852$-$00.059 &     29.852    &      $-$0.059  &      5.89     &      5.2          &      22.332      &      4.3         &     2.7          &  G034.243$+$00.132 &     34.243    &      0.132     &      51.86    &      1.6          &      22.732      &      4.5         &     2.5          \\
  G029.862$-$00.044 &     29.862    &      $-$0.044  &      13.1     &      5.2          &      22.462      &      4.7         &     3.0          &  G034.244$+$00.159 &     34.244    &      0.159     &      33.7     &      1.6          &      23.089      &      4.1         &     2.4          \\
  G029.887$+$00.004 &     29.887    &      0.004     &      9.49     &      5.2          &      22.224      &      4.5         &     2.9          &  G034.258$+$00.109 &     34.258    &      0.109     &      4.58     &      1.6          &      22.251      &      2.9         &     1.6          \\
  G029.889$-$00.009 &     29.889    &      $-$0.009  &      8.05     &      5.2          &      22.122      &      4.1         &     2.8          &  G034.258$+$00.154 &     34.258    &      0.154     &      217.0    &      1.6          &      23.917      &      4.8         &     3.2          \\
  G029.899$-$00.062 &     29.899    &      $-$0.062  &      7.21     &      5.2          &      22.213      &      3.9         &     2.9          &  G034.258$+$00.166 &     34.258    &      0.166     &      16.59    &      1.6          &      23.221      &      3.5         &     2.2          \\
  G029.911$-$00.042 &     29.911    &      $-$0.042  &      52.84    &      5.2          &      22.89       &      4.8         &     3.8          &  G034.261$+$00.176 &     34.261    &      0.176     &      30.85    &      1.6          &      $-$         &      $-$         &     $-$          \\
  G029.921$-$00.014 &     29.921    &      $-$0.014  &      6.38     &      5.2          &      22.415      &      3.8         &     2.8          &  G034.273$+$00.141 &     34.273    &      0.141     &      15.79    &      1.6          &      22.441      &      4.7         &     2.1          \\
  G029.931$-$00.064 &     29.931    &      $-$0.064  &      27.14    &      5.2          &      22.767      &      4.6         &     3.5          &  G034.284$+$00.184 &     34.284    &      0.184     &      3.77     &      1.6          &      22.61       &      2.0         &     1.8          \\
  G029.937$-$00.052 &     29.937    &      $-$0.052  &      23.85    &      5.2          &      22.472      &      4.8         &     3.3          &  G034.391$+$00.214 &     34.391    &      0.214     &      19.02    &      1.6          &      $-$         &      $-$         &     $-$          \\
  G029.941$-$00.012 &     29.941    &      $-$0.012  &      10.23    &      5.2          &      22.336      &      3.8         &     3.1          &  G034.411$+$00.234 &     34.411    &      0.234     &      34.74    &      1.6          &      23.378      &      3.5         &     2.6          \\
  G029.952$+$00.151 &     29.952    &      0.151     &      1.72     &      5.2          &      22.234      &      2.5         &     2.5          &  G034.454$+$00.006 &     34.454    &      0.006     &      4.78     &      5.3          &      22.63       &      3.4         &     2.9          \\
  G029.954$-$00.016 &     29.954    &      $-$0.016  &      57.01    &      5.2          &      23.142      &      5.7         &     3.6          &  G034.459$+$00.247 &     34.459    &      0.247     &      15.28    &      1.6          &      22.961      &      2.4         &     2.5          \\
  G029.959$-$00.067 &     29.959    &      $-$0.067  &      3.4      &      5.2          &      22.194      &      3.6         &     2.5          &  G035.026$+$00.349 &     35.026    &      0.349     &      17.42    &      2.3          &      22.854      &      4.1         &     2.4          \\
  G029.964$-$00.012 &     29.964    &      $-$0.012  &      3.31     &      5.2          &      22.476      &      4.8         &     2.4          &  G035.051$+$00.332 &     35.051    &      0.332     &      6.6      &      3.1          &      22.218      &      2.7         &     2.6          \\
  G029.964$-$00.414 &     29.964    &      $-$0.414  &      3.07     &      4.3          &      21.815      &      3.3         &     2.4          &  G035.344$+$00.347 &     35.344    &      0.347     &      1.77     &      5.9          &      22.082      &      3.6         &     2.4          \\
  G029.976$-$00.047 &     29.976    &      $-$0.047  &      41.23    &      5.2          &      22.83       &      4.4         &     3.7          &  G035.452$-$00.296 &     35.452    &      $-$0.296  &      1.94     &      10.3         &      22.334      &      3.9         &     3.0          \\
  G030.010$+$00.034 &     30.01     &      0.034     &      1.13     &      5.2          &      22.226      &      2.2         &     2.4          &  G035.457$-$00.179 &     35.457    &      $-$0.179  &      2.19     &      4.1          &      22.383      &      3.1         &     2.2          \\
  G030.013$-$00.031 &     30.013    &      $-$0.031  &      9.25     &      5.2          &      22.23       &      3.8         &     3.0          &  G035.466$+$00.141 &     35.466    &      0.141     &      23.2     &      4.7          &      22.855      &      4.7         &     3.3          \\
  G030.019$-$00.047 &     30.019    &      $-$0.047  &      27.56    &      5.2          &      22.526      &      4.6         &     3.5          &  G035.497$-$00.021 &     35.497    &      $-$0.021  &      13.45    &      10.4         &      22.336      &      4.4         &     3.9          \\
  G030.023$+$00.106 &     30.023    &      0.106     &      10.39    &      5.2          &      22.185      &      4.0         &     3.1          &  G035.517$-$00.034 &     35.517    &      $-$0.034  &      5.09     &      10.4         &      22.047      &      4.3         &     3.3          \\
  G030.029$+$00.117 &     30.029    &      0.117     &      9.4      &      5.2          &      22.179      &      4.1         &     3.0          &  G035.522$-$00.274 &     35.522    &      $-$0.274  &      8.35     &      2.7          &      22.866      &      2.1         &     3.0          \\
  G030.031$+$00.106 &     30.031    &      0.106     &      7.43     &      5.2          &      22.008      &      3.7         &     2.8          &  G035.577$+$00.047 &     35.577    &      0.047     &      10.09    &      10.4         &      22.393      &      4.5         &     3.7          \\
  G030.094$+$00.047 &     30.094    &      0.047     &      1.53     &      5.2          &      21.906      &      3.0         &     2.3          &  G035.577$+$00.067 &     35.577    &      0.067     &      22.04    &      10.4         &      22.69       &      5.0         &     4.0          \\
  G030.138$-$00.071 &     30.138    &      $-$0.071  &      4.98     &      5.2          &      22.505      &      2.3         &     3.3          &  G035.579$-$00.031 &     35.579    &      $-$0.031  &      18.05    &      10.4         &      22.906      &      5.3         &     3.8          \\
  G030.691+00.22   & 30.691 & 0.227  & 2.08      & 5.2   & 22.309          & 3.0               & 2.5   & 24d &G30.008-0.272      & 30.008 & -0.272 & 17.12   & 5.2   & 22.869           & 3.8              & 3.5\\ %add the two source at the second version of arxiv

           \hline
\end{tabular}
\end{table*}    
\addtocounter{table}{-1}
\begin{table*}
\setlength{\tabcolsep}{2pt}
\centering
 \scriptsize
\caption[]{{ \it \rm --continuum \rm Clump properties of all 325 ATLASGAL clumps to search for outflows}}
\begin{tabular}{llllllll|llllllllll}
\hline
\hline
ATLASGAL          & $\ell$  & $b$ & $\rm F_{int}$ & Dist. &  $\rm logN_{H_2}$   & $\rm logL_{bol}$& $\rm logM_{clump}$  & ATLASGAL          & $\ell$  & $b$ & $\rm F_{int}$ & Dist. &  $ N_{\rm H_2}$   & $\rm logL_{bol}$& $\rm logM_{clump}$ \\
CSC Gname            & ($\rm \degr$)  & ($\rm \degr$) & ($\rm Jy$) & (kpc) & ($\rm cm^{-2}$) & ($\rm logL_\odot$)&($\rm logM_{\odot}$) & CSC Gname            & ($\rm \degr$)  & ($\rm \degr$) & ($\rm Jy$) & (kpc) & ($\rm cm^{-2}$) & ($\rm L_\odot$)&($\rm M_{\odot}$) \\
\hline
  G030.198$-$00.169 &     30.198    &      $-$0.169  &      5.56     &      5.2          &      22.223      &      4.2         &     2.7          &  G035.602$+$00.222 &     35.602    &      0.222     &      2.38     &      3.0          &      22.31       &      2.0         &     2.2          \\
  G030.201$-$00.157 &     30.201    &      $-$0.157  &      6.61     &      5.2          &      21.896      &      4.3         &     2.7          &  G035.604$-$00.202 &     35.604    &      $-$0.202  &      3.04     &      3.0          &      22.141      &      2.8         &     2.1          \\
  G030.213$-$00.187 &     30.213    &      $-$0.187  &      23.18    &      5.2          &      22.592      &      4.5         &     3.4          &  G035.681$-$00.176 &     35.681    &      $-$0.176  &      6.97     &      2.1          &      22.758      &      2.3         &     2.4          \\
  G030.224$-$00.179 &     30.224    &      $-$0.179  &      7.89     &      5.2          &      22.315      &      4.2         &     2.9          &  G036.406$+$00.021 &     36.406    &      0.021     &      5.41     &      3.5          &      22.556      &      3.9         &     2.3          \\
  G030.251$+$00.054 &     30.251    &      0.054     &      7.66     &      5.2          &      22.325      &      4.0         &     2.9          &  G036.433$-$00.169 &     36.433    &      $-$0.169  &      5.4      &      4.6          &      22.242      &      3.0         &     2.9          \\
  G030.294$+$00.056 &     30.294    &      0.056     &      8.08     &      5.2          &      22.413      &      3.9         &     2.9          &  G036.794$-$00.204 &     36.794    &      $-$0.204  &      5.05     &      5.8          &      22.241      &      3.5         &     3.0          \\
  G030.299$-$00.202 &     30.299    &      $-$0.202  &      16.06    &      5.2          &      22.325      &      3.9         &     3.3          &  G036.826$-$00.039 &     36.826    &      $-$0.039  &      5.01     &      3.6          &      22.399      &      2.4         &     2.7          \\
  G030.341$-$00.116 &     30.341    &      $-$0.116  &      9.45     &      5.2          &      22.496      &      3.1         &     3.2          &  G037.043$-$00.036 &     37.043    &      $-$0.036  &      8.16     &      5.8          &      22.666      &      3.7         &     3.2          \\
  G030.348$+$00.097 &     30.348    &      0.097     &      4.16     &      5.2          &      22.007      &      3.4         &     2.6          &  G037.199$-$00.419 &     37.199    &      $-$0.419  &      2.92     &      2.2          &      22.424      &      2.4         &     1.9          \\
  G030.348$+$00.392 &     30.348    &      0.392     &      4.9      &      5.0          &      22.488      &      3.3         &     2.8          &  G037.268$+$00.081 &     37.268    &      0.081     &      10.35    &      5.8          &      22.728      &      3.9         &     3.3          \\
  G030.351$+$00.086 &     30.351    &      0.086     &      2.08     &      5.2          &      22.003      &      3.4         &     2.3          &  G037.341$-$00.062 &     37.341    &      $-$0.062  &      6.71     &      9.7          &      22.421      &      4.6         &     3.4          \\
  G030.386$-$00.104 &     30.386    &      $-$0.104  &      25.2     &      5.2          &      22.53       &      4.6         &     3.4          &  G037.374$-$00.236 &     37.374    &      $-$0.236  &      9.22     &      2.2          &      22.228      &      3.5         &     2.3          \\
  G030.399$-$00.102 &     30.399    &      $-$0.102  &      8.94     &      5.2          &      22.051      &      4.0         &     3.0          &  G037.444$+$00.137 &     37.444    &      0.137     &      2.82     &      2.2          &      22.109      &      1.5         &     2.0          \\
  G030.399$-$00.296 &     30.399    &      $-$0.296  &      3.6      &      5.2          &      22.457      &      3.3         &     2.7          &  G037.479$-$00.106 &     37.479    &      $-$0.106  &      3.7      &      9.7          &      22.35       &      4.0         &     3.2          \\
  G030.419$-$00.231 &     30.419    &      $-$0.231  &      23.55    &      5.2          &      23.055      &      4.3         &     3.5          &  G037.546$-$00.112 &     37.546    &      $-$0.112  &      9.43     &      9.7          &      22.55       &      5.1         &     3.4          \\
  G030.424$-$00.214 &     30.424    &      $-$0.214  &      7.53     &      5.2          &      22.624      &      3.2         &     3.1          &  G037.638$-$00.104 &     37.638    &      $-$0.104  &      1.58     &      9.7          &      21.945      &      4.2         &     2.7          \\
  G030.426$-$00.267 &     30.426    &      $-$0.267  &      11.63    &      5.2          &      21.929      &      4.1         &     3.0          &  G037.671$+$00.142 &     37.671    &      0.142     &      4.41     &      4.9          &      22.089      &      3.3         &     2.7          \\
  G030.488$-$00.301 &     30.488    &      $-$0.301  &      6.05     &      5.2          &      21.875      &      3.5         &     2.7          &  G037.672$-$00.091 &     37.672    &      $-$0.091  &      1.39     &      9.7          &      22.081      &      3.2         &     2.9          \\
  G030.493$-$00.391 &     30.493    &      $-$0.391  &      5.44     &      0.4          &      22.334      &      $-$         &     0.8          &  G037.734$-$00.112 &     37.734    &      $-$0.112  &      9.32     &      9.7          &      22.756      &      4.6         &     3.6          \\
  G030.513$+$00.031 &     30.513    &      0.031     &      2.29     &      2.7          &      22.014      &      2.5         &     1.9          &  G037.764$-$00.216 &     37.764    &      $-$0.216  &      21.78    &      9.7          &      22.765      &      5.0         &     3.9          \\
  G030.534$+$00.021 &     30.534    &      0.021     &      9.78     &      2.7          &      22.508      &      3.9         &     2.4          &  G037.819$+$00.412 &     37.819    &      0.412     &      7.57     &      12.3         &      22.744      &      4.8         &     3.7          \\
  G030.588$-$00.042 &     30.588    &      $-$0.042  &      25.61    &      2.7          &      22.904      &      4.0         &     2.9          &  G037.874$-$00.399 &     37.874    &      $-$0.399  &      18.45    &      9.7          &      22.799      &      5.7         &     3.6          \\
  G030.623$-$00.111 &     30.623    &      $-$0.111  &      4.37     &      5.2          &      22.303      &      3.1         &     2.9          &  G038.037$-$00.041 &     38.037    &      $-$0.041  &      2.28     &      3.3          &      22.059      &      3.0         &     2.0          \\
  G030.624$+$00.169 &     30.624    &      0.169     &      13.83    &      5.2          &      22.294      &      3.3         &     3.3          &  G038.119$-$00.229 &     38.119    &      $-$0.229  &      2.65     &      6.5          &      22.274      &      4.0         &     2.6          \\
  G030.641$-$00.117 &     30.641    &      $-$0.117  &      1.94     &      5.2          &      22.052      &      3.3         &     2.4          &  G038.646$-$00.226 &     38.646    &      $-$0.226  &      2.86     &      $-$          &      22.172      &      $-$         &     $-$          \\
  G030.648$-$00.119 &     30.648    &      $-$0.119  &      2.64     &      5.2          &      22.08       &      3.2         &     2.5          &  G038.694$-$00.452 &     38.694    &      $-$0.452  &      4.51     &      9.8          &      22.556      &      4.2         &     3.3          \\
  G030.651$-$00.204 &     30.651    &      $-$0.204  &      10.43    &      5.2          &      22.738      &      3.8         &     3.2          &  G038.909$-$00.462 &     38.909    &      $-$0.462  &      2.87     &      1.9          &      22.111      &      2.2         &     1.7          \\
  G030.659$+$00.229 &     30.659    &      0.229     &      3.41     &      5.2          &      22.315      &      3.4         &     2.7          &  G038.917$-$00.402 &     38.917    &      $-$0.402  &      3.96     &      1.9          &      21.982      &      2.3         &     1.8          \\
  G030.663$-$00.144 &     30.663    &      $-$0.144  &      5.71     &      5.2          &      22.317      &      3.7         &     2.9          &  G038.921$-$00.351 &     38.921    &      $-$0.351  &      20.17    &      1.9          &      22.91       &      3.3         &     2.6          \\
  G030.683$-$00.074 &     30.683    &      $-$0.074  &      17.24    &      5.2          &      22.744      &      4.7         &     3.2          &  G038.934$-$00.361 &     38.934    &      $-$0.361  &      12.79    &      1.9          &      22.55       &      3.2         &     2.4          \\
  G030.684$-$00.261 &     30.684    &      $-$0.261  &      5.48     &      5.2          &      22.378      &      4.4         &     2.7          &  G038.937$-$00.457 &     38.937    &      $-$0.457  &      9.3      &      1.9          &      22.521      &      2.5         &     2.3          \\
  G030.693$-$00.149 &     30.693    &      $-$0.149  &      3.39     &      5.2          &      22.267      &      3.0         &     2.8          &  G038.957$-$00.466 &     38.957    &      $-$0.466  &      14.71    &      1.9          &      22.844      &      2.1         &     2.8          \\
  G030.703$-$00.067 &     30.703    &      $-$0.067  &      99.54    &      5.2          &      23.424      &      5.2         &     4.0          &  G039.268$-$00.051 &     39.268    &      $-$0.051  &      4.81     &      11.8         &      22.401      &      4.4         &     3.5          \\
  G030.691$-$00.054 &     30.691    &      $-$0.054  &      60.1     &      5.2          &      23.424      &      5.2         &     3.8          &  G039.388$-$00.141 &     39.388    &      $-$0.141  &      5.51     &      3.4          &      22.541      &      3.8         &     2.3          \\
  G030.718$-$00.082 &     30.718    &      $-$0.082  &      50.64    &      5.2          &      23.271      &      4.7         &     3.8          &  G039.434$-$00.187 &     39.434    &      $-$0.187  &      2.23     &      3.3          &      22.015      &      2.8         &     2.0          \\
  G030.731$-$00.079 &     30.731    &      $-$0.079  &      14.74    &      5.2          &      22.903      &      3.9         &     3.3          &  G039.591$-$00.204 &     39.591    &      $-$0.204  &      5.26     &      $-$          &      22.346      &      $-$         &     $-$          \\
  G030.741$-$00.061 &     30.741    &      $-$0.061  &      75.23    &      5.2          &      22.904      &      5.4         &     3.8          &  G039.851$-$00.204 &     39.851    &      $-$0.204  &      4.4      &      9.3          &      22.249      &      3.4         &     3.4          \\
  G030.746$-$00.001 &     30.746    &      $-$0.001  &      78.93    &      5.2          &      $-$         &      $-$         &     $-$          &  G039.884$-$00.346 &     39.884    &      $-$0.346  &      7.05     &      9.3          &      22.469      &      4.6         &     3.3          \\
  G030.753$-$00.051 &     30.753    &      $-$0.051  &      33.49    &      5.2          &      22.76       &      5.5         &     3.4          &  G040.283$-$00.219 &     40.283    &      $-$0.219  &      13.24    &      6.4          &      23.079      &      4.5         &     3.3          \\
  G030.756$+$00.206 &     30.756    &      0.206     &      18.54    &      5.2          &      22.601      &      4.0         &     3.4          &  G040.622$-$00.137 &     40.622    &      $-$0.137  &      11.17    &      10.6         &      22.738      &      5.0         &     3.7          \\
  G030.763$-$00.031 &     30.763    &      $-$0.031  &      12.94    &      5.2          &      22.145      &      5.1         &     2.7          &  G040.814$-$00.416 &     40.814    &      $-$0.416  &      1.04     &      3.4          &      21.998      &      2.6         &     1.7          \\
  G030.766$-$00.046 &     30.766    &      $-$0.046  &      10.25    &      5.2          &      22.388      &      4.9         &     2.8          &  G041.031$-$00.226 &     41.031    &      $-$0.226  &      3.29     &      8.9          &      22.26       &      3.9         &     3.1          \\
  G030.769$-$00.087 &     30.769    &      $-$0.087  &      10.63    &      5.2          &      22.49       &      3.9         &     3.2          &  G041.077$-$00.124 &     41.077    &      $-$0.124  &      4.09     &      8.9          &      22.277      &      3.6         &     3.3          \\
  G030.773$-$00.216 &     30.773    &      $-$0.216  &      21.89    &      5.2          &      $-$         &      $-$         &     $-$          &  G041.099$-$00.237 &     41.099    &      $-$0.237  &      5.95     &      8.9          &      22.084      &      4.4         &     3.2          \\
  G030.783$-$00.262 &     30.783    &      $-$0.262  &      1.01     &      2.7          &      22.325      &      1.6         &     1.9          &  G041.122$-$00.219 &     41.122    &      $-$0.219  &      7.96     &      8.9          &      22.359      &      3.8         &     3.6          \\
  G030.784$-$00.021 &     30.784    &      $-$0.021  &      95.09    &      5.2          &      22.695      &      5.8         &     3.8          &  G041.161$-$00.184 &     41.161    &      $-$0.184  &      6.73     &      8.9          &      22.326      &      4.2         &     3.4          \\
  G030.786$+$00.204 &     30.786    &      0.204     &      9.99     &      5.2          &      22.921      &      3.9         &     3.1          &  G041.226$-$00.197 &     41.226    &      $-$0.197  &      3.59     &      8.9          &      22.275      &      3.9         &     3.2          \\
  G030.813$-$00.024 &     30.813    &      $-$0.024  &      17.21    &      5.2          &      22.825      &      4.4         &     3.3          &  G041.307$-$00.171 &     41.307    &      $-$0.171  &      4.1      &      8.9          &      22.361      &      3.3         &     3.4          \\
  G030.818$+$00.274 &     30.818    &      0.274     &      2.74     &      5.2          &      22.111      &      4.1         &     2.3          &  G041.377$+$00.037 &     41.377    &      0.037     &      3.35     &      8.9          &      22.159      &      4.3         &     3.0          \\
  G030.818$-$00.056 &     30.818    &      $-$0.056  &      113.85   &      5.2          &      23.669      &      5.4         &     4.1          &  G041.507$-$00.106 &     41.507    &      $-$0.106  &      0.92     &      8.9          &      22.133      &      3.4         &     2.6          \\
  G030.819$-$00.081 &     30.819    &      $-$0.081  &      1.78     &      5.2          &      22.27       &      3.2         &     2.4          &  G042.108$-$00.447 &     42.108    &      $-$0.447  &      6.0      &      3.4          &      22.303      &      3.7         &     2.4          \\
  G030.823$-$00.156 &     30.823    &      $-$0.156  &      16.48    &      5.2          &      22.605      &      4.6         &     3.5          &  G042.164$-$00.077 &     42.164    &      $-$0.077  &      1.19     &      9.9          &      22.238      &      3.1         &     2.9          \\
  G030.828$+$00.134 &     30.828    &      0.134     &      9.0      &      2.7          &      22.172      &      3.0         &     2.5          &  G042.421$-$00.259 &     42.421    &      $-$0.259  &      7.26     &      4.4          &      22.138      &      4.0         &     2.7          \\
  G030.828$-$00.122 &     30.828    &      $-$0.122  &      1.18     &      2.7          &      22.029      &      2.9         &     1.5          &  G043.038$-$00.452 &     43.038    &      $-$0.452  &      7.32     &      $-$          &      22.803      &      $-$         &     $-$          \\
  G030.839$-$00.019 &     30.839    &      $-$0.019  &      12.82    &      5.2          &      22.555      &      3.3         &     3.4          &  G043.108$+$00.044 &     43.108    &      0.044     &      13.22    &      11.1         &      22.326      &      4.8         &     3.9          \\
  G030.853$-$00.109 &     30.853    &      $-$0.109  &      9.65     &      5.2          &      22.329      &      3.9         &     3.1          &  G043.124$+$00.031 &     43.124    &      0.031     &      22.3     &      11.1         &      22.419      &      5.1         &     4.1          \\
  G030.866$+$00.114 &     30.866    &      0.114     &      13.16    &      2.7          &      22.641      &      4.1         &     2.5          &  G043.148$+$00.014 &     43.148    &      0.014     &      48.87    &      11.1         &      22.894      &      5.9         &     4.3          \\
  G030.866$-$00.119 &     30.866    &      $-$0.119  &      11.92    &      5.2          &      22.338      &      3.9         &     3.3          &  G043.164$-$00.029 &     43.164    &      $-$0.029  &      86.15    &      11.1         &      23.265      &      6.2         &     4.5          \\
  G030.874$-$00.094 &     30.874    &      $-$0.094  &      4.61     &      5.2          &      22.314      &      3.7         &     2.7          &  G043.166$+$00.011 &     43.166    &      0.011     &      319.98   &      11.1         &      23.892      &      6.9         &     5.0          \\
  G030.886$-$00.231 &     30.886    &      $-$0.231  &      2.8      &      5.2          &      21.917      &      3.2         &     2.5          &  G043.178$-$00.011 &     43.178    &      $-$0.011  &      47.31    &      11.1         &      22.882      &      6.6         &     4.2          \\
  G030.893$+$00.139 &     30.901    &      0.147     &      1.8      &      5.2          &      23.023      &      3.3         &     2.5          &  G043.236$-$00.047 &     43.236    &      $-$0.047  &      18.83    &      11.1         &      22.904      &      5.1         &     4.0          \\
  G030.898$+$00.162 &     30.898    &      0.162     &      17.69    &      5.2          &      22.694      &      4.0         &     3.3          &  G043.306$-$00.212 &     43.306    &      $-$0.212  &      7.95     &      4.2          &      22.78       &      4.1         &     2.7          \\
  G030.901$-$00.034 &     30.901    &      $-$0.034  &      1.23     &      5.2          &      22.26       &      2.6         &     2.3          &  G043.519$+$00.016 &     43.519    &      0.016     &      2.75     &      4.3          &      22.483      &      2.5         &     2.6          \\
  G030.908$+$00.027 &     30.908    &      0.027     &      7.06     &      5.2          &      22.049      &      3.2         &     3.0          &  G043.528$+$00.017 &     43.528    &      0.017     &      3.32     &      4.3          &      22.385      &      2.6         &     2.6          \\
  G030.919$+$00.091 &     30.919    &      0.091     &      4.68     &      5.2          &      22.55       &      3.3         &     2.9          &  G043.794$-$00.127 &     43.794    &      $-$0.127  &      13.83    &      6.0          &      22.876      &      5.1         &     3.1          \\
  G030.959$+$00.086 &     30.959    &      0.086     &      8.77     &      2.7          &      22.459      &      3.6         &     2.4          &  G043.817$-$00.119 &     43.817    &      $-$0.119  &      1.86     &      6.0          &      22.106      &      3.2         &     2.5          \\
  G030.969$-$00.044 &     30.969    &      $-$0.044  &      1.44     &      5.2          &      22.133      &      2.5         &     2.4          &  G043.994$-$00.012 &     43.994    &      $-$0.012  &      4.23     &      6.0          &      22.311      &      4.1         &     2.7          \\
  G030.971$-$00.141 &     30.971    &      $-$0.141  &      17.45    &      5.2          &      22.841      &      3.8         &     3.6          &  G044.309$+$00.041 &     44.309    &      0.041     &      12.23    &      8.1          &      22.698      &      4.5         &     3.5          \\
  G030.978$+$00.216 &     30.978    &      0.216     &      8.34     &      5.2          &      22.445      &      3.7         &     3.1          &  G045.071$+$00.132 &     45.071    &      0.132     &      20.13    &      8.0          &      22.885      &      5.7         &     3.5          \\
  G030.994$+$00.236 &     30.994    &      0.236     &      19.85    &      5.2          &      22.378      &      3.3         &     3.5          &  G045.086$+$00.132 &     45.086    &      0.132     &      1.24     &      8.0          &      22.206      &      3.7         &     2.6          \\
  G030.996$-$00.076 &     30.996    &      $-$0.076  &      5.46     &      5.2          &      22.465      &      3.8         &     2.8          &  G045.121$+$00.131 &     45.121    &      0.131     &      42.78    &      8.0          &      22.958      &      6.0         &     3.9          \\
  G031.024$+$00.262 &     31.024    &      0.262     &      24.42    &      5.2          &      22.627      &      3.6         &     3.8          &  G045.454$+$00.061 &     45.454    &      0.061     &      35.12    &      8.4          &      22.702      &      5.8         &     3.8          \\
  G031.046$+$00.357 &     31.046    &      0.357     &      8.83     &      5.2          &      22.436      &      4.0         &     3.0          &  G045.463$+$00.027 &     45.463    &      0.027     &      7.2      &      8.4          &      22.293      &      4.7         &     3.3          \\
  G031.054$+$00.469 &     31.054    &      0.469     &      7.43     &      2.0          &      22.295      &      3.5         &     2.1          &  G045.474$+$00.134 &     45.474    &      0.134     &      20.08    &      8.4          &      22.641      &      5.5         &     3.6          \\
  G031.071$+$00.049 &     31.071    &      0.049     &      2.76     &      2.7          &      21.911      &      3.8         &     1.7          &  G045.543$-$00.007 &     45.543    &      $-$0.007  &      2.51     &      8.4          &      22.015      &      4.3         &     2.8          \\
  G031.121$+$00.062 &     31.121    &      0.062     &      3.35     &      2.7          &      22.031      &      2.8         &     2.0          &  G045.544$-$00.032 &     45.544    &      $-$0.032  &      3.37     &      8.4          &      22.122      &      3.7         &     3.0          \\
  G031.148$-$00.149 &     31.148    &      $-$0.149  &      4.22     &      2.7          &      21.988      &      2.8         &     2.1          &  G045.568$-$00.121 &     45.568    &      $-$0.121  &      1.49     &      0.3          &      22.063      &      0.5         &     -0.3         \\
  G031.158$+$00.047 &     31.158    &      0.047     &      8.87     &      2.7          &      22.564      &      3.3         &     2.5          &  G045.776$-$00.254 &     45.776    &      $-$0.254  &      2.95     &      5.8          &      22.305      &      3.0         &     2.8          \\
  G031.208$+$00.101 &     31.208    &      0.101     &      3.0      &      5.2          &      21.998      &      3.7         &     2.5          &  G045.804$-$00.356 &     45.804    &      $-$0.356  &      3.65     &      5.8          &      22.458      &      4.0         &     2.7          \\
  G031.239$+$00.062 &     31.239    &      0.062     &      7.17     &      5.2          &      22.515      &      3.2         &     3.2          &  G045.821$-$00.284 &     45.821    &      $-$0.284  &      4.09     &      5.8          &      22.181      &      4.2         &     2.7          \\
  G031.239$-$00.057 &     31.239    &      $-$0.057  &      5.06     &      2.7          &      22.298      &      2.1         &     2.5          &  G045.829$-$00.292 &     45.829    &      $-$0.292  &      3.48     &      5.8          &      22.063      &      4.0         &     2.5          \\
  G031.243$-$00.111 &     31.243    &      $-$0.111  &      8.0      &      12.9         &      22.691      &      5.3         &     3.6          &  G045.936$-$00.402 &     45.936    &      $-$0.402  &      5.6      &      5.8          &      22.246      &      4.5         &     2.8          \\
  G045.466+00.046 & 45.466 & 0.046  & 30.79    & 8.4   & 22.915          & 5.6               & 3.8& G046.118$+$00.399   &     46.118    &      0.399      &      5.74     &      7.5          &      22.348      &      4.0         &     3.3          \\
%%the source G045.466+00.046 has been added at the second submitted version of arxiv.
\hline
\hline
\end{tabular}
\begin{tablenotes}  
\item
$\star$ Noted: we add three clumps in table 8 after publication, which won't influence the discussion and results in the paper.
\end{tablenotes}
\label{total_clump_properties}
\end{table*}
%%%%%end table1%%%%

%%%table2%%%%
\begin{table*}
%\scriptsize
\centering
\caption[]{{\it \rm $\rm{^{13}CO}$ outflow calculations of all blue and red wings for 225 ATLASGAL clumps$^{\star}$: 
 observed peak $\rm{^{13}CO}$ and $\rm C^{18}O$ velocities, the antenna temperatures 
 are corrected for main-beam efficiency (0.72), the velocity range $\rm \Delta{V_{b/r}}$ for blue and red wings of $\rm{^{13}CO}$ spectra, 
 the maximum projected velocity for blue and red shifted $\rm V_{max_{b/r}}$ relative to the peak $\rm C^{18}O$ velocity.}}
\begin{tabular}{lllllllll}
\hline
\hline
ATLASGAL            &$\rm{^{13}CO}\,v_p$ & $\rm{^{13}CO}\,T_{mb}$ & $\rm C^{18}O\,v_p$ & $\rm C^{18}O\,T_{mb}$ & $\rm \Delta{V_b}$ &  $\rm \Delta{V_r}$ &  $\rm V_{max_{b}}$ & $\rm V_{max_{r}}$ \\
CSC Gname          &   ($\rm km\,s^{-1}$)    & (K)                                      &   ($\rm km\,s^{-1}$)    & (K) &   ($\rm km\,s^{-1}$) &   ($\rm km\,s^{-1}$)  &   ($\rm km\,s^{-1}$)    &   ($\rm km\,s^{-1}$) \\ 
\hline
  G027.784$+$00.057   &    101.2      &    5.9     &    100.8      &    1.8     &    [96.3,100.8]    &    [103.8,104.8]   &    4.5             &    4.0             \\
  G027.903$-$00.012   &    97.9       &    6.3     &    97.5       &    4.9     &    [95.3,96.8]     &    [98.8,100.3]    &    2.2             &    2.8             \\
  G027.919$-$00.031   &    47.6       &    6.1     &    47.7       &    3.7     &    [46.3,46.8]     &    [48.3,49.8]     &    1.4             &    2.1             \\
  G027.936$+$00.206   &    42.3       &    6.2     &    42.0       &    2.3     &    [37.3,40.3]     &    [43.8,46.8]     &    4.7             &    4.8             \\
  G027.978$+$00.077   &    74.7       &    4.2     &    75.3       &    2.9     &    [71.8,73.3]     &    [76.8,79.3]     &    3.5             &    4.0             \\
  G028.148$-$00.004   &    98.6       &    4.0     &    98.5       &    3.1     &    [96.3,97.8]     &    [99.8,100.8]    &    2.2             &    2.3             \\
  G028.151$+$00.171   &    89.7       &    4.8     &    89.6       &    2.1     &    [86.8,88.8]     &    [90.8,92.3]     &    2.8             &    2.7             \\
  G028.199$-$00.049   &    96.3       &    6.8     &    95.6       &    3.6     &    [89.3,95.8]     &    [98.3,107.3]   &    6.3             &    11.7            \\%
  G028.231$+$00.041   &    107.0      &    3.3     &    107.0      &    1.2     &    [104.8,105.8]   &    [107.3,110.3]   &    2.2             &    3.3             \\
  G028.234$+$00.062   &    107.1      &    4.9     &    107.0      &    1.8     &    [104.8,105.8]   &    [107.8,108.8]   &    2.2             &    1.8             \\
  G028.244$+$00.012   &    106.0      &    6.7     &    106.6      &    2.4     &    [103.8,104.8]   &    [106.8,109.3]   &    2.8             &    2.7             \\
  G028.288$-$00.362   &    48.3       &    7.8     &    47.8       &    4.4     &    [42.3,47.8]     &    [50.3,53.3]     &    5.5             &    5.5             \\%
  G028.301$-$00.382   &    84.8       &    11.4    &    84.6       &    3.8     &    [80.8,84.3]     &    [86.3,88.8]     &    3.8             &    4.2             \\
  G028.321$-$00.009   &    100.0      &    5.1     &    99.6       &    2.0     &    [96.3,98.8]     &    [100.8,101.8]   &    3.3             &    2.2             \\
  G028.388$+$00.451   &    83.9       &    7.9     &    83.7       &    3.6     &    [81.8,82.3]     &    [84.3,87.3]     &    1.9             &    3.6             \\
  G028.398$+$00.081   &    77.6       &    4.1     &    78.5       &    2.9     &    $-$               &    [79.8,81.8]     &    $-$               &    3.3             \\
  G028.438$+$00.036   &    83.1       &    12.2    &    83.1       &    5.0     &    [80.3,82.3]     &    [83.8,85.3]     &    2.8             &    2.2             \\
  G028.469$-$00.282   &    48.3       &    5.5     &    48.3       &    4.4    &    [46.3,47.3]     &    [48.8,50.3]     &    2.0             &    2.0             \\%
  G028.608$+$00.019   &    102.1      &    8.3     &    101.7      &    3。8     &    [96.3,99.8]     &    [102.3,107.3]   &    5.4             &    5.6             \\
  G028.608$-$00.027   &    45.2       &    8.9     &    45.6       &    3.7     &    [41.8,44.3]     &    [46.3,46.8]     &    3.8             &    1.2             \\
  G028.649$+$00.027   &    103.1      &    7.8     &    103.4      &    3.7     &    $-$               &    [104.3,108.8]   &    $-$               &    5.4             \\%
  G028.707$-$00.294   &    89.2       &    4.4     &    88.7       &    2.3     &    [86.8,88.3]     &    [89.8,90.8]     &    1.9             &    2.1             \\
  G028.802$-$00.022   &    100.7      &    3.1     &    99.6       &    1.8     &    [96.8,98.3]     &    [101.8,104.3]   &    2.8             &    4.7             \\
  G028.812$+$00.169   &    105.4      &    8.1     &    105.1      &    2.9     &    [101.3,104.3]   &    [106.8,109.3]   &    3.8             &    4.2             \\
  G028.831$-$00.252   &    87.4       &    6.0     &    87.2       &    3.0     &    [82.3,85.8]     &    [87.8,96.8]     &    4.9             &    9.6             \\
  G028.861$+$00.066   &    102.8      &    7.7     &    103.2      &    3.6     &    [97.3,99.3]     &    [103.8,109.8]   &    5.9             &    6.6             \\
  G028.881$-$00.021   &    101.1      &    7.7     &    101.0      &    3.5     &    [98.3,100.8]    &    [102.3,103.3]   &    2.7             &    2.3             \\
  G028.974$+$00.081   &    72.0       &    5.5     &    72.1       &    2.5     &    [69.3,71.3]     &    [72.8,73.8]     &    2.8             &    1.7             \\
  G029.002$+$00.067   &    70.0       &    18.1    &    71.2       &    6.0     &    [68.0,69.5]     &    [70.5,71.5]     &    3.2             &    0.3             \\
  G029.476$-$00.179   &    105.3      &    9.7     &    105.4      &    4.6     &    [103.2,104.2]   &    [106.2,109.2]   &    2.2             &    3.8             \\
  G029.852$-$00.059   &    99.4       &    26.9    &    99.5       &    10.2    &    [97.7,98.2]     &    [100.7,102.7]   &    1.8             &    3.2             \\
  G029.862$-$00.044   &    101.2      &    11.8    &    100.8      &    4.4     &    [94.7,100.2]    &    [103.2,106.2]   &    6.1             &    5.4             \\
  G029.889$-$00.009   &    97.3       &    12.1    &    95.2       &    3.8     &    [91.7,95.2]     &    [100.2,102.7]   &    3.5             &    7.5             \\
  G029.899$-$00.062   &    100.6      &    8.3     &    100.9      &    3.5     &    [97.2,100.2]     &    [102.2,104.2]   &    3.7             &    3.3             \\
  G029.911$-$00.042   &    99.5       &    14.8    &    99.8       &    3.9     &    [92.2,96.7]     &    [101.7,105.7]   &    7.6             &    5.9             \\
  G029.931$-$00.064   &    98.8       &    6.8     &    99.3       &    3.2     &    [90.7,96.7]     &    [99.7,107.2]    &    8.6             &    7.9             \\
  G029.937$-$00.052   &    99.9       &    6.6     &    99.9       &    2.8     &    [92.7,97.7]     &    [102.7,109.2]   &    7.2             &    9.3             \\%
  G029.954$-$00.016   &    97.8       &    13.8    &    97.5       &    14.4    &    [92.1,96.6]     &    [101.1,104.6]   &    5.4             &    7.1             \\
  G029.959$-$00.067   &    101.1      &    7.9     &    102.6      &    4.7     &    [97.2,100.2]    &    [102.2,105.2]   &    5.4             &    2.6             \\
  G029.964$-$00.012   &    98.2       &    14.8    &    98.6       &    8.1     &    [91.7,95.7]     &    [100.2,104.7]   &    6.9             &    6.1             \\
  G029.976$-$00.047   &    99.1       &    7.4     &    101.6      &    1.9     &    [89.7,97.2]     &    [100.7,105.7]   &    11.9            &    4.1             \\
  G030.008$-$00.272   &    103.1      &    2.4    &    103.1      &    1.6     &    [100.6,102.6]    &    [105.6,107.6]   &    2.5             &    4.5             \\
  G030.010$+$00.034   &    103.0      &    14.4    &    105.6      &    1.8     &    [99.6,100.1]    &    [105.1,106.1]   &    6.0             &    0.5             \\
  G030.019$-$00.047   &    95.3       &    10.6    &    92.6       &    2.3     &    [87.1,93.1]     &    [99.6,103.6]    &    5.5             &    11.0            \\
  G030.023$+$00.106   &    106.2      &    7.5     &    106.2      &    2.8     &    [100.6,105.1]   &    [106.6,109.6]   &    5.6             &    3.4             \\
  G030.029$+$00.117   &    106.3      &    6.0     &    106.0      &    2.8     &    [103.6,105.6]   &    [107.1,109.1]   &    2.4             &    3.1             \\
  G030.094$+$00.047   &    105.4      &    5.1     &    106.3      &    1.3     &    [102.1,105.1]   &    [106.1,107.1]   &    4.2             &    0.8             \\
  G030.198$-$00.169   &    103.2      &    6.5     &    103.1      &    3.2     &    [99.6,101.6]    &    [103.6,109.1]   &    3.5             &    6.0             \\
  G030.201$-$00.157   &    103.1      &    7.0     &    103.3      &    2.3     &    [101.1,102.6]   &    [103.6,104.6]   &    2.2             &    1.3             \\
  G030.213$-$00.187   &    104.9      &    7.7     &    104.8      &    3.7     &    [101.1,104.1]   &    [106.1,109.1]   &    3.7             &    4.3             \\
  G030.224$-$00.179   &    103.8      &    16.7    &    103.8      &    7.3     &    [100.1,102.6]   &    [104.1,109.6]   &    3.7             &    5.8             \\
  G030.251$+$00.054   &    71.0       &    3.4     &    71.0       &    1.5     &    $-$               &    [72.4,73.9]     &    $-$               &    2.9             \\%
  G030.299$-$00.202   &    102.4      &    7.6     &    102.1      &    2.6     &    [99.6,101.6]     &    [103.6,106.6]    &    2.5             &    4.5             \\
  G030.348$+$00.392   &    92.9       &    6.2     &    92.8       &    2.9     &    [90.9,91.9]     &    [93.9,95.4]     &    1.9             &    2.6             \\
  G030.351$+$00.086   &    96.3       &    4.6     &    96.8       &    3.0     &    [94.4,95.4]     &    [97.4,98.9]     &    2.4             &    2.1             \\
  G030.386$-$00.104   &    86.9       &    11.5    &    86.9       &    5.8     &    [84.1,86.1]     &    [88.6,89.1]     &    2.8             &    2.2             \\
  G030.399$-$00.102   &    87.4       &    7.0     &    87.9       &    3.3     &    [85.6,87.1]     &    [89.1,90.1]     &    2.3             &    2.2             \\
  G030.399$-$00.296   &    101.8      &    7.3     &    102.2      &    2.4     &    [96.1,99.6]     &    [102.6,107.1]   &    6.1             &    4.9             \\
  G030.419$-$00.231   &    104.8      &    10.8    &    104.8      &    3.9     &    [95.9,103.9]    &    [105.9,113.4]   &    8.9             &    8.6             \\
  G030.426$-$00.267   &    103.3      &    9.3     &    103.0      &    3.0     &    [101.4,102.9]   &    [104.9,105.4]   &    1.6             &    2.4             \\
  G030.534$+$00.021   &    48.1       &    11.6    &    48.0       &    3.1    &    [39.9,46.9]     &    [49.8,54.4]     &    8.1             &    6.4             \\%
  G030.588$-$00.042   &    42.1       &    5.2     &    41.9       &    2.3     &    [33.9,41.4]     &    [44.4,49.4]     &    8.0            &    7.5             \\
  G030.623$-$00.111   &    113.8      &    8.0     &    113.9      &    2.5     &    [111.4,113.4]   &    [114.9,115.9]   &    2.5             &    2.0             \\
  G030.624$+$00.169   &    105.3      &    7.0     &    105.5      &    2.9     &    [102.4,104.4]   &    [106.4,107.4]   &    3.1             &    1.9             \\
  G030.641$-$00.117   &    114.2      &    11.7    &    114.5      &    4.3     &    [112.4,112.9]   &    [114.9,115.9]   &    2.1             &    1.4             \\
  G030.648$-$00.119   &    114.1      &    7.7     &    114.4      &    2.8     &    [112.4,113.4]   &    [114.9,115.9]   &    2.0             &    1.5             \\
  G030.651$-$00.204   &    90.7       &    5.0     &    90.5       &    1.9     &    [83.4,89.4]     &    [93.4,100.9]    &    7.1             &    10.4            \\
  G030.659$+$00.229   &    100.5      &    5.1     &    100.4      &    2.9     &    [98.4,99.4]     &    [101.4,101.9]   &    2.0             &    1.5             \\
  G030.663$-$00.144   &    116.2      &    5.4     &    116.0      &    1.7     &    [113.4,115.9]   &    [117.9,118.4]   &    2.6             &    2.4             \\
  G030.683$-$00.074   &    91.7       &    7.5     &    92.0       &    6.1     &    [84.4,91.4]     &    [93.9,98.9]     &    7.6             &    6.9             \\
  G030.684$-$00.261   &    103.2      &    8.2     &    103.7      &    4.5     &    [98.9,101.4]    &    [104.9,107.4]   &    4.8             &    3.7             \\
  G030.693$-$00.149   &    91.5       &    4.9     &    91.5       &    1.7     &    [88.4,89.4]     &    [92.4,95.4]     &    3.1             &    3.9             \\
  G030.691$-$00.05   &    91.5       &    8.9     &    90.9       &    2.6     &    [76.4,88.4]     &    [96.4,105.9]    &    14.5            &    15.0            \\
  G030.703$-$00.067   &    92.2       &    15.1    &    91.0       &    4.2     &    [82.9,89.4]     &    [96.4,104.4]    &    8.1             &    13.4            \\
  G030.691$+$00.22    & 104.6    & 6.8  &104.8   & 4.8   & [102.8,103.8] &     $-$ &  2.0 &  - \\  
  G030.718$-$00.082   &    93.2       &    7.5     &    93.1       &    4.3     &    [85.9,91.9]     &    [96.4,102.4]    &    7.2             &    9.3             \\%
  G030.731$-$00.079   &    92.4       &    7.5     &    91.1       &    3.3     &    [83.4,90.4]     &    [94.9,99.4]     &    7.7             &    8.3             \\%%
                 \hline
\end{tabular}
\end{table*}    
\addtocounter{table}{-1}
\begin{table*}
%\scriptsize
\centering
\caption[]{{ \it \rm --continuum $\rm{^{13}CO}$ outflow calculations of all blue and red wings for 225 ATLASGAL clumps}}
\begin{tabular}{lllllllll}
\hline
\hline
ATLASGAL            &$\rm{^{13}CO}\,v_p$ & $\rm{^{13}CO}\,T_{mb}$ & $\rm C^{18}O\,v_p$ & $\rm C^{18}O\,T_{mb}$ & $\rm \Delta{V_b}$ &  $\rm \Delta{V_r}$ &  $\rm V_{max_{b}}$ & $\rm V_{max_{r}}$ \\
CSC Gname          &   ($\rm km\,s^{-1}$)    & (K)                                      &   ($\rm km\,s^{-1}$)    & (K) &   ($\rm km\,s^{-1}$) &   ($\rm km\,s^{-1}$)  &   ($\rm km\,s^{-1}$)    &   ($\rm km\,s^{-1}$) \\ 
\hline

  G030.741$-$00.061   &    93.6       &    8.5     &    93.1       &    3.6     &    [80.9,90.9]     &    [96.9,106.4]    &    12.2            &    13.3            \\
  G030.746$-$00.001   &    91.0       &    3.1     &    91.9       &    1.1     &    [75.9,87.4]     &    [92.4,107.9]    &    16.0            &    16.0            \\
  G030.753$-$00.051   &    93.6       &    13.6    &    91.5       &    6.6    &    [85.9,91.4]     &    [95.9,100.4]    &    5.6             &    8.9             \\%
  G030.756$+$00.206   &    99.0       &    3.0     &    99.5       &    2.1     &    [95.9,97.4]     &    [100.4,101.9]   &    3.6             &    2.4             \\
  G030.763$-$00.031   &    94.3       &    8.2     &    94.1       &    3.6     &    [77.9,92.9]     &    [95.9,110.4]    &    16.2            &    16.3            \\
  G030.766$-$00.046   &    92.3       &    15.5    &    89.8       &    3.9     &    [77.4,89.9]     &    [95.9,106.4]    &    12.4            &    16.6            \\
  G030.769$-$00.087   &    94.2       &    7.0     &    94.2       &    2.8     &    [88.9,93.4]     &    [96.4,103.9]    &    5.3             &    9.7             \\
  G030.773$-$00.216   &    103.7      &    3.6     &    103.8      &    1.5     &    [96.9,102.9]    &    [105.9,111.9]   &    6.9             &    8.1             \\
  G030.784$-$00.021   &    94.4       &    6.6     &    94.3       &    2.0     &    [77.4,90.9]     &    [97.4,111.9]    &    16.9            &    17.6            \\
  G030.786$+$00.204   &    81.8       &    5.9     &    81.9       &    3.8     &    [73.9,80.4]     &    [83.4,89.4]     &    8.0             &    7.5             \\
  G030.813$-$00.024   &    95.6       &    8.3     &    95.4       &    3.5     &    [89.4,94.9]     &    [96.9,100.4]    &    6.0             &    5.0             \\
  G030.818$+$00.274   &    97.9       &    6.3     &    97.9       &    3.2     &    [94.9,97.4]     &    [98.9,100.4]    &    3.0             &    2.5             \\
  G030.818$-$00.056   &    97.3       &    9.2     &    96.9       &    2.5     &    [85.4,95.9]     &    [100.9,106.4]   &    11.5            &    9.5             \\
  G030.819$-$00.081   &    94.9       &    5.4     &    94.8       &    2.0     &    [92.4,94.4]     &    [98.4,100.4]     &    2.4             &    5.6             \\
  G030.823$-$00.156   &    104.3      &    3.7     &    104.2      &    1.9     &    [88.9,103.4]    &    [106.9,113.9]   &    15.3            &    9.7             \\%
  G030.828$+$00.134   &    38.0       &    5.7     &    37.8       &    2.4     &    [35.4,36.9]     &    [39.4,40.4]     &    2.4             &    2.6             \\
  G030.828$-$00.122   &    51.3       &    7.2     &    51.6       &    3.8     &    [48.4,50.4]     &    [51.9,54.4]     &    3.2             &    2.8             \\
  G030.839$-$00.019   &    93.0       &    4.9     &    92.4       &    1.7     &    [84.9,92.4]     &    [94.9,97.9]     &    7.5             &    5.5             \\
  G030.853$-$00.109   &    99.4       &    2.9     &    100.2      &    2.2     &    [93.4,96.4]     &    [100.9,105.4]   &    6.8             &    5.2             \\
  G030.866$+$00.114   &    39.4       &    10.3     &    39.3       &    3.1    &    [35.9,37.4]     &    [41.9,44.9]     &    3.4             &    5.6             \\%
  G030.866$-$00.119   &    99.8       &    2.5     &    100.7      &    1.0     &    [84.9,98.4]     &    [101.4,106.9]   &    15.8            &    6.2             \\
  G030.874$-$00.094   &    100.8      &    9.6     &    101.2      &    5.3     &    [95.9,99.4]     &    [101.9,104.4]   &    5.3             &    3.2             \\
  G030.886$-$00.231   &    111.0      &    8.9     &    111.2      &    3.0     &    [106.4,110.4]   &    [111.4,113.9]   &    4.8             &    2.7             \\
  G030.898$+$00.162   &    105.6      &    6.2     &    105.6      &    1.9     &    [91.4,103.9]    &    [105.9,109.9]   &    14.2            &    4.3             \\
  G030.901$-$00.034   &    75.4       &    7.4     &    74.9       &    3.8     &    [73.4,74.4]     &    [76.9,77.4]     &    1.5             &    2.5             \\
  G030.959$+$00.086   &    39.8       &    6.9     &    39.6       &    2.4     &    [34.7,39.2]     &    [40.7,45.2]     &    4.9             &    5.6             \\%
  G030.971$-$00.141   &    77.8       &    2.8     &    77.6       &    0.9     &    [75.4,77.4]     &    [78.9,81.4]     &    2.2            &    3.8             \\
  G030.978$+$00.216   &    107.8      &    4.9     &    108.0      &    3.5     &    [105.7,106.2]   &    [109.2,110.7]   &    2.3             &    2.7             \\
  G030.996$-$00.076   &    81.6       &    14.3    &    81.7       &    5.2     &    [75.9,79.9]     &    [81.9,84.4]     &    5.8             &    2.7             \\
  G031.024$+$00.262   &    96.2       &    3.0     &    96.1       &    1.2     &    [91.7,95.7]     &    [96.7,98.2]     &    4.4             &    2.1             \\
  G031.046$+$00.357   &    77.0       &    7.7     &    77.0       &    4.0     &    [72.7,76.2]     &    [78.2,79.2]     &    4.3             &    2.2             \\
  G031.071$+$00.049   &    38.2       &    9.7     &    37.9       &    2.6     &    [34.7,37.2]     &    [39.2,41.2]     &    3.2             &    3.3             \\
  G031.121$+$00.062   &    42.1       &    7.5     &    42.1       &    3.3     &    [36.2,41.7]     &    [43.2,46.2]     &    5.9             &    4.1             \\
  G031.148$-$00.149   &    41.7       &    2.6     &    42.2       &    0.9     &    [39.7,40.7]     &    [43.2,44.2]     &    2.5             &    2.0             \\
  G031.158$+$00.047   &    39.0       &    5.7     &    39.2       &    2.7     &    [34.2,38.2]     &    [42.7,43.7]     &    5.0             &    4.5             \\%
  G031.208$+$00.101   &    108.1      &    7.1     &    108.1      &    3.3     &    [106.2,107.7]   &    [108.7,110.2]   &    1.9             &    2.1             \\
  G031.243$-$00.111   &    20.6       &    10.5     &    21.5       &    4.1    &    [16.7,18.7]     &    [23.7,25.7]     &    4.8             &    4.2             \\%
  G031.281$+$00.062   &    108.3      &    6.5     &    108.8      &    2.9     &    [101.2,105.7]   &    [110.7,113.7]   &    7.6             &    4.9             \\%
  G031.386$-$00.269   &    87.4       &    7.1     &    86.5       &    4.6     &    [84.7,86.7]     &    [88.7,92.2]     &    1.8             &    5.7             \\
  G031.396$-$00.257   &    87.1       &    19.2    &    86.6       &    8.4    &    [81.7,85.7]     &    [87.7,92.7]     &    4.9             &    6.1             \\%
  G031.412$+$00.307   &    97.7       &    5.9     &    97.4       &    3.4     &    [92.7,95.7]     &    [100.2,102.2]   &    4.7             &    4.8             \\%
  G031.542$-$00.039   &    44.8       &    3.5     &    44.5       &    1.1     &    [43.2,44.2]     &    [45.7,46.7]     &    1.3             &    2.2             \\
  G031.568$+$00.092   &    96.2       &    6.5     &    96.2       &    3.0     &    [94.2,95.2]     &    [96.7,97.7]     &    2.0             &    1.5             \\
  G031.581$+$00.077   &    96.0       &    9.4     &    95.8       &    6.9     &    [91.7,95.2]     &    [98.2,101.7]    &    4.1             &    5.9             \\%
 G031.596$-$00.33   &    99.7       &    4.6     &    99.7       &    1.3     &    [95.2,98.7]     &    [101.2,104.7]    &    4.5             &    5.0            \\%the name of this source is wrong to be writted as G031.596+00.33, change it after the second version of arxiv.
  G031.644$-$00.266   &    43.9       &    6.5     &    43.9       &    2.8     &    [41.7,43.2]     &    [44.7,46.2]     &    2.2             &    2.3             \\
  G032.019$+$00.064   &    98.6       &    4.7     &    99.2       &    2.1     &    [90.4,96.9]     &    [99.4,101.4]    &    8.8             &    2.2             \\
  G032.044$+$00.059   &    95.1       &    7.0     &    95.3       &    1.8     &    $-$               &    [97.9,99.9]     &    $-$               &    4.6             \\%
  G032.117$+$00.091   &    96.5       &    10.6    &    96.2       &    5.5     &    [90.9,94.9]     &    [97.4,101.4]    &    5.3             &    5.2             \\
  G032.149$+$00.134   &    94.4       &    9.8    &    94.4       &    3.1     &    [90.4,91.9]     &    [95.4,98.4]     &    4.0             &    4.0             \\%
  G032.456$+$00.387   &    48.9       &    8.0     &    48.6       &    3.8     &    [46.3,48.3]     &    [49.3,52.3]     &    2.3             &    3.7             \\
  G032.471$+$00.204   &    49.2       &    6.1     &    49.4       &    2.1     &    [45.3,48.8]     &    [51.3,53.8]     &    4.1             &    4.4             \\%
  G032.604$-$00.256   &    90.2       &    7.5     &    90.4       &    3.4     &    [88.3,89.8]     &    [91.8,92.8]     &    2.1             &    2.4             \\
  G032.739$+$00.192   &    19.0       &    4.3     &    19.0       &    1.3     &    [14.3,17.3]     &    [19.8,23.3]     &    4.7             &    4.3             \\%
  G032.744$-$00.076   &    37.4       &    5.7     &    37.5       &    2.0     &    [30.3,35.8]     &    [39.3,42.8]     &    7.2             &    5.3             \\%
  G032.797$+$00.191   &    14.4       &    13.8   &    14.7       &    5.4    &    [6.8,11.8]      &    [16.3,21.8]     &    7.9             &    7.1             \\%
  G032.821$-$00.331   &    79.4       &    3.8     &    78.8       &    2.4     &    [75.3,78.3]     &    [80.8,85.3]     &    3.5             &    6.5            \\
  G032.990$+$00.034   &    82.6       &    5.3     &    82.5       &    3.2     &    [78.3,81.3]     &    [83.3,88.3]     &    4.2             &    5.8             \\
  G033.133$-$00.092   &    76.5       &    8.1     &    76.3       &    3.9     &    [69.8,74.3]     &    [77.8,82.8]     &    6.5             &    6.5             \\%
  G033.203$+$00.019   &    101.0      &    4.7     &    101.1      &    2.5     &    [98.3,99.3]     &    [101.3,102.8]   &    2.8             &    1.7             \\
  G033.206$-$00.009   &    99.7       &    5.7     &    99.6       &    2.8     &    [96.3,97.8]     &    [100.3,102.8]   &    3.3             &    3.2             \\
  G033.264$+$00.067   &    98.6       &    5.1     &    98.8       &    3.7     &    [96.3,97.8]     &    [100.3,100.8]   &    2.5             &    2.0             \\
  G033.288$-$00.019   &    99.4       &    7.5     &    99.4       &    5.3     &    [97.8,98.3]     &    [100.3,102.8]   &    1.6             &    3.4             \\
  G033.338$+$00.164   &    85.1       &    11.7    &    85.2       &    4.7     &    [82.8,84.8]     &    [85.8,87.3]     &    2.4             &    2.1             \\
  G033.388$+$00.199   &    85.4       &    8.1     &    85.1       &    3.9     &    [83.3,84.3]     &    [85.8,87.3]     &    1.8             &    2.2             \\
  G033.389$+$00.167   &    9.3        &    5.6     &    9.4        &    2.9     &    [6.8,8.8]       &    [9.8,11.8]      &    2.6             &    2.4             \\
  G033.393$+$00.011   &    103.2      &    5.1     &    103.5      &    1.8     &    [94.3,102.8]    &    [104.8,107.8]   &    9.2             &    4.3             \\
  G033.416$-$00.002   &    74.6       &    7.1     &    74.4       &    1.5     &    [71.2,73.7]     &    [76.7,77.7]     &    3.2             &    3.3             \\%
  G033.418$+$00.032   &    103.4      &    8.8     &    103.4      &    4.3     &    [101.2,102.7]   &    [104.2,105.7]   &    2.2             &    2.3             \\
  G033.651$-$00.026   &    103.9      &    5.0     &    104.2      &    2.8     &    [102.7,103.2]   &    [105.7,107.7]   &    1.5             &    3.5             \\
  G033.739$-$00.021   &    105.7      &    2.1     &    105.8      &    1.0     &    [102.7,104.7]   &    [107.2,108.7]   &    3.1             &    2.9             \\
  G033.809$-$00.159   &    52.2       &    10.2    &    52.7       &    3.4     &    [50.2,51.2]     &    [52.7,54.2]     &    2.5             &    1.5             \\
  G033.914$+$00.109   &    107.7      &    11.4    &    107.6      &    4.9    &    [102.9,106.9]   &    [108.9,112.9]   &    4.7             &    5.3             \\%
  G034.096$+$00.017   &    57.1       &    3.4     &    57.6       &    2.1     &    [52.9,55.4]     &    [57.9,61.4]     &    4.7             &    3.8             \\%
  G034.221$+$00.164   &    57.5       &    8.8     &    57.7       &    3.4     &    [54.9,56.4]     &    [59.4,62.4]     &    2.8             &    4.7             \\
  G034.229$+$00.134   &    57.4       &    14.4    &    57.7       &    6.0     &    [51.9,56.4]     &    [58.4,64.9]     &    5.8             &    7.2             \\
  G034.241$+$00.107   &    56.1       &    18.7    &    56.3       &    7.5     &    [52.4,54.9]     &    [56.4,58.9]     &    3.9             &    2.6             \\
  G034.243$+$00.132   &    56.9       &    16.4    &    57.0       &    6.9     &    [51.9,55.4]     &    [57.4,62.9]     &    5.1             &    5.9             \\
  G034.244$+$00.159   &    58.3       &    19.3    &    58.0       &    5.7     &    [49.9,56.4]     &    [61.4,67.4]     &    8.1             &    9.4             \\
               \hline
\end{tabular}
\end{table*}    
\addtocounter{table}{-1}
\begin{table*}
%\scriptsize
\centering
\caption[]{{ \it \rm --continuum $\rm{^{13}CO}$ outflow calculations of all blue and red wings for 225 ATLASGAL clumps}}
\begin{tabular}{lllllllll}
\hline
\hline
ATLASGAL            &$\rm{^{13}CO}\,v_p$ & $\rm{^{13}CO}\,T_{mb}$ & $\rm C^{18}O\,v_p$ & $\rm C^{18}O\,T_{mb}$ & $\rm \Delta{V_b}$ &  $\rm \Delta{V_r}$ &  $\rm V_{max_{b}}$ & $\rm V_{max_{r}}$ \\
CSC Gname          &   ($\rm km\,s^{-1}$)    & (K)                                      &   ($\rm km\,s^{-1}$)    & (K) &   ($\rm km\,s^{-1}$) &   ($\rm km\,s^{-1}$)  &   ($\rm km\,s^{-1}$)    &   ($\rm km\,s^{-1}$) \\ 
\hline

  G034.258$+$00.109   &    55.2       &    17.2    &    54.5       &    7.0     &    [52.4,54.4]     &    [56.4,59.4]     &    2.1             &    4.9             \\%
  G034.258$+$00.154   &    58.5       &    30.6    &    57.7       &    10.6    &    [49.9,56.4]     &    [62.9,66.4]     &    7.8             &    8.7             \\%
  G034.258$+$00.166   &    58.2       &    13.0    &    58.8       &    6.0     &    [52.9,56.9]     &    [62.4,65.4]     &    5.9             &    6.6             \\
  G034.261$+$00.176   &    58.7       &    5.8     &    58.6       &    2.5     &    [51.4,57.9]     &    [60.4,64.9]     &    7.2             &    6.3             \\
  G034.273$+$00.141   &    58.6       &    10.0    &    58.9       &    3.5     &    [54.9,57.9]     &    [59.4,62.4]     &    4.0             &    3.5             \\
  G034.391$+$00.214   &    57.3       &    4.8     &    57.5       &    1.5     &    $-$               &    [59.4,60.4]     &    $-$               &    2.9             \\
  G034.411$+$00.234   &    57.8       &    7.6     &    58.1       &    2.7     &    [53.4,56.4]     &    [58.9,61.4]     &    4.7             &    3.3             \\
  G034.459$+$00.247   &    58.7       &    3.8     &    58.8       &    1.0     &    [55.9,58.4]     &    [59.9,61.4]     &    2.9             &    2.6             \\
  G035.026$+$00.349   &    53.1       &    8.7     &    52.6       &    5.4     &    [49.1,51.1]     &    [53.6,58.1]     &    3.5             &    5.5             \\%
  G035.344$+$00.347   &    94.5       &    8.9     &    94.7       &    3.7     &    [92.1,93.6]     &    [95.6,96.6]     &    2.6             &    1.9             \\
  G035.457$-$00.179   &    65.1       &    7.4     &    65.2       &    2.5     &    [62.2,63.7]     &    [66.7,67.2]     &    3.0             &    2.0             \\
  G035.466$+$00.141   &    76.8       &    8.9     &    77.3       &    3.7     &    [74.2,76.2]     &    [78.2,80.2]     &    3.1             &    2.9             \\%
  G035.497$-$00.021   &    58.0       &    5.6     &    58.0       &    1.6     &    [52.7,57.2]     &    [59.2,62.2]     &    5.3             &    4.2             \\
  G035.522$-$00.274   &    45.4       &    3.7     &    45.1       &    1.1     &    [43.7,45.2]     &    [46.2,47.7]     &    1.4             &    2.6             \\
  G035.577$+$00.047   &    50.0       &    7.9     &    48.5       &    1.8     &    [40.2,48.7]     &    [54.7,58.2]     &    8.3             &    9.7             \\
  G035.577$+$00.067   &    49.8       &    6.5     &    50.1       &    2.2     &    [45.2,48.2]     &    [50.7,57.2]     &    4.9             &    7.1             \\
  G035.579$-$00.031   &    52.9       &    11.3    &    53.0       &    2.6    &    [45.7,48.7]     &    [53.7,59.2]     &    7.3             &    6.2             \\%
  G035.602$+$00.222   &    49.6       &    4.7     &    49.6       &    1.7     &    [47.7,48.7]     &    [51.2,52.2]     &    1.9             &    2.6             \\
  G035.681$-$00.176   &    28.3       &    3.8     &    28.1       &    1.1     &    [26.7,27.2]     &    [29.2,30.2]     &    1.4             &    2.1             \\
  G036.406$+$00.021   &    57.8       &    11.2    &    57.8       &    5.4    &    [54.2,56.2]     &    [60.2,62.7]     &    3.6             &    4.9             \\%
  G036.794$-$00.204   &    78.3       &    4.1     &    78.1       &    2.1     &    [75.7,77.7]     &    [78.7,80.2]     &    2.4             &    2.1             \\
  G036.826$-$00.039   &    60.2       &    5.8     &    60.6       &    2.8     &    [58.2,59.2]     &    [61.2,62.2]     &    2.4             &    1.6             \\
  G037.043$-$00.036   &    81.5       &    4.8     &    81.3       &    2.2     &    [79.5,80.5]     &    [82.5,83.0]     &    1.8             &    1.7             \\
  G037.268$+$00.081   &    91.1       &    6.0     &    91.5       &    3.3     &    $-$               &    [92.5,94.5]     &    $-$               &    3.0             \\
  G037.374$-$00.236   &    39.2       &    4.9     &    40.0       &    2.5     &    [35.5,37.5]     &    [40.5,43.5]     &    4.5             &    3.5             \\
  G037.546$-$00.112   &    52.7       &    12.8    &    52.7       &    2.7    &    [49.0,50.5]     &    [53.5,57.0]     &    3.7             &    4.3             \\%
  G037.672$-$00.091   &    47.9       &    4.4     &    48.2       &    2.2     &    [46.5,47.5]     &    [48.5,49.0]     &    1.7             &    0.8             \\
  G037.734$-$00.112   &    46.1       &    7.5     &    45.6       &    2.7     &    [43.5,44.0]     &    [47.5,49.5]     &    2.1             &    3.9             \\%
  G037.819$+$00.412   &    17.5       &    8.9     &    17.2       &    3.1     &    [14.6,16.1]     & [19.6,21.1]     &    2.6             &    3.9               \\%
  G037.874$-$00.399   &    60.7       &    11.9    &    60.8       &    3.6    &    [52.1,55.1]     &    [66.1,68.6]     &    8.7             &    7.8             \\%
  G038.037$-$00.041   &    55.9       &    10.3    &    56.0       &    3.5     &    [53.1,55.6]     &    [56.6,58.1]     &    2.9             &    2.1             \\
  G038.119$-$00.229   &    83.5       &    9.5     &    83.1       &    5.0     &    [80.1,82.6]     &    [85.1,87.1]     &    3.0             &    4.0             \\
  G038.646$-$00.226   &    69.1       &    11.8    &    68.7       &    6.7    &    [66.6,68.6]     &    [69.6,72.1]     &    2.1             &    3.4             \\%
  G038.917$-$00.402   &    40.8       &    11.4    &    40.7       &    3.4     &    [39.3,39.8]     &    [41.3,41.8]     &    1.4             &    1.1             \\
  G038.921$-$00.351   &    38.8       &    11.9    &    39.0       &    3.1     &    [35.3,37.8]     &    [39.8,40.8]     &    3.7             &    1.8             \\
  G038.934$-$00.361   &    39.7       &    11.8    &    40.0       &    3.2     &    [37.3,37.8]     &    [40.3,42.3]     &    2.7             &    2.3             \\
  G038.937$-$00.457   &    41.6       &    11.5    &    41.6       &    4.9     &    [38.8,41.3]     &    [42.3,43.8]     &    2.8             &    2.2             \\
  G038.957$-$00.466   &    42.2       &    5.7     &    42.0       &    2.4     &    [40.3,41.3]     &    [43.3,44.3]     &    1.7             &    2.3             \\
  G039.591$-$00.204   &    64.5       &    8.9     &    64.5       &    2.7     &    [63.1,63.6]     &    [65.1,66.6]     &    1.4             &    2.1             \\
  G039.851$-$00.204   &    57.3       &    4.0     &    56.6       &    2.3     &    [55.1,56.6]     &    [58.6,59.6]     &    1.5             &    3.0             \\
  G039.884$-$00.346   &    58.2       &    7.7     &    58.3       &    2.4     &    [54.1,57.1]     &    [60.6,62.6]     &    4.2             &    4.3             \\%
  G040.283$-$00.219   &    73.9       &    9.8     &    73.8       &    3.9     &    [66.1,72.6]     &    [77.1,77.6]     &    7.7             &    3.8             \\%
  G040.622$-$00.137   &    32.8       &    5.5     &    32.6       &    1.6     &    [26.2,31.2]     &    [33.7,38.7]     &    6.4             &    6.1             \\
  G040.814$-$00.416   &    80.3       &    3.1     &    80.2       &    2.3     &    [78.7,79.2]     &    [81.7,83.2]     &    1.5             &    3.0             \\
  G041.031$-$00.226   &    60.8       &    10.8    &    60.9       &    3.7     &    [57.2,60.2]     &    [62.2,63.7]     &    3.7             &    2.8             \\
  G041.226$-$00.197   &    59.5       &    8.0     &    59.5       &    2.0     &    [52.2,57.7]     &    [59.7,68.7]     &    7.3             &    9.2             \\
  G041.307$-$00.171   &    57.7       &    3.4     &    57.1       &    1.0     &    [51.2,56.7]     &    $-$               &    5.9             &    $-$               \\
  G041.507$-$00.106   &    63.0       &    4.0     &    62.8       &    1.5     &    [61.6,62.1]     &    [64.1,64.6]     &    1.2             &    1.8             \\
  G042.108$-$00.447   &    55.2       &    8.9     &    54.7       &    2.7     &    [52.3,54.8]     &    [56.3,59.3]     &    2.4             &    4.6             \\%
  G043.038$-$00.452   &    57.8       &    9.2     &    57.8       &    4.0     &    [54.3,56.8]     &    [59.3,63.3]     &    3.5             &    5.5             \\
  G043.108$+$00.044   &    12.5       &    3.7     &    12.8       &    1.1     &    [8.3,10.3]       &    [13.8,16.3]     &    4.5             &    3.5             \\
  G043.124$+$00.031   &    9.3        &    5.8     &    6.8        &    1.1     &    [0.8,8.3]       &    [10.3,17.3]     &    6.0             &    10.5            \\
  G043.148$+$00.014   &    7.0        &    16.8    &    9.5        &    1.7    &    [$-$2.2,$-$0.2]     &    [10.3,16.3]     &    11.7            &    6.8             \\%
  G043.164$-$00.029   &    13.7       &    11.2    &    13.7       &    2.3    &    [2.3,11.3]      &    [17.3,25.3]     &    11.4            &    11.6            \\%
  G043.236$-$00.047   &    6.3        &    7.6     &    6.8        &    1.9     &    [3.3,4.3]       &    [9.8,10.8]      &    3.5             &    4.0             \\%
  G043.306$-$00.212   &    59.2       &    7.8     &    59.4       &    4.4     &    [57.3,58.3]     &    [61.3,62.8]     &    2.1             &    3.4             \\%%
  G043.519$+$00.016   &    62.9       &    5.3     &    62.9       &    2.5     &    [61.3,61.8]     &    [64.3,64.8]     &    1.6             &    1.9             \\
  G043.528$+$00.017   &    61.6       &    6.2     &    62.5       &    3.4     &    $-$               &    [63.8,64.3]     &    $-$               &    1.8             \\
  G043.794$-$00.127   &    44.1       &    13.5    &    43.7       &    6.9    &    [36.8,42.3]     &    [45.8,52.3]     &    6.9             &    8.6             \\%
  G043.817$-$00.119   &    46.3       &    3.6     &    46.4       &    2.0     &    [42.8,45.3]     &    [47.3,48.8]     &    3.6             &    2.4             \\
  G044.309$+$00.041   &    56.9       &    4.8     &    56.8       &    1.9     &    [53.2,54.7]     &    [58.2,61.2]     &    3.6             &    4.4             \\%
  G045.071$+$00.132   &    58.4       &    13.3    &    58.6       &    3.0    &    [50.9,53.4]     &    [61.9,63.9]     &    7.7             &    5.3             \\%
  G045.121$+$00.131   &    58.7       &    14.0    &    58.8       &    3.5    &    [52.9,57.4]     &    [60.9,64.9]     &    5.9             &    6.1             \\%
  G045.454$+$00.061   &    58.5       &    9.1     &    58.7       &    2.8     &    [52.1,57.6]     &    [60.6,63.6]     &    6.6             &    4.9             \\%
  G045.463$+$00.027   &    58.3       &    3.5     &    58.0       &    1.3     &    [54.1,57.1]     &    [59.1,61.6]     &    3.9             &    3.6             \\
  G045.466$+$00.046   &    60.7       &    7.9     &    61.5       &    2.2     &    [54.1,58.6]     &    [62.1,67.6]     &    7.4             &    6.1             \\%
  G045.474$+$00.134   &    62.1       &    11.5    &    61.5       &    3.1     &    [54.6,60.1]     &    $-$               &    6.9             &    $-$               \\
  G045.543$-$00.007   &    55.5       &    10.9    &    55.8       &    2.5     &    [53.1,53.6]     &    [56.6,59.1]     &    2.7             &    3.3             \\
  G045.544$-$00.032   &    55.5       &    9.8     &    55.3       &    3.7     &    [53.1,55.1]     &    [56.1,57.1]     &    2.2             &    1.8             \\
  G045.804$-$00.356   &    59.2       &    6.5     &    58.4       &    4.5     &    [54.9,58.4]     &    [60.4,63.4]     &    3.5             &    5.0             \\
  G045.829$-$00.292   &    60.8       &    11.5    &    60.9       &    6.1     &    [59.4,60.4]     &    [61.4,62.4]     &    1.5             &    1.5             \\
  G046.118$+$00.399   &    55.3       &    4.9     &    55.6       &    3.1     &    [53.9,54.4]     &    [56.9,57.9]     &    1.7             &    2.3             \\
\hline
\hline
\end{tabular}
\begin{tablenotes} 
\item 
$\star$Noted: we add three clumps in table 9, which won't influence the discussion and results in the paper.
\end{tablenotes}
\label{total_outflow_wings}
\end{table*}    
%%%%table2 end%%%%

%%%%table3%%%%
\begin{table*}
\centering
\caption[]{{ \it  $\rm{^{13}CO}$ Outflow Properties of All Blue and Red Lobes for 153 ATLASGAL Clumps: Blue/Red Lobe Length $ l_{b/r}[\rm pc]$, 
 Masses $ M_{\rm b}(\rm blue)$,  $ M_{\rm r}(\rm red)$, $ M_{\rm out}(M_{\rm out} = M_{\rm b} + M_{\rm r}) [M_\odot]$, momentum $ p [10\,M_\odot\,\rm km\,s^{-1}]$, energy $\rm E [10^{39}\,J]$, dynamic time $ t_{\rm d}[\rm 10^4\,yr]$, 
 mass entrainment rates $ \dot{M}_{\rm out} [10^{-4}\,M_\odot/yr]$, mechanical force $ F_{\rm CO} [10^{-3}\,M_\odot\,km\,s^{-1}/yr]$, 
 and mechanical luminosity $ L_{\rm CO} [L_\odot]$.  }}
 \scriptsize
\begin{tabular}{llllllllllll}
\hline
\hline
ATLASGAL            &$ \ell_{\rm b}$ & $ \ell_{\rm r} $ & $ M_{\rm b}$ & $ M_{\rm r}$ & $ M_{\rm out}$ &  $p$ &  E & $ t_{\rm d}$ & $\dot{ M}_{\rm out}$ &  $   F_{\rm CO}$ & $   L_{\rm CO}$  \\
 \footnotesize CSC Gname            &  ($\rm pc$) & ($\rm pc) $ &\scriptsize ($ M_\odot$) &\scriptsize ($ M_\odot$) &\scriptsize ($ M_\odot$) &\scriptsize  ($ 10\,M_\odot\,\rm km\,s^{-1}$) 
 &\scriptsize  ($\rm 10^{39}\,J$) &\scriptsize ($\rm 10^4 yr$) &\scriptsize ($ 10^{-4}M_\odot \rm yr^{-1}$) &\scriptsize ($ 10^{-3}M_\odot\,\rm km\,s^{-1}\,yr^{-1}$) &\scriptsize ($ L_\odot$)  \\
\hline
G027.784$+$00.057 & 1.1 & 0.6 & 39.4 & 5.4 & 44.8 & 20.8 & 2.4 & 14.8 & 2.9 & 1.2 & 1.2 \\
G027.903$-$00.012 & 0.8 & 1.0 & 18.8 & 18.8 & 37.6 & 14.0 & 0.8 & 24.5 & 1.5 & 0.6 & 0.28 \\
G027.919$-$00.031 & 0.5 & 0.5 & 3.0 & 9.5 & 12.4 & 4.6 & 0.16 & 16.0 & 0.7 & 0.2 & 0.08 \\
G027.936$+$00.206 & 0.2 & 0.2 & 1.9 & 3.2 & 5.1 & 3.8 & 0.4 & 3.1 & 1.6 & 1.2 & 0.8 \\
G027.978$+$00.077 & 0.5 & 1.0 & 7.4 & 13.3 & 20.7 & 12.6 & 1.2 & 16.1 & 1.2 & 0.8 & 0.4 \\
G028.148$-$00.004 & 0.5 & 0.6 & 5.4 & 8.5 & 13.9 & 5.0 & 0.32 & 14.6 & 0.9 & 0.4 & 0.16 \\
G028.151$+$00.171 & 0.6 & 1.2 & 6.0 & 2.7 & 8.7 & 4.0 & 0.28 & 25.5 & 0.3 & 0.14 & 0.08 \\
G028.199$-$00.049 & 0.8 & 1.5 & 83.5 & 86.0 & 169.5 & 176.0 & 38.8 & 9.7 & 16.8 & 16.6 & 30.8 \\
G028.234$+$00.062 & 0.4 & 0.5 & 8.2 & 6.9 & 15.0 & 6.6 & 0.36 & 15.9 & 0.9 & 0.4 & 0.2 \\
G028.244$+$00.012 & 1.0 & 1.0 & 26.0 & 25.5 & 51.5 & 28.2 & 2.0 & 21.8 & 2.3 & 1.2 & 0.8 \\
G028.288$-$00.362 & 1.9 & 1.5 & 312.5 & 193.9 & 506.5 & 353.4 & 54.8 & 18.3 & 26.6 & 17.8 & 23.2 \\
G028.301$-$00.382 & 1.7 & 2.2 & 118.0 & 118.0 & 236.0 & 117.0 & 9.2 & 32.6 & 7.0 & 3.2 & 2.0 \\
G028.321$-$00.009 & 0.5 & 0.9 & 13.9 & 10.7 & 24.6 & 10.6 & 0.8 & 18.3 & 1.3 & 0.6 & 0.36 \\
G028.438$+$00.036 & 0.6 & 0.9 & 12.7 & 14.2 & 26.9 & 10.2 & 0.4 & 21.2 & 1.2 & 0.4 & 0.2 \\
G028.469$-$00.282 & 1.4 & 1.2 & 37.5 & 54.7 & 92.2 & 29.8 & 1.2 & 40.9 & 2.2 & 0.6 & 0.24 \\
G028.608$+$00.019 & 0.7 & 0.7 & 53.9 & 59.0 & 112.8 & 99.8 & 11.6 & 6.3 & 17.2 & 14.6 & 14.0 \\
G028.608$-$00.027 & 0.2 & 0.2 & 2.6 & 1.3 & 3.9 & 0.6 & 0.04 & 5.3 & 0.7 & 0.1 & 0.08 \\
G028.802$-$00.022 & 2.0 & 1.5 & 5.6 & 12.1 & 17.7 & 16.6 & 1.2 & 31.6 & 0.5 & 0.4 & 0.28 \\
G028.831$-$00.252 & 0.5 & 0.8 & 18.0 & 21.4 & 39.4 & 44.8 & 4.0 & 6.4 & 5.9 & 6.4 & 4.8 \\
G028.861$+$00.066 & 1.0 & 0.6 & 4.1 & 96.7 & 100.8 & 245.4 & 6.8 & 9.1 & 10.6 & 24.8 & 5.6 \\
G028.881$-$00.021 & 0.8 & 0.7 & 33.2 & 10.3 & 43.6 & 15.2 & 1.2 & 19.7 & 2.1 & 0.8 & 0.4 \\
G028.974$+$00.081 & 1.3 & 0.9 & 28.7 & 21.3 & 50.1 & 6.4 & 0.4 & 33.0 & 1.5 & 0.18 & 0.08 \\
G029.002$+$00.067 & 1.6 & 1.4 & 101.3 & 92.7 & 194.1 & 89.2 & 7.2 & 47.4 & 3.9 & 1.8 & 1.2 \\
G029.476$-$00.179 & 0.8 & 0.8 & 15.9 & 30.3 & 46.2 & 27.4 & 1.6 & 15.2 & 2.9 & 1.6 & 0.8 \\
G029.937$-$00.052 & 1.6 & 1.6 & 45.8 & 32.9 & 78.7 & 104.0 & 17.2 & 11.4 & 6.6 & 8.4 & 11.6 \\
G029.954$-$00.016 & 0.7 & 1.3 & 143.5 & 118.7 & 262.2 & 177.8 & 28.4 & 11.6 & 21.7 & 14.0 & 18.8 \\
G029.959$-$00.067 & 0.9 & 0.6 & 8.7 & 18.2 & 27.0 & 6.8 & 0.8 & 12.5 & 2.1 & 0.4 & 0.4 \\
G030.010$+$00.034 & 1.3 & 1.1 & 6.2 & 10.6 & 16.9 & 14.8 & 2.8 & 23.9 & 0.7 & 0.6 & 0.8 \\
G030.019$-$00.047 & 0.4 & 0.5 & 53.7 & 10.1 & 63.8 & 17.0 & 4.0 & 3.5 & 17.5 & 4.4 & 8.8 \\
G030.023$+$00.106 & 0.8 & 1.2 & 10.6 & 16.7 & 27.3 & 16.0 & 0.8 & 16.0 & 1.6 & 1.0 & 0.4 \\
G030.029$+$00.117 & 1.1 & 0.9 & 8.4 & 11.6 & 20.0 & 7.2 & 0.4 & 22.9 & 0.8 & 0.2 & 0.12 \\
G030.094$+$00.047 & 0.8 & 0.9 & 18.2 & 9.5 & 27.7 & 14.0 & 1.2 & 20.1 & 1.3 & 0.6 & 0.4 \\
G030.224$-$00.179 & 1.5 & 1.2 & 12.2 & 91.8 & 103.9 & 106.2 & 3.2 & 17.9 & 5.6 & 5.4 & 1.6 \\
G030.299$-$00.202 & 0.9 & 0.4 & 2.2 & 49.5 & 51.7 & 234.6 & 2.4 & 8.7 & 5.7 & 24.6 & 2.0 \\
G030.348$+$00.392 & 0.6 & 1.3 & 6.4 & 4.9 & 11.3 & 5.0 & 0.28 & 34.6 & 0.3 & 0.14 & 0.04 \\
G030.386$-$00.104 & 0.8 & 0.5 & 38.7 & 12.6 & 51.3 & 26.0 & 2.0 & 18.1 & 2.7 & 1.4 & 0.8 \\
G030.399$-$00.296 & 1.0 & 0.7 & 6.2 & 15.6 & 21.8 & 18.8 & 2.0 & 11.1 & 1.9 & 1.6 & 1.2 \\
G030.419$-$00.231 & 0.7 & 0.7 & 14.7 & 27.9 & 42.6 & 26.0 & 3.2 & 4.7 & 8.7 & 5.0 & 5.2 \\
G030.534$+$00.021 & 1.7 & 1.3 & 491.2 & 108.5 & 599.6 & 391.8 & 60.8 & 13.4 & 43.0 & 26.8 & 34.8 \\
G030.588$-$00.042 & 1.6 & 1.0 & 313.3 & 148.1 & 461.3 & 484.4 & 82.4 & 12.3 & 35.9 & 36.0 & 51.6 \\
G030.624$+$00.169 & 0.8 & 1.0 & 11.7 & 6.2 & 17.9 & 9.8 & 0.8 & 23.2 & 0.7 & 0.4 & 0.24 \\
G030.648$-$00.119 & 1.0 & 1.1 & 7.9 & 6.8 & 14.7 & 5.8 & 0.24 & 37.6 & 0.4 & 0.14 & 0.04 \\
G030.651$-$00.204 & 0.5 & 0.9 & 14.2 & 8.8 & 23.0 & 19.2 & 3.6 & 5.9 & 3.7 & 3.0 & 4.8 \\
G030.659$+$00.229 & 1.1 & 1.1 & 11.7 & 13.8 & 25.5 & 10.0 & 0.4 & 37.2 & 0.7 & 0.2 & 0.08 \\
G030.663$-$00.144 & 0.9 & 0.5 & 23.4 & 3.7 & 27.2 & 8.2 & 0.4 & 22.1 & 1.2 & 0.4 & 0.16 \\
G030.684$-$00.261 & 0.6 & 0.7 & 23.5 & 23.7 & 47.2 & 36.0 & 4.4 & 10.0 & 4.5 & 3.4 & 3.2 \\
G030.693$-$00.149 & 1.4 & 1.1 & 10.2 & 18.0 & 28.2 & 26.8 & 1.6 & 23.5 & 1.1 & 1.0 & 0.4 \\
G030.703$-$00.067 & 1.9 & 1.0 & 45.2 & 23.0 & 68.1 & 71.6 & 16.0 & 10.2 & 6.4 & 6.4 & 12.4 \\
G030.753$-$00.051 & 1.0 & 0.8 & 46.6 & 33.4 & 80.0 & 148.6 & 16.8 & 7.7 & 10.0 & 17.8 & 16.8 \\
G030.756$+$00.206 & 0.6 & 0.9 & 7.1 & 16.6 & 23.7 & 13.2 & 1.2 & 18.0 & 1.3 & 0.6 & 0.4 \\
G030.763$-$00.031 & 0.8 & 1.3 & 65.8 & 68.1 & 133.9 & 252.4 & 80.4 & 4.6 & 27.9 & 50.0 & 134.4 \\
G030.773$-$00.216 & 1.0 & 0.5 & 17.7 & 27.6 & 45.3 & 32.0 & 4.8 & 7.5 & 5.8 & 4.0 & 5.2 \\
G030.784$-$00.021 & 0.9 & 1.4 & 16.8 & 34.8 & 51.6 & 73.4 & 19.2 & 4.8 & 10.4 & 14.2 & 31.2 \\
G030.786$+$00.204 & 0.8 & 0.5 & 27.2 & 25.2 & 52.4 & 51.2 & 8.0 & 5.7 & 8.8 & 8.2 & 10.8 \\
G030.818$+$00.274 & 0.6 & 0.6 & 14.6 & 9.4 & 24.0 & 9.4 & 0.8 & 13.1 & 1.8 & 0.6 & 0.4 \\
G030.828$+$00.134 & 0.3 & 0.7 & 4.9 & 3.1 & 8.0 & 3.8 & 0.28 & 15.8 & 0.5 & 0.2 & 0.12 \\
G030.828$-$00.122 & 0.2 & 0.3 & 3.0 & 3.9 & 6.9 & 3.2 & 0.2 & 6.3 & 1.0 & 0.4 & 0.24 \\
G030.839$-$00.019 & 1.2 & 1.5 & 6.9 & 57.9 & 64.8 & 193.6 & 6.0 & 13.9 & 4.5 & 12.8 & 3.2 \\
G030.866$+$00.114 & 0.9 & 1.6 & 146.3 & 244.2 & 390.5 & 277.2 & 29.6 & 15.9 & 23.6 & 16.0 & 14.4 \\
G030.866$-$00.119 & 1.3 & 1.1 & 30.4 & 18.6 & 49.0 & 70.6 & 18.0 & 7.2 & 6.5 & 9.0 & 19.2 \\
G030.874$-$00.094 & 0.5 & 0.7 & 27.6 & 24.3 & 51.9 & 41.4 & 5.2 & 9.7 & 5.1 & 4.0 & 4.0 \\
G030.886$-$00.231 & 2.8 & 1.2 & 80.2 & 54.3 & 134.5 & 85.2 & 8.4 & 43.9 & 2.9 & 1.8 & 1.6 \\
G030.898$+$00.162 & 0.6 & 0.7 & 13.0 & 12.1 & 25.2 & 31.4 & 8.0 & 4.3 & 5.7 & 6.8 & 14.4 \\
G030.901$-$00.034 & 0.5 & 0.6 & 4.7 & 1.7 & 6.3 & 1.8 & 0.12 & 16.2 & 0.4 & 0.1 & 0.04 \\
G030.959$+$00.086 & 1.6 & 1.7 & 202.2 & 225.6 & 427.8 & 237.0 & 24.0 & 15.1 & 27.2 & 14.4 & 12.0 \\
G030.971$-$00.141 & 0.5 & 0.9 & 9.7 & 12.2 & 21.9 & 9.0 & 0.8 & 17.0 & 1.2 & 0.4 & 0.32 \\
G030.978$+$00.216 & 1.2 & 0.4 & 7.5 & 8.6 & 16.1 & 11.2 & 0.4 & 28.9 & 0.5 & 0.4 & 0.16 \\
G030.996$-$00.076 & 0.8 & 1.0 & 17.6 & 57.3 & 74.9 & 41.8 & 3.2 & 13.6 & 5.3 & 2.8 & 1.6 \\
G031.024$+$00.262 & 0.8 & 1.5 & 7.8 & 9.4 & 17.2 & 5.8 & 0.4 & 26.7 & 0.6 & 0.2 & 0.12 \\
G031.046$+$00.357 & 1.0 & 1.0 & 19.1 & 21.3 & 40.5 & 17.6 & 1.2 & 18.7 & 2.1 & 0.8 & 0.4 \\
G031.071$+$00.049 & 0.4 & 0.5 & 7.7 & 9.5 & 17.2 & 8.2 & 0.4 & 8.8 & 1.9 & 0.8 & 0.4 \\
G031.121$+$00.062 & 0.5 & 0.5 & 10.3 & 8.4 & 18.7 & 11.0 & 1.2 & 6.4 & 2.8 & 1.6 & 1.6 \\
G031.148$-$00.149 & 0.5 & 0.6 & 1.5 & 1.9 & 3.4 & 1.4 & 0.08 & 14.7 & 0.2 & 0.1 & 0.04 \\
G031.158$+$00.047 & 0.3 & 0.4 & 2.6 & 3.5 & 6.2 & 5.8 & 0.8 & 4.6 & 1.3 & 1.2 & 1.2 \\
G031.208$+$00.101 & 1.2 & 0.8 & 33.4 & 7.9 & 41.3 & 22.2 & 2.0 & 22.6 & 1.8 & 1.0 & 0.8 \\
G031.243$-$00.111 & 1.7 & 2.1 & 270.2 & 79.5 & 349.7 & 321.4 & 38.0 & 26.8 & 12.5 & 11.0 & 10.8 \\
G031.281$+$00.062 & 1.5 & 0.9 & 70.5 & 33.1 & 103.6 & 99.2 & 13.6 & 14.3 & 7.0 & 6.4 & 7.6 \\
G031.396$-$00.257 & 1.2 & 1.7 & 173.0 & 253.4 & 426.4 & 320.6 & 31.2 & 14.7 & 27.9 & 20.0 & 16.4 \\
G031.412$+$00.307 & 0.6 & 0.9 & 42.7 & 29.7 & 72.3 & 61.0 & 8.8 & 8.0 & 8.7 & 7.0 & 8.4 \\
   \hline
\end{tabular}
\end{table*}    
\addtocounter{table}{-1}
\begin{table*}
\centering
\caption[]{{ \it \rm --continuum$\rm{^{13}CO}$ outflow properties of all blue and red lobes for 153 ATLASGAL clumps }}
 \scriptsize
\begin{tabular}{llllllllllll}
\hline
\hline
ATLASGAL            &$ l_b$ & $ l_r $ & $ M_{\rm b}$ & $ M_{\rm r}$ & $ M_{\rm out}$ &  {  p} &  {  E} & $ t_{\rm d}$ & $\dot{ M}_{\rm out}$ &  $\rm   F_{CO}$ & $\rm   L_{CO}$  \\
 \footnotesize CSC Gname            &  ($\rm pc$) & ($\rm pc) $ &\scriptsize ($\rm M_\odot$) &\scriptsize ($\rm M_\odot$) &\scriptsize ($\rm M_\odot$) &\scriptsize  ($\rm 10\,M_\odot\,km\,s^{-1}$) 
 &\scriptsize  ($\rm 10^{39}\,J$) &\scriptsize ($\rm 10^4\,yr$) &\scriptsize ($\rm 10^{-4}M_\odot/yr$) &\scriptsize ($\rm 10^{-3}M_\odot\,km\,s^{-1}/yr$) &\scriptsize ($\rm L_\odot$)  \\
\hline
G031.542$-$00.039 & 0.3 & 0.3 & 1.1 & 1.1 & 2.2 & 0.6 & 0.028 & 10.5 & 0.2 & 0.06 & 0.024 \\
G031.568$+$00.092 & 0.8 & 0.8 & 9.2 & 12.2 & 21.4 & 7.6 & 0.32 & 26.6 & 0.8 & 0.2 & 0.08 \\
G031.581$+$00.077 & 0.7 & 0.7 & 72.0 & 32.1 & 104.2 & 53.8 & 6.4 & 7.3 & 13.7 & 6.8 & 6.8 \\
G031.644$-$00.266 & 0.3 & 0.3 & 2.8 & 1.9 & 4.7 & 1.6 & 0.08 & 8.3 & 0.5 & 0.18 & 0.08 \\
G032.117$+$00.091 & 0.6 & 0.8 & 42.5 & 68.0 & 110.5 & 77.2 & 7.2 & 9.3 & 11.4 & 7.6 & 6.0 \\
G032.149$+$00.134 & 0.6 & 0.8 & 39.4 & 57.6 & 97.0 & 84.8 & 6.4 & 12.1 & 7.7 & 6.4 & 4.0 \\
G032.456$+$00.387 & 0.4 & 0.4 & 5.3 & 5.1 & 10.5 & 3.8 & 0.16 & 8.4 & 1.2 & 0.4 & 0.16 \\
G032.471$+$00.204 & 1.0 & 2.0 & 66.8 & 33.4 & 100.2 & 76.6 & 9.6 & 21.1 & 4.6 & 3.4 & 3.6 \\
G032.604$-$00.256 & 1.3 & 0.8 & 24.2 & 4.9 & 29.1 & 11.0 & 0.8 & 34.6 & 0.8 & 0.2 & 0.12 \\
G032.744$-$00.076 & 1.1 & 0.9 & 110.4 & 52.5 & 162.9 & 164.4 & 25.2 & 11.4 & 13.7 & 13.2 & 16.8 \\
G032.797$+$00.191 & 1.3 & 1.6 & 841.3 & 950.3 & 1791.6 & 2288.6 & 385.2 & 10.0 & 172.3 & 210.0 & 297.2 \\
G032.821$-$00.331 & 0.3 & 0.4 & 7.9 & 7.4 & 15.3 & 14.0 & 2.8 & 2.7 & 5.4 & 4.8 & 7.6 \\
G032.990$+$00.034 & 1.2 & 1.7 & 75.3 & 95.7 & 170.9 & 136.0 & 10.8 & 19.6 & 8.4 & 6.4 & 4.4 \\
G033.133$-$00.092 & 1.0 & 0.8 & 86.9 & 104.9 & 191.7 & 180.4 & 20.8 & 8.4 & 21.8 & 19.6 & 19.2 \\
G033.264$+$00.067 & 1.1 & 0.9 & 25.5 & 10.1 & 35.6 & 16.8 & 1.2 & 27.7 & 1.2 & 0.6 & 0.32 \\
G033.288$-$00.019 & 0.7 & 0.5 & 11.1 & 13.7 & 24.8 & 11.2 & 0.8 & 16.8 & 1.4 & 0.6 & 0.28 \\
G033.338$+$00.164 & 0.5 & 0.7 & 23.8 & 14.1 & 37.9 & 9.8 & 0.4 & 18.3 & 2.0 & 0.4 & 0.2 \\
G033.388$+$00.199 & 0.7 & 0.7 & 10.7 & 19.6 & 30.3 & 15.0 & 0.4 & 20.3 & 1.4 & 0.6 & 0.16 \\
G033.389$+$00.167 & 1.8 & 2.0 & 42.9 & 32.0 & 74.9 & 23.4 & 1.2 & 46.0 & 1.6 & 0.4 & 0.24 \\
G033.393$+$00.011 & 1.0 & 1.3 & 39.3 & 25.8 & 65.1 & 45.8 & 6.8 & 11.3 & 5.5 & 3.8 & 4.8 \\
G033.416$-$00.002 & 0.8 & 2.0 & 43.3 & 79.3 & 122.6 & 62.0 & 4.0 & 35.6 & 3.3 & 1.6 & 0.8 \\
G033.418$+$00.032 & 0.6 & 0.9 & 16.8 & 25.1 & 41.8 & 14.0 & 0.8 & 24.6 & 1.6 & 0.6 & 0.24 \\
G033.651$-$00.026 & 0.8 & 1.4 & 18.8 & 8.8 & 27.6 & 12.8 & 0.8 & 32.5 & 0.8 & 0.4 & 0.2 \\
G033.809$-$00.159 & 0.5 & 0.4 & 5.7 & 10.6 & 16.3 & 5.8 & 0.32 & 16.0 & 1.0 & 0.4 & 0.16 \\
G033.914$+$00.109 & 0.5 & 0.6 & 49.1 & 41.9 & 91.1 & 51.0 & 4.8 & 6.6 & 13.3 & 7.2 & 6.0 \\
G034.096$+$00.017 & 0.4 & 0.2 & 2.0 & 3.2 & 5.2 & 3.8 & 0.4 & 5.4 & 0.9 & 0.6 & 0.4 \\
G034.258$+$00.154 & 0.3 & 0.4 & 36.9 & 9.0 & 45.9 & 44.4 & 8.0 & 2.5 & 17.5 & 16.2 & 24.4 \\
G034.459$+$00.247 & 0.1 & 0.2 & 1.4 & 1.4 & 2.8 & 1.0 & 0.08 & 4.2 & 0.6 & 0.2 & 0.12 \\
G035.026$+$00.349 & 1.1 & 1.8 & 71.7 & 236.6 & 308.4 & 433.2 & 22.4 & 20.3 & 14.6 & 19.6 & 8.4 \\
G035.344$+$00.347 & 0.5 & 0.4 & 7.2 & 4.4 & 11.6 & 5.0 & 0.32 & 13.4 & 0.8 & 0.4 & 0.2 \\
G035.457$-$00.179 & 0.8 & 0.5 & 8.9 & 2.2 & 11.1 & 7.8 & 0.8 & 18.7 & 0.6 & 0.4 & 0.28 \\
G035.466$+$00.141 & 0.9 & 1.8 & 62.5 & 50.1 & 112.6 & 55.4 & 4.0 & 34.9 & 3.1 & 1.4 & 0.8 \\
G035.577$+$00.047 & 1.3 & 2.1 & 284.5 & 57.7 & 342.2 & 238.6 & 48.4 & 13.8 & 23.7 & 15.8 & 26.8 \\
G035.579$-$00.031 & 0.8 & 0.7 & 67.8 & 102.8 & 170.5 & 297.8 & 40.4 & 6.2 & 26.5 & 44.2 & 50.8 \\
G035.602$+$00.222 & 0.3 & 0.3 & 1.1 & 1.0 & 2.1 & 0.8 & 0.04 & 6.9 & 0.3 & 0.12 & 0.08 \\
G035.681$-$00.176 & 0.5 & 0.5 & 0.5 & 0.9 & 1.4 & 0.6 & 0.024 & 15.6 & 0.08 & 0.04 & 0.012 \\
G036.406$+$00.021 & 0.5 & 0.4 & 18.2 & 10.4 & 28.6 & 21.4 & 2.4 & 7.1 & 3.9 & 2.8 & 2.4 \\
G036.826$-$00.039 & 0.6 & 0.4 & 3.3 & 2.3 & 5.6 & 2.6 & 0.16 & 16.7 & 0.3 & 0.14 & 0.08 \\
G037.043$-$00.036 & 0.7 & 0.3 & 9.2 & 3.0 & 12.2 & 5.0 & 0.24 & 23.1 & 0.5 & 0.2 & 0.08 \\
G037.374$-$00.236 & 0.2 & 0.4 & 5.0 & 6.1 & 11.2 & 8.8 & 0.8 & 6.2 & 1.7 & 1.4 & 1.2 \\
G037.546$-$00.112 & 1.3 & 0.9 & 56.1 & 156.1 & 212.2 & 251.0 & 8.8 & 21.8 & 9.3 & 10.6 & 3.2 \\
G037.672$-$00.091 & 1.6 & 1.2 & 34.9 & 20.6 & 55.5 & 15.6 & 0.8 & 75.4 & 0.7 & 0.2 & 0.08 \\
G037.734$-$00.112 & 1.2 & 1.8 & 74.7 & 97.7 & 172.4 & 85.0 & 8.4 & 33.0 & 5.0 & 2.4 & 2.0 \\
G037.819$+$00.412 & 0.9 & 1.1 & 74.4 & 42.0 & 116.3 & 81.4 & 6.8 & 19.5 & 5.7 & 3.8 & 2.8 \\
G037.874$-$00.399 & 1.1 & 1.1 & 87.1 & 48.2 & 135.3 & 305.8 & 94.0 & 6.0 & 21.7 & 46.8 & 120.8 \\
G038.037$-$00.041 & 0.6 & 0.6 & 16.5 & 9.1 & 25.6 & 10.4 & 0.8 & 14.1 & 1.7 & 0.6 & 0.4 \\
G038.119$-$00.229 & 0.7 & 0.4 & 38.9 & 12.4 & 51.3 & 25.4 & 2.0 & 11.9 & 4.2 & 2.0 & 1.2 \\
G038.646$-$00.226 & 0.7 & 0.5 & 19.5 & 21.6 & 41.0 & 13.0 & 0.8 & 15.7 & 2.5 & 0.8 & 0.32 \\
G038.917$-$00.402 & 0.4 & 0.4 & 4.2 & 5.4 & 9.6 & 3.4 & 0.08 & 18.9 & 0.5 & 0.16 & 0.04 \\
G038.921$-$00.351 & 0.5 & 0.4 & 8.0 & 4.3 & 12.3 & 7.0 & 0.4 & 10.7 & 1.1 & 0.6 & 0.36 \\
G038.937$-$00.457 & 0.5 & 0.6 & 6.3 & 2.8 & 9.1 & 2.8 & 0.16 & 14.8 & 0.6 & 0.18 & 0.08 \\
G038.957$-$00.466 & 0.4 & 0.3 & 4.1 & 1.5 & 5.7 & 2.4 & 0.12 & 10.3 & 0.5 & 0.2 & 0.08 \\
G039.851$-$00.204 & 1.9 & 1.9 & 74.2 & 52.5 & 126.7 & 29.4 & 2.8 & 50.8 & 2.4 & 0.6 & 0.4 \\
G039.884$-$00.346 & 1.1 & 1.6 & 66.6 & 42.1 & 108.7 & 71.0 & 7.6 & 19.3 & 5.4 & 3.4 & 3.2 \\
G040.283$-$00.219 & 1.1 & 1.6 & 79.6 & 6.3 & 85.9 & 79.2 & 10.4 & 16.3 & 5.1 & 4.4 & 4.8 \\
G040.622$-$00.137 & 1.5 & 0.9 & 120.3 & 125.9 & 246.2 & 202.2 & 21.2 & 13.9 & 17.0 & 13.4 & 12.0 \\
G040.814$-$00.416 & 0.2 & 0.3 & 1.5 & 1.4 & 2.9 & 1.4 & 0.08 & 7.9 & 0.3 & 0.16 & 0.08 \\
G041.031$-$00.226 & 1.4 & 1.3 & 67.5 & 12.0 & 79.5 & 31.4 & 2.0 & 25.0 & 3.1 & 1.2 & 0.4 \\
G041.507$-$00.106 & 1.8 & 2.3 & 25.6 & 8.6 & 34.2 & 10.6 & 0.4 & 89.0 & 0.4 & 0.1 & 0.036 \\
G042.108$-$00.447 & 1.1 & 1.6 & 122.0 & 109.8 & 231.8 & 77.2 & 8.4 & 22.7 & 9.8 & 3.2 & 2.8 \\
G043.124$+$00.031 & 2.8 & 2.8 & 376.0 & 228.2 & 604.2 & 460.2 & 82.4 & 19.8 & 29.4 & 21.4 & 32.0 \\
G043.148$+$00.014 & 2.2 & 2.0 & 287.6 & 478.3 & 765.8 & 2246.4 & 440.4 & 14.0 & 52.6 & 147.2 & 242.8 \\
G043.164$-$00.029 & 2.5 & 1.6 & 1145.1 & 920.1 & 2065.3 & 2964.6 & 786.4 & 12.1 & 164.5 & 225.2 & 502.8 \\
G043.236$-$00.047 & 1.1 & 3.0 & 336.8 & 245.3 & 582.1 & 488.6 & 74.8 & 32.2 & 17.3 & 13.8 & 18.0 \\
G043.306$-$00.212 & 0.4 & 0.3 & 9.3 & 4.3 & 13.6 & 6.6 & 0.4 & 8.0 & 1.6 & 0.8 & 0.4 \\
G043.794$-$00.127 & 1.1 & 1.0 & 268.5 & 235.0 & 503.5 & 513.0 & 76.8 & 7.7 & 62.6 & 60.8 & 76.8 \\
G044.309$+$00.041 & 0.8 & 0.5 & 22.3 & 12.8 & 35.1 & 23.8 & 2.8 & 11.1 & 3.0 & 2.0 & 2.0 \\
G045.071$+$00.132 & 0.3 & 0.5 & 21.7 & 43.8 & 65.5 & 114.4 & 21.2 & 3.1 & 20.2 & 33.6 & 52.0 \\
G045.121$+$00.131 & 1.0 & 1.0 & 123.1 & 90.2 & 213.3 & 247.0 & 46.4 & 5.8 & 35.2 & 39.0 & 61.6 \\
G045.454$+$00.061 & 0.9 & 1.9 & 146.6 & 156.1 & 302.8 & 256.2 & 35.2 & 16.9 & 17.2 & 14.0 & 16.0 \\
G045.543$-$00.007 & 0.8 & 1.1 & 16.3 & 41.2 & 57.5 & 53.2 & 2.0 & 22.1 & 2.5 & 2.2 & 0.8 \\
G045.544$-$00.032 & 1.1 & 1.1 & 38.2 & 18.9 & 57.1 & 16.0 & 0.8 & 32.6 & 1.7 & 0.4 & 0.16 \\
G045.804$-$00.356 & 1.0 & 0.7 & 35.3 & 30.9 & 66.2 & 37.4 & 3.2 & 13.4 & 4.7 & 2.6 & 2.0 \\
G045.829$-$00.292 & 1.3 & 1.6 & 29.4 & 29.2 & 58.6 & 14.2 & 0.4 & 63.7 & 0.9 & 0.2 & 0.08 \\

\hline
\hline
\end{tabular}
\label{total_outflow_phyparam}
\end{table*}    
%%%%end table3%%%

%\section{APPENDIX A} measured from clump coordinates to each outflow's extreme; derived from 870$\mu m$ flux density

\end{document}